\documentclass[prd,aps,twocolumn,a4paper,showkeys,nofootinbib,showpacs]{revtex4-1}

\usepackage[colorlinks=true,backref=true,linkcolor=blue,linktocpage=true,anchorcolor=black,citecolor=blue,filecolor=black,menucolor=black,pagecolor=black,urlcolor=blue]{hyperref}

\usepackage{graphicx,psfrag}
\usepackage{mathrsfs}
\usepackage{amsmath,amsfonts,amssymb}
\usepackage{multirow}
\usepackage{comment}
\usepackage{bm}
\usepackage{ulem}
\usepackage{hyperref}

\newcommand{\be}{\begin{equation}}
\newcommand{\ee}{\end{equation}}
\newcommand{\bea}{\begin{eqnarray}}
\newcommand{\eea}{\end{eqnarray}}
\newcommand{\bel}{\begin{align}}
\newcommand{\eel}{\end{align}}

\def\CC{{C\nolinebreak[4]\hspace{-.05em}\raisebox{.4ex}{\tiny\bf ++}}}

\def\p{\partial}

\def\Msun{M_{\odot}}
\def\GMc2{G M_{\odot} c^{-2}}

\def\M{\mathcal{M}}
\def\O{\mathcal{O}}
\def\vareps{\varepsilon}

\def\F{{\cal F}}
\def\lm{{\ell m}}

\def\lm{{\ell m}}
\def\p{\partial}

\def\de{\partial}
\def\lm{{\ell m}}

\def\ii{{\rm i}}
\def\l{{\ell }}

\def\F{{\cal F}}

\def\M{{\cal M}}

\def\O{{\cal O}}

\def\ha{{\hat{a}}}

\def\Msun{M_\odot}
\def\TEOBResum{\texttt{TEOBResum}}
\def\TEOBResumS{\texttt{TEOBResumS}}
\def\IMRPhenomD{\texttt{IMRPhenomD}}

\def\TEOBNNLO{\texttt{TEOBNNLO}}
\def\TEOBResumROM{\texttt{TEOBResum\_ROM}}
\def\SEOBNRvq{{\texttt{SEOBNRv4}}}
\def\SEOBNRvqT{{\texttt{SEOBNRv4T}}}

\newcommand\SEOBNRv[1]{\texttt{SEOBNRv{#1}}}
\def\BAM{{\texttt{BAM}}}
\def\THC{{\texttt{THC}}}

\DeclareSymbolFontAlphabet{\mathrsfs}{rsfs}
\DeclareMathAlphabet{\mathcal}{OMS}{cmsy}{m}{n}
\DeclareSymbolFontAlphabet{\mathrsfs}{rsfs}
\DeclareMathAlphabet\mathbfcal{OMS}{cmsy}{b}{n}

\usepackage{color}
\definecolor{cyan}{rgb}{0,0.9,0.9}
\definecolor{orange}{rgb}{0.9,0.5,0}
\definecolor{magenta}{rgb}{1,0,1}
\definecolor{purple}{rgb}{0.8,0.4,0.8}
\definecolor{gray}{rgb}{0.8242,0.8242,0.8242}
\definecolor{dodgerblue}{rgb}{0.12, 0.56, 1.0}

\begin{document}
\title{Time-domain effective-one-body gravitational waveforms \\
  for coalescing compact binaries with nonprecessing spins, tides and self-spin effects}
\author{Alessandro \surname{Nagar}$^{1,2,3}$}
\author{Sebastiano \surname{Bernuzzi}${}^{4,5,6}$}
\author{Walter \surname{Del Pozzo}$^{7}$}

\author{Gunnar \surname{Riemenschneider}$^{2,8}$}

\author{Sarp \surname{Akcay}$^{4}$}
\author{Gregorio \surname{Carullo}$^{7}$}
\author{Philipp \surname{Fleig}$^9$}
\author{Stanislav \surname{Babak}$^{10}$}
\author{Ka Wa \surname{Tsang}$^{12}$}

\author{Marta \surname{Colleoni}$^{13}$}
\author{Francesco \surname{Messina}$^{14,15}$}
\author{Geraint \surname{Pratten}$^{13}$}
\author{David  \surname{Radice}$^{16,17}$}
\author{Piero \surname{Rettegno}$^{2,8}$}
\author{Michalis \surname{Agathos}$^{18}$}
\author{Edward \surname{Fauchon-Jones}$^{19}$}
\author{Mark \surname{Hannam}$^{19}$}
\author{Sascha \surname{Husa}$^{13}$}
\author{Tim \surname{Dietrich}$^{12,20}$ }
\author{Pablo \surname{Cerd\'a-Duran$^{21}$}}
\author{Jos\'e A. \surname{Font}$^{21,22}$}
\author{Francesco \surname{Pannarale}$^{19,23}$}
\author{Patricia \surname{Schmidt}$^{24}$}
\author{Thibault \surname{Damour}$^3$}

%
\affiliation{${}^1$Centro Fermi - Museo Storico della Fisica e Centro Studi e Ricerche Enrico Fermi, Rome, Italy}
\affiliation{${}^2$INFN Sezione di Torino, Via P. Giuria 1, 10125 Torino, Italy}
\affiliation{${}^3$Institut des Hautes Etudes Scientifiques, 91440 Bures-sur-Yvette, France}
\affiliation{${}^4$Theoretisch-Physikalisches Institut, Friedrich-Schiller-Universit{\"a}t Jena, 07743, Jena, Germany}
\affiliation{${}^5$Istituto Nazionale di Fisica Nucleare, Sezione Milano Bicocca, gruppo collegato di Parma, I-43124 Parma, Italy}  
\affiliation{${}^6$Department of Mathematical, Physical and Computer Sciences, University of Parma, I-43124 Parma, Italy}  
\affiliation{${}^7$Dipartimento di Fisica ``Enrico Fermi'', Universit\`a di Pisa,  and INFN sezione di Pisa, Pisa I-56127, Italy}
\affiliation{${}^{8}$Dipartimento di Fisica, Universit\`a di Torino, via P. Giuria 1, I-10125 Torino, Italy}
\affiliation{${}^9$Max Planck Institute for Dynamics and Self-Organization, 37077 G\"ottingen, Germany}
\affiliation{${}^{10}$APC, CNRS-Universit\'e Paris 7, 75205 Paris CEDEX 13, France}
\affiliation{${}^{11}$Moscow Institute of Physics and Technology, Dolgoprudny, Moscow region, Russia}
\affiliation{${}^{12}$Nikhef, Science Park, 1098 XG Amsterdam, The Netherlands}
\affiliation{${}^{13}$Universitat de les Illes Balears, IAC3-IEEC, 07122 Palma de Mallorca, Spain}
\affiliation{${}^{14}$Dipartimento di Fisica, Universit\`a degli studi di Milano Bicocca, Piazza della Scienza 3, 20126 Milano, Italy}
\affiliation{${}^{15}$INFN, Sezione di Milano Bicocca, Piazza della Scienza 3, 20126 Milano, Italy}
\affiliation{${}^{16}$Department of Astrophysical Sciences, Princeton University, 4 Ivy Lane, Princeton NJ 08544, USA}
\affiliation{${}^{17}$Institute for Advanced Study, 1 Einstein Drive, Princeton NJ 08540, USA}
\affiliation{${}^{18}$DAMTP, Centre for Mathematical Sciences, University of Cambridge, Wilberforce Road, Cambridge CB3 0WA, UK}
\affiliation{${}^{19}$Gravity Exploration Institute, School of Physics and Astronomy, Cardiff University, The Parade, Cardiff CF24 3AA, UK}      
\affiliation{${}^{20}$Max Planck Institute for Gravitational Physics (Albert Einstein Institute), Am M\"uhlenberg 1, Potsdam 14476, Germany}
\affiliation{${}^{21}$Departament d'Astronomia i Astrof\'sica, Universitat de Val\`encia, Dr. Moline 50, 46100 Burjassot (Valencia)}
\affiliation{${}^{22}$Observatori Astron\`omic, Universitat de Val\`encia, C/ Catedr\'atico Jos\'e Beltr\'an 2, 46980, Paterna (Val\`encia), Spain} 
\affiliation{${}^{23}$Dipartimento di Fisica, Universit\`a di Roma ``Sapienza'' \& Sezione INFN Roma1, P.A.\,Moro 5, 00185, Roma, Italy}
\affiliation{${}^{24}$Department of Astrophysics/IMAPP, Radboud University Nijmegen, P.O. Box 9010, 6525 GL Nijmegen, The Netherlands}

\date{\today}

\begin{abstract}
We present \TEOBResumS{}, a new effective-one-body (EOB) 
waveform model for nonprecessing (spin-aligned) and tidally 
interacting compact binaries.
Spin-orbit and spin-spin effects are blended together by making use of
the concept of centrifugal EOB radius.
The point-mass sector through merger and
ringdown is informed by numerical relativity (NR) simulations of
binary black holes (BBH) computed with the {\tt SpEC} and \BAM{} codes.
An improved, NR-based phenomenological description of the postmerger
waveform is developed.
The tidal sector of \TEOBResumS{} describes the dynamics of neutron 
star binaries up to merger and incorporates a resummed attractive 
potential motivated by recent advances in the post-Newtonian and 
gravitational self-force description of relativistic tidal interactions.
Equation-of-state dependent self-spin interactions
(monopole-quadrupole effects) are incorporated in the model using
leading-order post-Newtonian results in a new expression of the centrifugal radius.
\TEOBResumS{} is compared to 135 {\tt SpEC} and 19 \BAM{} BBH waveforms.
The maximum unfaithfulness to {\tt SpEC} data 
$\bar{F}$ -- at design Advanced-LIGO sensitivity and evaluated with
total mass $M$ varying between $10M_\odot \leq M \leq 200 M_\odot$ --
is always below $2.5 \times 10^{-3}$ except for a single outlier that
grazes the $7.1 \times 10^{-3}$ level. When compared to \BAM{} data,
$\bar{F}$ is smaller than $0.01$ except for a single  
outlier in one of the corners of the NR-covered parameter space, 
that reaches the $0.052$ level.
\TEOBResumS{} is also compatible, up to merger, to high end NR waveforms from binary
neutron stars with spin effects and reduced initial eccentricity computed with the \BAM{}
and \THC{} codes. The data quality of binary neutron star waveforms 
is assessed via rigorous convergence tests from multiple resolution runs and takes
into account systematic effects estimated by using
the two independent high-order NR codes.
The model is designed to generate accurate templates for the analysis
of LIGO-Virgo data through merger and ringdown. We demonstrate its use
by analyzing the publicly available data for GW150914. 
\end{abstract}


\maketitle


\section{Introduction}

Analytical waveform models informed by (or calibrated to) numerical relativity (NR)
simulations are essential for the analysis of gravitational wave (GW)
events~\cite{Abbott:2016blz,Abbott:2016nmj,Abbott:2017vtc,TheLIGOScientific:2017qsa}.
The effective-one-body (EOB) approach to the general relativistic two-body
problem~\cite{Buonanno:1998gg,Buonanno:2000ef,Damour:2001tu,Damour:2000we} 
is a powerful analytical tool that reliably describes both the 
dynamics and gravitational waveform through inspiral, merger and ringdown
for BBHs~\cite{Bohe:2016gbl,Nagar:2017jdw,Cotesta:2018fcv}
and up to merger for BNSs~\cite{Dietrich:2017feu}. The analytical model is
crucially improved in the late-inspiral, strong-field, fast-velocity regime
by NR information, that allows one to properly represent
the merger and ringdown part of the waveform~\cite{Damour:2014yha,Bohe:2016gbl,Nagar:2017jdw}.
The synergy between EOB and NR creates EOBNR models, whose more recent
avatars implemented in publicly available LIGO Scientific Collaboration Algorithm Library (LAL)~\cite{LALSimulation} 
are~\SEOBNRvq/\SEOBNRvqT~\cite{Bohe:2016gbl,Cotesta:2018fcv},  
that describe nonprecessing binaries (both BNSs and BBHs) 
and {\tt SEOBNRv3}~\cite{Babak:2016tgq}, that incorporates precession for BBHs.
The purpose of this paper is to introduce \TEOBResumS{}, a state-of-the art
EOB model, informed by BBH NR simulations, that is fit to describe the dynamics
and waveforms from nonprecessing coalescing binaries, both black holes and neutron
stars. For BBH binaries, \TEOBResumS{} is an improvement of the model of
Refs.~\cite{Nagar:2017jdw,Nagar:2015xqa,Damour:2014sva} 
implementing a refined phenomenological representation of the 
postmerger waveform (ringdown). The latter is built from an effective 
fit of many spin-aligned NR waveform data available in the
SXS catalog~\cite{SXS:catalog} obtained with the {\tt SpEC}
code~\cite{Buchman:2012dw,Chu:2009md,Hemberger:2013hsa,
Scheel:2014ina,Blackman:2015pia,Lovelace:2011nu,Lovelace:2010ne,
Lovelace:2014twa,Mroue:2013xna,Kumar:2015tha,Chu:2015kft}
and, notably, also incorporates test-particle
results\footnote{In doing so, we corrected 
a minor coding error in the numerical implementation that had affected 
the $\ell=5$, $m=\text{odd}$ flux modes from Ref.~\cite{Damour:2014sva}.}. 
We show here the performance of the model over the SXS~\cite{SXS:catalog}
and \BAM{} waveform catalogs (the latter consisting of simulations produced using
the \BAM{} code~\cite{Bruegmann:2006at,Husa:2007hp}), and check its robustness also outside NR-covered
regions of the parameter space.

For BNSs, we built on our previous efforts~\cite{Bernuzzi:2014owa}
(see also~\cite{Kiuchi:2017pte,Kawaguchi:2018gvj})
and merged together into a single EOB code, tidal and spin effects,
so as to produce a complete waveform model of spinning BNSs. We show that
the EOB waveform is accurate up to BNS merger by comparing with
state-of-the art, high end, NR simulations. The tidal-and-spin model
uses most of the existing analytical knowledge. In particular, we
incorporate in the EOB model equation-of-state (EOS) dependent
self-spin effects at leading-order (also known as spin-induced quadrupole
moment or monopole-quadrupole couplings~\cite{Poisson:1997ha}).
\TEOBResumS{} has been the first EOB model to have these effects.
As such, it was used for validating  the phenomenological waveform model,
{\tt PhenomPv2\_NRTidal}, that incorporates similar self-spin effects~\cite{Dietrich:2018uni}
and that was recently used for a detailed study of the parameters
of GW170817~\cite{Abbott:2018wiz,Abbott:2018exr}. However,
while \TEOBResumS{} was under internal LVC review, leading-order self-spin
effects were also included in~\SEOBNRvqT, though in a different fashion for
what concerns the Hamiltonian~\cite{Bohe:2016gbl,Hinderer:2016eia,Steinhoff:2016rfi,Marsat-Vines-LIGO-tech-note}.
A targeted comparison between the two models for BNS configurations
is described in Sec.~\ref{sec:vsSEOBNRv4}.

This paper is organized as follows: in Sec.~\ref{sec:bbh} we
remind the reader the main theoretical features of the EOB model for
BBHs, compare its performance against the SXS~\cite{SXS:catalog}
and \BAM{} NR data, test its robustness over a large portion of
the parameter space; in Sec.~\ref{sec:bns} we discuss the BNS case,
focusing on our analytical strategy to incorporate in a
consistent, and resummed, way both tidal and spin effects,
including the self-spin ones. In this respect, Sec.~\ref{sec:MonQuad}
compares the EOB description with the corresponding nonresummed PN-based expressions.
Section~\ref{sec:vsSEOBNRv4} collects selected comparisons (photon
potential and, notably, faithfulness) between \TEOBResumS, \SEOBNRv4
and \SEOBNRv4{\tt T}.
To probe our model, that is implemented as publicly 
available C codes (see Appendix~\ref{sec:code}),
for production runs,  we also present, in Sec.~\ref{sec:pe},
a case study done on the GW150914 event~\cite{Abbott:2016blz}.
Conclusions are in Sec.~\ref{end}. The paper is complemented
by several technical appendices. Among these, the case of mixed
black-hole and neutron-star binaries is discussed in Appendix~\ref{sec:BHNS}.

We use units with $G=c=1$. In the following, the gravitational mass of
the binary is $M=M_A+M_B$, with the two bodies labeled as $(A,B)$.
We adopt the convention that that $M_A\geq M_B$, so as to define
the mass ratio $q\equiv M_A/M_B\geq 1$, the reduced mass
$\mu\equiv M_A M_B/M$, and the symmetric mass ratio
$\nu\equiv \mu/M$, that ranges from $0$ (test-particle limit) to
$\nu=1/4$ (equal-mass case). The dimensionless spins are addressed
as $\chi_{A,B}=S_{A,B}/M_{A,B}^2$. We also define the quantities $X_{A} =
M_{A}/M$ and $X_{AB}\equiv X_{A}-X_{B}=\sqrt{1-4\nu}$ (with $X_A\geq X_B$). 
As convenient spin variables we shall also use 
$\tilde{a}_{A,B}\equiv X_{A,B}\chi_{A,B}=S_{A,B}/(M_{A,B} M)$.

\section{Binary Black Holes}
\label{sec:bbh}

General relativity predicts that the GW signal from quasi-circular
inspiral-merger of BBHs is chirp-like~\cite{Maggiore:1900zz}.
The GW phase evolution at Newtonian
order, i.e. at large separations and low orbital frequencies, is driven by the value of
the chirp mass, ${\cal M}_c = (M_A M_B)^{3/5}/(M_A + M_B )^{1/5}$.
Higher post-Newtonian (PN) corrections depend on the
symmetric mass ratio $\nu$ as well as spin-orbit and spin-spin
couplings. The analytic description of the dynamics and waveform
for coalescing binaries is based on PN theory~\cite{Damour:2016abl,Bernard:2017ktp,Blanchet:2013haa}.
However, PN results, an expansion in
the small parameter $(v/c)^2$, where $v$ is the orbital velocity of the system,
are not apt to reliably describe the dynamics and waveform emitted
by the binary in the strong-field, fast-velocity regime typical of
the binary while it approaches the merger. The effective-one-body (EOB)
approach to the two-body general relativistic
dynamics~\cite{Buonanno:2000ef,Buonanno:1998gg,Damour:2000we,Damour:2001tu,
  Damour:2008qf,Barausse:2009xi,Damour:2009sm,Damour:2009wj,Damour:2016gwp,Vines:2017hyw,Damour:2017zjx}
builds upon post-Newtonian results, properly resummed,
so as to deliver a representation of the dynamics (and gravitational waveform)
that is reliable and predictive also close to this extreme dynamical
regime. Such a resummed description of the binary dynamics is further improved
by informing the analytical model with NR simulations.

\subsection{Main features}
\label{sec:mainfeats}

The EOB approach delivers a resummation of
the standard PN-expanded relative dynamics that is reliable
and predictive also in the strong-field, fast-velocity regime,
i.e. up to merger.
The relative dynamics is described by a Hamiltonian for
the conservative part and an angular momentum flux, that
accounts for the loss of angular momentum through gravitational
radiation. Both functions are given as special resummations of
the PN-expanded ones.
At a more technical level, it is worth
remembering that the comparable-mass EOB Hamiltonian is
a {\it continuous deformation}, $\nu$ being the deformation parameter,
of the Hamiltonian of a (spinning) particle in Kerr background.
For instance, for nonspinning binaries, it is a $\nu$-deformation
of the standard Hamiltonian of a test-particle on a Schwarzschild
metric. The effect of the $\nu$-dependent corrections is to make
the interaction potential more repulsive than in the simple
Schwarzschild case, allowing the system to inspiral and merge at higher
frequencies. This explains why a system of equal-mass BBHs merges
at frequencies that are higher than the case of a test-particle
plunging into a nonrotating black hole~\cite{Buonanno:1998gg}.
Spin-orbit and spin-spin couplings are similarly included in
the EOB Hamiltonian mimicking the structure they have in the
test-particle case~\cite{Damour:2014sva}.

Let us briefly review the structure of the \TEOBResumS{} model, more details can be found in 
Ref.~\cite{Damour:2014sva,Nagar:2015xqa,Nagar:2017jdw}.

The EOB {\it Hamiltonian} describes the conservative
part of the binary dynamics. The crucial functions that enter the Hamiltonian and that
mostly determine the attraction between the bodies are the EOB orbital interaction
potential $A(r)$, that is a $\nu$-dependent deformation of the Schwarzschild
potential $A^{\rm Schw}=1-2/r$ (where $r = c^{2} R/(GM)$ is the dimensionless
relative separation), and the gyro-gravitomagnetic functions $G_{S}$ and $G_{S_{*}}$,
that account for the spin-orbit interaction and are $\nu$-dependent deformations,
properly resummed, of the corresponding functions entering the Hamiltonian of a
spinning particle in Kerr background~\cite{Damour:2014sva}.
The spin-spin coupling was inserted, at next-to-leading order, in a special resummed
form involving the centrifugal radius $r_{c}$~\cite{Damour:2014sva} that mimics the
same structure present in the Hamiltonian of a test particle on a Kerr spacetime.

The relative dynamics is evolved using phase space dimensionless
variables $(r,p_r,\varphi,p_\varphi)$, associated to polar coordinates
in the equatorial plane $\theta=\pi/2$. We denote by $r$ the
relative separation. Its conjugate momentum, $p_r$ is replaced by
$p_{r_*} = (A/B)^{1/2} \ p_r$, with respect to the ``tortoise"
(dimensionless) radial coordinate $r_*=\int dr(A/B)^{-1/2}$,
where $A$ and $B$ are the EOB potentials. Their explicit expressions,
in the general spinning case, are given in Ref.~\cite{Damour:2014sva},
though we shall recall a few important elements below.
The dimensionless phase-space variables are related to
the dimensional ones $(R,P_R,\varphi,P_\varphi)$ by
\begin{equation}
r = \frac{R}{GM},\ p_{r_*}=\frac{P_{R_*}}{\mu},\ p_\varphi = \frac{P_\varphi}{\mu GM},\ t = \frac{T}{GM}. 
\end{equation}
The spin dependence in the spin-orbit sector of the EOB dynamics
is expressed using the following combinations of the individual spins
\begin{align}
S = &\ S_A + S_B,\\
S_* = & \frac{M_B}{M_A}S_A + \frac{M_A}{M_B}S_B. 
\end{align}
The $\mu$-rescaled EOB Hamiltonian is given by
\begin{equation}
\hat{H}_{\rm EOB} = \frac{H_{\rm EOB}}{\mu} = \frac1\nu \sqrt{1+2\nu\big(\hat{H}_{\rm eff}-1\big)},
\end{equation}
with
\begin{align}
\hat{H}_{\rm eff} = &\hat{H}_{\rm eff}^{\rm orb} + p_\varphi \big(G_S \hat{S} + G_{S_*} \hat{S}_* \big),\\
\hat{H}_{\rm eff}^{\rm orb} = &\sqrt{p_{r_*}^2+A\left(1+\frac{p_\varphi^2}{r_c^2}+z_3\frac{p_{r_*}^4}{r_c^2}\right)}.
\end{align}
Here, we introduced the dimensionless spin variables
$\hat{S}\equiv S/M^2$, $\hat{S}_*\equiv S_*/M^2$, 
$z_3 = 2\nu(4-3\nu)$ and $r_c$ is the centrifugal
radius~\cite{Damour:2014sva} that incorporates next-to-leading
(NLO) spin-spin terms~\cite{Hartung:2010jg}. It formally reads
\begin{equation}
\label{eq:rc2_nlo}
r_c^2 = r^2 + \ha_0^2\left(1+\dfrac{2}{r}\right) + \delta\ha^2,
\end{equation}
where $\hat{a}_0$ is the dimensionless effective Kerr parameter 
\begin{equation}
  \label{eq:hata0}
\ha_0 \equiv \hat{S}+\hat{S}_*=X_A\chi_A+X_B\chi_B=\tilde{a}_A + \tilde{a}_B \ ,\\
\end{equation}
and the NLO spin-spin contribution is included in the function $\delta \ha^2$ that
explicitly reads~\cite{Balmelli:2015lva,Damour:2014sva}
\begin{align}
  \label{eq:deltaa2}
\delta \ha^2 =& \dfrac{1}{r}\Bigg\{\dfrac{5}{4}(\tilde{a}_{A}-\tilde{a}_B)\hat{a}_0X_{AB}-\left(\dfrac{5}{4}+\dfrac{\nu}{2}\right)\hat{a}_0^2\nonumber\\
                 &+ \left(\dfrac{1}{2}+2\nu\right)\tilde{a}_A \tilde{a}_B\Bigg\}.
\end{align}
The quantities $G_S$ and $G_{S_*}$ entering the spin-orbit sector of the model
are the gyro-gravitomagnetic functions and determine the strength of the
spin-orbit coupling. Following Refs.~\cite{Nagar:2011fx,Damour:2014sva},
we work at next-to-next-to-leading order (NNLO)~\cite{Hartung:2011te}
in the spin-orbit coupling and we fix the Damour-Jaranowski-Sch{\"a}fer
gauge~\cite{Damour:2008qf,Nagar:2011fx},
so that $(G_S, G_{S_*})$ are only functions of $(r,p_{r_*}^2)$
and {\it not} of the angular momentum $p_\varphi$. 
This simplifies Hamilton's equations\footnote{Note that this gauge choice is not made
in \SEOBNRv4{\tt T}, that follows Ref.~\cite{Barausse:2011ys}.}, 
which formally read
\begin{subequations}
\begin{align}
\frac{d\varphi}{dt}=& \ \Omega = \frac{\p \hat{H}_{\rm EOB}}{\p p_\varphi},\\
\frac{dr}{dt}=& \ \Big(\frac{A}{B} \Big)^{1/2} \ \frac{\p \hat{H}_{\rm EOB}}{\p p_{r_*}},\\
\frac{dp_\varphi}{dt}=& \ \hat{\F}_\varphi,\\
\label{eq:dprdt}
\frac{dp_{r_*}}{dt}=& -\left(\frac{A}{B} \right)^{1/2} \ \frac{\p \hat{H}_{\rm EOB}}{\p r},
\end{align}
\end{subequations}
and explicitly become 
\begin{subequations}
\begin{align}
\label{eq:Omg-Omgorb}
\frac{d\varphi}{dt}=&\Omega= \ \frac{1}{\nu \hat{H}_{\rm EOB} \hat{H}_{\rm eff}^{\rm orb} }
\Big[A \frac{p_\varphi}{r_c^2}+\hat{H}_{\rm eff}^{\rm orb}\big(G_S {\hat S} + G_{S_*} {\hat S}_*\big)\Big],\\
\label{eq:drdt}
\frac{dr}{dt}=& \ \left(\frac{A}{B} \right)^{1/2} \frac{1}{\nu \hat{H}_{\rm EOB} \hat{H}_{\rm eff}^{\rm orb}} \Big[ p_{r_*} \Big(1+2z_3 \frac{A}{r_c^2} p_{r_*}^2\Big) +\nonumber \\
&+\hat{H}_{\rm eff}^{\rm orb} p_\varphi \Big(\frac{\p G_S}{\p p_{r_*}} \hat{S} + \frac{\p G_{S_*}}{\p p_{r_*}} \hat{S}_*\Big) \Big],\\
\frac{dp_\varphi}{dt}=& \ \hat{\F}_\varphi,\\
\label{eq:dprdt2}
\frac{dp_{r_*}}{dt}=& -\left(\frac{A}{B} \right)^{1/2} \frac{1}{2 \nu \hat{H}_{\rm EOB} \hat{H}_{\rm eff}^{\rm orb} } \Big[A'+ p_\varphi^2 \Big(\frac{A}{r_c^2}\Big)'+\nonumber \\
& + z_3 \ p_{r_*}^4\Big(\frac{A}{r_c^2}\Big)' +2 \hat{H}_{\rm eff}^{\rm orb} p_\varphi \Big(G_S' \hat{S} + G_{S_*}'\hat{S}_*\Big) \Big],
\end{align}
\end{subequations}
where the prime indicates the partial derivative with respect to $r$, i.e. $(\cdots)'\equiv \de_r(\cdots)$.
Above, $\hat{\F}_\varphi \equiv \F_\varphi / \mu$ denotes the {\it radiation reaction}
force entering the equation of motion of the angular momentum (that is not conserved)
and that relies on a special factorization and resummation of the multipolar
waveform~\cite{Damour:2008gu} (see below). Following the choice made in
previous work~\cite{Damour:2014sva}, we set $\hat{\F}_{r_*}=0$ explicitly, so that the radial flux does not appear in the r.h.s. of Eq.~\eqref{eq:dprdt}.
Note that the effect of the absorption due to the horizon is explicitly
included in the model at leading order (see Eqs.~(97)-(98) of~\cite{Damour:2014sva}).
The relative dynamics is initiated using post-post-adiabatic (2PA)
initial data~\cite{Damour:2007yf,Damour:2012ky}, as explicitly detailed
in Appendix~\ref{sec:ID}.

The multipolar waveform strain is computed out of the dynamics with
the following convention
\be
h_+-{\rm i}h_\times = \dfrac{1}{\cal R}\sum_{\ell=2}^{\ell_{\rm max}}\sum_{\ell = -m}^{\ell +m}h_{\lm}\,{}_{-2}Y_{\lm}(\theta,\phi),
\ee
where ${}_{-2}Y_{\lm}(\theta,\phi)$ are the $s= - 2$ spin-weighted spherical harmonics.
In the following text, for consistency with previous work, we shall often use the
Regge-Wheeler-Zerilli~\cite{Nagar:2005ea} normalized
waveform $\Psi_{\ell m}=h_{\ell m}/\sqrt{(\ell +2)(\ell+1)\ell(\ell -1)}$.
The strain multipoles $h_{\lm}$ are written in special factorized and resummed
form~\cite{Damour:2008gu,Pan:2010hz,Damour:2014sva}. Following the notation
of~\cite{Damour:2014sva}, they read
\be
h_\lm = h_\lm^{(N,\epsilon)}\hat{S}^{(\epsilon)}_{\rm eff} \hat{h}^{\rm tail}_{\lm} f_\lm \hat{h}^{\rm NQC}_\lm,
\ee
where $\epsilon$ denotes the parity of $\ell+m$, $h_\lm^{(N,\epsilon)}$ is the Newtonian (or leading-order)
contribution, $\hat{S}^{(\epsilon)}_{\rm eff}$ the effective source, $\hat{h}^{\rm tail}_{\lm}$ the
tail factor, $f_\lm$ the residual amplitude correction and $\hat{h}^{\rm NQC}_\lm$ the next-to-quasi-circular
(NQC) correction factor. We recall that $\hat{h}_\lm^{\rm NQC}$ accounts for corrections to
the circularized EOB waveform that explicitly depend on the radial momentum and that
are relevant during the plunge up to merger~\cite{Damour:2007xr}. For each $(\ell,m)$,
$\hat{h}_\lm^{\rm NQC}$ depends on 4 parameters that are NR-informed
by requiring osculation between the NR amplitude and frequency (and their first time derivatives)
close to merger (see Sec.~IIIA of~\cite{Nagar:2017jdw} and below for additional detail).
Then, for consistency between waveform and flux, the NQC factor also enters the radiation reaction
and one iterates the procedure a few times until the procedure converges.
We focus here only on the $\ell=m=2$ waveform mode. In this case, the NQC factor reads
\begin{equation}
  \label{eq:nqc_factor}
  \hat{h}_{22}^{\rm NQC} = (1 + a_1 n_1 + a_2 n_2)e^{{\rm i}(b_1 n'_1 + b_2 n'_2)},
\end{equation}
where $(a_1,a_2,b_1,b_2)$ are the free parameters while $(n_1,n_2,n_1',n_2')$ are
explicit functions of the radial momentum and its time derivative that are listed
in Eq.~(96) of Ref.~\cite{Damour:2014sva}. 
On the EOB time axis, $t$, the NQC parameters are determined at a time defined as
\be
\label{eq:tNQC}
t_{\rm NQC}^{\rm EOB} = t^{\rm peak}_{\rm \Omega_{\rm orb}}-\Delta t_{\rm NQC} \ ,
\ee
where $\Omega_{\rm orb}$ was called the {\it pure orbital frequency}
in Ref.~\cite{Damour:2014sva} (see Eq.~(100) there) and is defined,
from Eq.~\eqref{eq:Omg-Omgorb} above, as
\be
\Omega_{\rm orb}\equiv \dfrac{1}{H_{\rm EOB}}\dfrac{\de\hat{H}^{\rm eff}_{\rm orb}}{\de p_\varphi}=\dfrac{p_\varphi u_c^2 A}{H_{\rm EOB}\hat{H}^{\rm eff}_{\rm orb}},
\ee
where $u_c=1/r_c$. In previous work~\cite{Damour:2014sva,Nagar:2015xqa,Nagar:2017jdw},
it was found that $\Delta t_{\rm NQC}$ needed to be informed by NR simulations for large,
positive spins. In Sec.~\ref{sec:error} below we point out that this
was the result of a small, though nonnegligible, implementation mistake,
so that we fix $\Delta t_{\rm NQC}=1$ always except for some extreme
corners of the parameter space defined by Eqs.~\eqref{eq:m09}-\eqref{eq:m08} below,
where it is helpful to change $\Delta t_{\rm NQC}$ to obtain a qualitatively sane waveform.

On top of the NQC corrections to the waveform, \TEOBResumS{} is also
NR-informed in the nonspinning and spinning sector of the dynamics.
Section IIIA of Ref.~\cite{Nagar:2017jdw} gives a comprehensive summary
of the {\it analytical flexibility} of the model, while Sec.~IIIB and
IIIC of~\cite{Nagar:2017jdw} illustrate how the NR information is injected
in the model. The nonspinning sector of \TEOBResumS{} fully coincides with
Sec.~IIIB of Ref.~\cite{Nagar:2017jdw}: the orbital interaction potential $A$,
taken at formal 5PN order, is Pad\'e resummed with a $(1,5)$ Pad\'e approximant
and it incorporates an ``effective'' 5PN parameter $a_6^c(\nu)=3097.3\nu^2 - 1330.6\nu+81.38$
that was determined by EOB/NR comparisons with a set of nonspinning SXS simulations.
More precisely this specific functional form dates back to Ref.~\cite{Nagar:2015xqa},
it was based on the SXS NR simulations publicly available at the time
(see Table~I of~\cite{Nagar:2015xqa}) and never changed after. We address the
reader to Sec.~III of Ref.~\cite{Nagar:2015xqa} for details and in particular to
Eq.~(1) there for the explicit analytical form of the orbital interaction potential.

The spinning sector of the model is flexed by a single NNNLO 
effective spin-orbit parameter $c_3$ that enters both $G_S$ and $G_{S_*}$
(see e.g. Eqs.~(19)-(20) of~\cite{Nagar:2017jdw}). 
Finally, the factorized waveform is then complemented by a description of the
post-merger and ringdown phase~\cite{Damour:2014yha,Nagar:2016iwa}.
The model of~\cite{Nagar:2017jdw}, though informed by a rather sparse number of
NR simulations, proved to be rather accurate, reliable, and robust against a set
of 149 public NR simulations by the SXS collaboration~\cite{SXS:catalog}
(see specifically Tables~V-IX therein). It also showed, however, its drawbacks, mostly
restricted to the merger and post-merger part that was obtained through fit of
only a sparse number ($\approx 40$) of NR simulations, most of them clustered
around the equal-mass, equal-spin case. Here these problems are overcome by making
crucial use of {\it all} the NR information available in order to devise better
fits of the NR data to describe the post-merger-ringdown part of the waveform.
This will be discussed in the forthcoming section.

\subsection{Improvement over previous work}
\label{sec:error}

The BBH sector of the \TEOBResumS{} model improves the version
of the one discussed in Ref.~\cite{Damour:2014sva,Nagar:2015xqa,Nagar:2017jdw}
on the following aspects: (i) improved (and corrected) $\ell=5$ flux; (ii) related
new determination of the NNNLO spin-orbit parameter $c_3$; (iii) more robust
description of the postmerger and ringdown waveform; (iv) more robust and accurate
fits of the NR point used to determine the NQC waveform corrections.

\subsubsection{Flux multipoles: the $\ell=5$, $m=odd$  modes} 
We start the technical discussion of the BBH sector of \TEOBResumS{} by
pointing out a coding error in its {\tt Matlab} numerical implementation
that has affected (though marginally) the spin-dependent sector of the model
as soon as it was conceived back in 2013~\cite{Damour:2014sva}, 
with effects on Refs.~\cite{Damour:2014sva,Nagar:2015xqa,Nagar:2017jdw,Dietrich:2018uni}.
We found that there was a missing overall factor $X_{AB}=\sqrt{1-4\nu}$ in the $\ell=5$,
$m=$~odd multipolar waveform amplitudes that, once squared, contributed to 
the radiation reaction force $\hat{\cal F}_\varphi$.
Such small, though non-negligible, difference in the radiation reaction resulted
in an inconsistency between the nonspinning and spinning sector of the model,
that are implemented through a different set of routines. The effect of this
error was more important for spins of large amplitude, both aligned with the
angular momentum. Once this error was corrected, we had to redetermine, through
comparison with NR waveform data, the function $c_3(\tilde{a}_A,\tilde{a}_B,\nu)$,
that describes the NNNLO spin-orbit effective
correction~\cite{Damour:2014sva,Nagar:2015xqa,Nagar:2017jdw}. In doing so, 
we found that the correct implementation of the $\ell=5$ modes brings
a simplification to the model: there is {\it no need} of ad-hoc NR-calibrating
the additional parameter $\Delta t_{\rm NQC}$ when $\chi_A=\chi_B>0.85$,
as it was necessary to do in Ref.~\cite{Nagar:2015xqa} [see also Sec.~IIIC of
Ref.~\cite{Nagar:2017jdw}, Eqs.~(24)-(25) therein]. As in the nonspinning case,
we can choose $\Delta t_{\rm NQC}=1$ for all configurations,
without any special tweaks needed for the high-spin case.

\subsubsection{New determination of $c_3$}
It was possible to inform a new function $c_3(\tilde{a}_A,\tilde{a}_B,\nu)$
with the limited set of 27 SXS NR simulations (see Table~\ref{tab:c3}), most
of which are the same used in Ref.~\cite{Nagar:2017jdw}.
The determination of $c_3(\tilde{a}_A,\tilde{a}_B,\nu)$ is based on two steps.
First, for each of the 27 SXS configurations of Table~\ref{tab:c3} one determines,
tuning it by hand, a value of $c_3$ such that the dephasing between EOB and NR
waveform at merger is within the NR uncertainty. Such, first-guess, values of
$c_3$ are then globally fitted with a suitable functional form that,
as in~\cite{Nagar:2017jdw}, is chosen to represent a quasi-linear behavior
in the spins. More precisely, the new representation of $c_3$ is given by
 \begin{align}
   \label{Eq:c3}
   c_3(\tilde{a}_A,\tilde{a}_B,\nu)=&\ p_0\dfrac{1+n_1\hat{a}_0 + n_2\hat{a}_0^2}{1+d_1\hat{a}_0}\nonumber\\
   &+\left(p_1\nu +p_2\nu^2 + p_3\nu^3\right)\hat{a}_0\sqrt{1-4\nu}\nonumber\\
   &+p_4\left(\tilde{a}_A-\tilde{a}_B\right)\nu^2,
 \end{align}
 where 
\begin{subequations}
 \begin{align}
   p_0 &= 43.371638,\\
   n_1 &=-1.174839,\\
   n_2 &= 0.354064,\\
   d_1 &= -0.151961,\\
   p_1 &= 929.579,\\
   p_2 &= -9178.87,\\
   p_3 &=  23632.3,\\
   p_4 &= -104.891.
 \end{align}
 \end{subequations}
 Table~\ref{tab:c3} lists, for the configuration chosen, both the
 first-guess value of $c_3$, that yields an EOB/NR phase agreement
 within the NR error at merger, as well as the value obtained
 from the fit~\eqref{Eq:c3}. The last column lists the relative error 
 $(c_3^{\rm first\;guess}-c_3^{\rm fit})/c_3^{\rm fit}$.
 As it will be shown below, despite the fact that for some
 configurations the first-guess value and the corresponding one obtained
 from the fit are significantly different, the EOB/NR unfaithfulness (see below)
 is still considerably smaller than the usually accepted limit of $1\%$.
 We note however that the global fit can be further improved, if needed,
 by incorporating more NR datasets and/or changing the functional form of
 Eq.~\eqref{Eq:c3}. We shall briefly discuss an example at the end of next section.

 \subsubsection{Post-merger and ringdown}
 Let us come now to discussing the improved representation of the post-merger and
ringdown, that in~\cite{Nagar:2017jdw}  relied on the, rather simplified, 
fits presented in~\cite{Nagar:2016iwa}. For completeness, we also recall
that the NR-based phenomenological description of the waveform is attached
at the inspiral part, NQC-modified, at $t_{\rm NQC}^{\rm EOB}$ given
by Eq.~\eqref{eq:tNQC} above. The new fits for the $\ell=m=2$ merger and
postmerger waveform are detailed in Appendix~\ref{sec:postmerger_NQC}
Let us briefly summarize their new features.
First, the major novelty behind the fitting procedure is that it
is done by exploiting the rather simple behavior that the
merger\footnote{As in previous work, the merger time is 
defined as the peak of the waveform strain amplitude $A_{22}\equiv |h_{22}|$.} 
waveform strain amplitude and frequency $\left(A^{\rm mrg}_{22},\omega^{\rm mrg}_{22}\right)$
show when plotted versus the spin-variable $\hat{S}=(S_A+S_B)/M^2$.
This allows one to capture the full dependence on mass ratio and
spins by means of rather simple two-dimensional fits versus  $(\nu,\hat{S})$.
In addition, we use a larger set of NR waveforms than in previous work:
more precisely, we use 135 spin-aligned NR waveforms\footnote{Out of the 149 waveforms 
listed in Ref.~\cite{Nagar:2017jdw}, 14 are older simulations whose parameters are covered by 
simulations more recently released. These 14 waveforms were not used in the determination 
of the new merger and postmerger parameters.} from the publicly available SXS catalog~\cite{SXS:catalog}
obtained with the {\tt SpEC} code~\cite{Buchman:2012dw,Chu:2009md,Hemberger:2013hsa,
Scheel:2014ina,Blackman:2015pia,Lovelace:2011nu,Lovelace:2010ne,
Lovelace:2014twa,Mroue:2013xna,Kumar:2015tha,Chu:2015kft} 
whose parameters are  summarized in the Tables~V-IX of Ref.~\cite{Nagar:2017jdw}.
These waveforms replace and update the set of 39 waveforms used in~\cite{Nagar:2016iwa}.  
In particular, the SXS waveforms used  are corrected for the effect of the
spurious motion of the center of mass, as pointed out in Ref.~\cite{Boyle:2015nqa}
as well as in Sec.~V~\cite{Nagar:2017jdw}. These SXS waveform data are
complemented by 5 \BAM{} waveforms with mass ratio $q=18$, where the
heavier BH is spinning with $\chi_A=(-0.8,-0.4,0,+0.4, +0.8)$ and
by test-mass waveform\footnote{Note that the phenomenological representation
  of the fit with the template proposed in Refs.~\cite{Damour:2014yha,Nagar:2016iwa}
  is not accurate for high-spin and larger-mass ratio limit waveforms,
  and needs to be modified, including more parameters, to be more flexible.
  That is the reason why in the current representation test-mass data are
  only used to improve the representation of {\it merger} quantities
  $\left(A^{\rm mrg}_{22},\omega^{\rm mrg}_{22}\right)$, and not of the postmerger ones.}
data~\cite{Harms:2014dqa} obtained from new simulations with an improved version
of the test-particle radiation reaction, now resummed according to Refs.~\cite{Nagar:2016ayt,Messina:2018ghh}.
The model is completed by the fit of the spin and mass of the remnant BH of Ref.~\cite{Jimenez-Forteza:2016oae},
and by accurate fits of the quasi-normal-mode (QNM) frequency and inverse-damping
times versus the dimensionless spin of the remnant BH. These are fits of the corresponding
data extracted from the publicly available tables of Berti~{\it et al.}~\cite{Berti:2005ys,Berti:2009kk}.
This is an improvement with respect to previous work, where the final QNM frequencies
were obtained simply by interpolation the publicly available data of
Ref.~\cite{Berti:2005ys,Berti:2009kk}.
We address the reader to Appendix~\ref{sec:postmerger_NQC} for precise technical details.

\subsubsection{The NR waveform point used to obtain NQC parameters}
Using all available information listed above, it was also possible to obtain
more accurate fits of the NR waveform point $\left(A^{\rm NQC}_{22},\dot{A}^{\rm NQC}_{22}
,\omega^{\rm NQC}_{22},\dot{\omega}^{\rm NQC}_{22}\right)$, used to compute the NQC    
parameters $(a_1,a_2,b_1,b_2)$ entering the $\ell=m=2$ NQC waveform correction
factor discussed above. These fits replace those of Sec.~IVB of~\cite{Nagar:2017jdw}
for $q\geq 4$ and are listed together with the details of the new improved postmerger 
fits in Appendix~\ref{sec:postmerger_NQC}.

 \begin{table}[t]
   \caption{\label{tab:c3}First-guess values of $c_3$ compared with the values
     obtained from the interpolating fit for the sample of 27 SXS NR datasets
     used to construct the fit itself. The last column also lists the 
     spin combination $\hat{S}$, helpful in characterizing the
     gravitational wave frequency at merger, see Appendix~\ref{sec:postmerger_NQC}.}
   \begin{center}
 \begin{ruledtabular}
   \begin{tabular}{llllcc}
     $\#$ & $(q,\chi_A,\chi_B)$ & $c_3^{\rm first\;guess}$ & $c_3^{\rm fit}$ & $\Delta c_3/c_3^{\rm fit} [\%]$ & $\hat{S}$\\
     \hline
    1 & $(1,-0.95,-0.95)$  & 93.0 & 92.31 & 0.75 & $-0.4750$\\
    2 & $(1,-0.90,-0.90)$ & 89.0 & 89.44 & -0.49 & $-0.4500$\\
    3 & $(1,-0.80,-0.80)$ & 83.0 & 83.78 & -0.93 & $-0.4000$\\
    4 & $(1,-0.60,-0.60)$ & 73.5 & 72.83 & 0.92 & $-0.3000$\\
    5 & $(1,-0.44,-0.44)$ & 64 & 64.45 & -0.70 & $-0.2200$\\
    6 & $(1,+0.20,+0.20)$ & 35 & 34.85 & 0.43 & $+0.1000$\\
    7 & $(1,+0.60,+0.60)$ & 20.5 & 20.17 & 1.64 & $+0.3000$\\
    8 & $(1,+0.80,+0.80)$ & 13.5 & 14.15 & -4.59 & $+0.4000$\\
    9 & $(1,+0.90,+0.90)$ & 11.5 & 11.52 & -0.17 & $+0.4500$\\
    10 & $(1,+0.99,+0.99)$ & 9.5 & 9.39 & 1.17& $+0.4950$\\
    11 & $(1,+0.994,+0.994)$ & 9.5 & 9.30  & 2.15& $+0.4970$\\
    12 & $(1,-0.50,0)$ & 61.5 & 56.62 & 8.62& $-0.1250$\\
    13 & $(1,+0.90,0)$ & 25.5 & 22.33  & 14.20& $+0.2250$\\
    14 & $(1,+0.90,+0.50)$ & 17.0 & 15.73  & 8.07& $+0.3500$\\
    15 & $(1,+0.50,0)$ & 32.0 & 31.20 & 2.56& $+0.1250$\\
    16 & $(1.5,-0.50,0$ & 62.0 & 57.97  & 6.95& $-0.1800$\\
    17 & $(2,+0.60,0)$ & 29.0 & 26.71  & 8.57& $+0.2\bar{6}$\\
    18 & $(2,+0.85,+0.85)$ & 15.0 & 14.92  & 0.54& $+0.47\bar{2}$\\
    19 & $(3,-0.50,0)$ & 63.0 & 61.15 & 3.03& $-0.28125$\\
    20 & $(3,-0.50,-0.50)$ & 70.5 & 66.63  & 5.81& $-0.3125$\\
    21 & $(3,+0.50,0)$ & 28.0 & 28.02 & -0.07& $+0.28125$\\
    22 & $(3,+0.50,+0.50)$ & 26.5 & 24.44  & 8.43& $+0.3125$\\
    23 & $(3,+0.85,+0.85)$ & 16.5 & 14.38 & 14.74& $+0.53125$\\
    24 & $(5,-0.50,0)$ & 62.0 & 59.84 & 3.61& $-0.347\bar{2}$\\
    25 & $(5,+0.50,0)$ & 30.5 & 29.01 & 5.14& $+0.347\bar{2}$\\
    26 & $(8,-0.50,0)$ & 57.0 & 56.48 & 0.92& $-0.3951$\\
    27 & $(8,+0.50,0)$ & 35.0 & 33.68 & 3.92& $+0.3951$\\
 \end{tabular}
 \end{ruledtabular}
 \end{center}
 \end{table}
\begin{figure}[t]
\center
\includegraphics[width=0.45\textwidth]{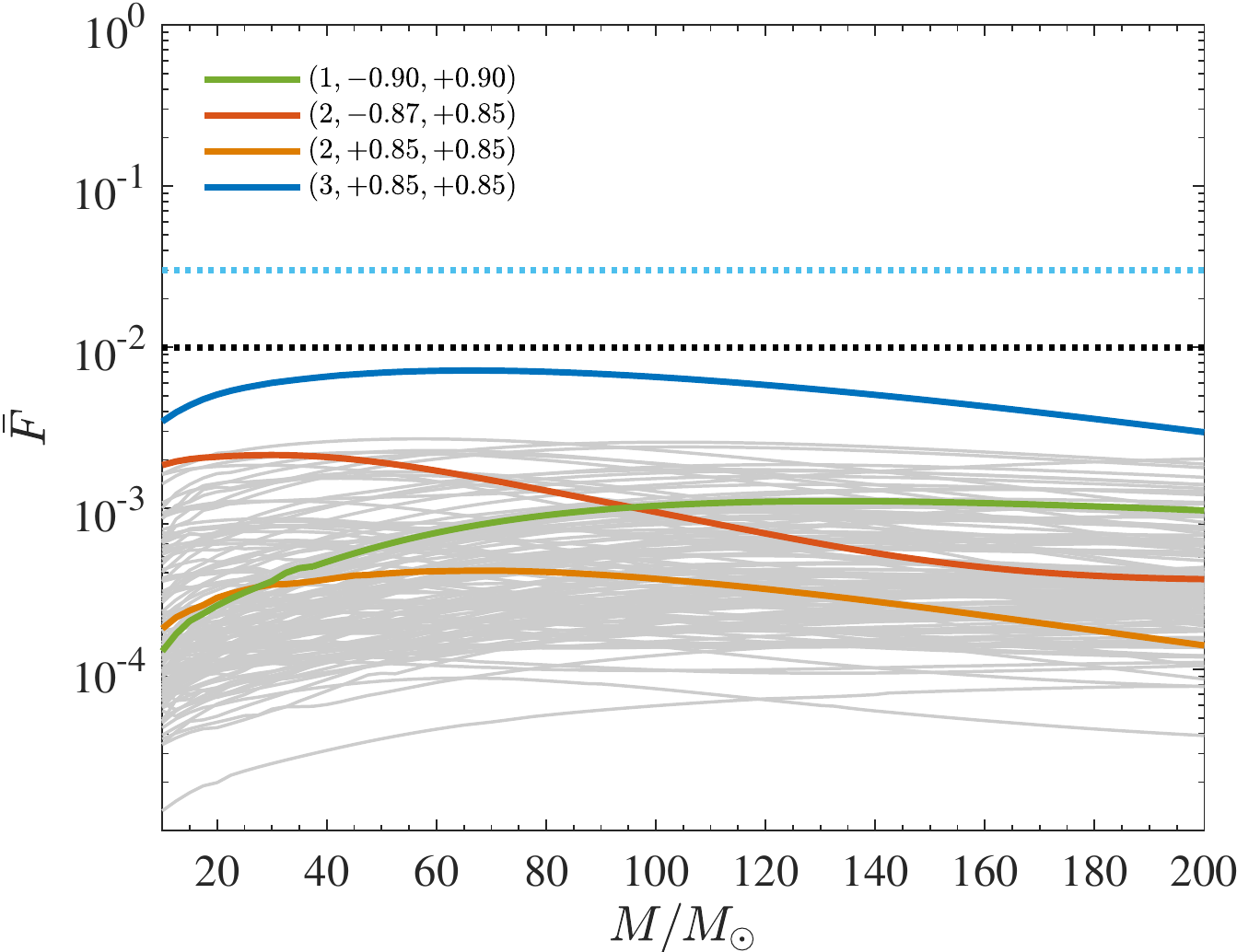}
\caption{\label{fig:SXS}Unfaithfulness, Eq.~\eqref{eq:barF}, comparison between \TEOBResumS{}
  and SXS waveforms, using the design-sensitivity noise curve of Advanced LIGO. 
  This figure is the updated version of Fig.~7 of Ref.~\cite{Nagar:2017jdw}.
  Thanks to the joint action of (i) the correct implementation of the $\ell=5$, $m={\rm odd}$
  modes of the radiation reaction and the related new determination of the NNNLO effective
  spin orbit parameter $c_3$ and (ii) the improved treatment of the postmerger part of
  the signal as well as of the improved NQC determination, there are no outliers above
  the $1\%$ limit. Remarkably, it is found $\max{(\bar{F})}\lesssim 2.5\times 10^{-3}$
  all over the SXS catalog except for a single outlier, $(q, \chi_A,\chi_B)=(3,+0.85,+0.85)$,
  with $\max{(\bar{F})}\simeq 7.1\times 10^{-3}$.}
\end{figure}

\subsection{Comparison with NR data}
\label{sec:barF}
Let us evaluate the global accuracy of the BBH model
that incorporates the new fit for $c_3$, Eq.~\eqref{Eq:c3}, as well as the new fits for
the NQC point and post-merger part. We do this by computing the
usual EOB/NR unfaithfulness $\bar{F}$ defined as
\be
\label{eq:barF}
\bar{F}(M) \equiv 1-F = 1 -\max_{t_0,\phi_0}\dfrac{\langle h_{22}^{\rm EOB},h_{22}^{\rm NR}\rangle}{||h_{22}^{\rm EOB}||\:||h_{22}^{\rm NR}||},
\ee
where $(t_0,\phi_0)$ are the arbitrary initial time and phase and $||h||\equiv \sqrt{\langle h,h\rangle}$.
The inner product between two waveforms is defined as 
$\langle h_1,h_2\rangle\equiv 4\Re \int_{f_{\rm min}^{\rm NR}(M)}^\infty \tilde{h}_1(f)\tilde{h}_2^*(f)/S_n(f)\, df$,
where $\tilde{h}(f)$ denotes the Fourier transform of $h(t)$, $S_n(f)$ is the zero-detuned,
high-power noise spectral density of Advanced LIGO~\cite{dcc:2974} and
$f_{\rm min}^{\rm NR}(M)=\hat{f}^{\rm NR}_{\rm min}/M$ is the {\it starting frequency of the NR waveform}
(after the junk radiation initial transient). Both EOB and NR waveforms are tapered in the
time domain so as to reduce high-frequency oscillations in the corresponding Fourier transforms.
We display $\bar{F}(M)$, for $10M_\odot\leq M\leq 200M_\odot$, in Fig.~\ref{fig:SXS}
for the 171 SXS waveform data and in Fig.~\ref{fig:BAM} for
the 18 \BAM{} datasets.
Let us discuss first the \TEOBResumS{}/SXS comparison, Fig.~\ref{fig:SXS}. To better appreciate
the improvement brought by the correct implementation of the $\ell=5$, $m=$odd flux modes
and the post-merger fits, this figure should be compared with Fig.~7 of~\cite{Nagar:2017jdw}.
Figure~\ref{fig:SXS} illustrates that $\max(\bar{F})\lesssim 2.7\times 10^{-3}$ all over
the waveform database except for a single outlier, $(3,+0.85,+0.85)$, where $\max(\bar{F})=7.1\times 10^{-3}$.
Note however that the performance is much better than the minimal accepted limit of $3\%$ (light-blue, dotted,
horizontal line) or the more stringent $1\%$ limit (black, dotted, horizontal line) that
is taken as a goal by \texttt{SEOBNRv4} (see Fig.~2 in ~\cite{Bohe:2016gbl}); in fact,
it is the {\it lowest ever} value of $\max\left[\max{(\bar{F})}\right]$ obtained from SXS/EOB comparisons.
We note that the reason why $\bar{F}\simeq 7.1\times 10^{-3}$ for $(3,+0.85,+0.85)$ is entirely
due to the fact that the global representation of $c_3$ yielded by Eq.~\eqref{Eq:c3} is not that
accurate in that corner of the parameter space, and yields the value 14.38 instead of 16.5
(see line \#23 of Table~\ref{tab:c3}). Interestingly, we have verified that,
by using the value 16.5, the value of $\bar{F}(M)$ significantly drops, being smaller
than $10^{-4}$ at $M=10M_\odot$ and just growing up to $2\times 10^{-4}$ at $M=200M_\odot$. 
This illustrates that our analytical representation of $c_3$ is actually very conservative.
It would be easy, by either incorporating more datasets in the global fit and/or
improving the functional form of~\eqref{Eq:c3} to reduce the discrepancy between
the first-guess and fitted value of $c_3$. As a simple attempt to do so, we
slightly changed the functional form of $c_3(\tilde{a}_A,\tilde{a}_B,\nu)$ so as to introduce
nonlinear spin-dependence away from the equal-mass, equal-spin case.
\begin{figure}[t]
\center
\includegraphics[width=0.45\textwidth]{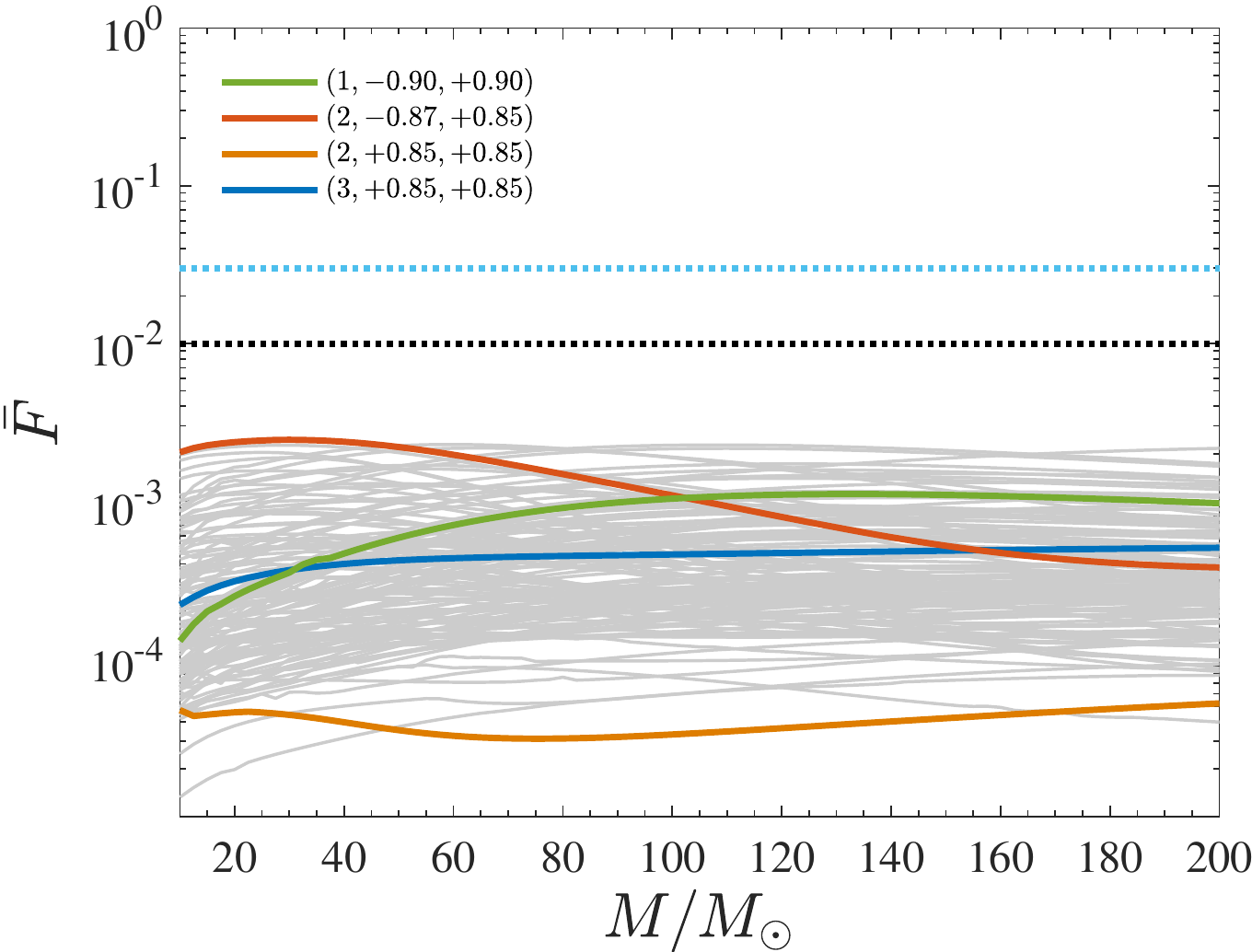}
\caption{\label{fig:SXS_a02}Same as Fig.~\ref{fig:SXS}, but including an additional term
  proportional to $\nu\ha_0^2\sqrt{1-4\nu}$ in the functional form Eq.~\eqref{Eq:c3} using to
  fit the $c_3^{\rm first\;guess}$ values of Table~\ref{tab:c3}. One has $\max{(\bar{F})}<2.5\times 10^{-3}$
  all over the SXS catalog of public NR waveforms.}
\end{figure}
For example, to introduce such nonlinearities in spin in a simple way, one easily checks
that the addition to Eq.~\eqref{Eq:c3} of only one term {\it quadratic}
in $\hat{a}_0$ of the form $p_5\nu\hat{a}_0^2\sqrt{1-4\nu}$, where $p_5$ is a further
fitting coefficient, is by itself sufficient to obtain $c_3=17.28$ for $(3,+0.85,+0.85)$,
with a corresponding value of $\max(\bar{F})=5\times 10^{-4}$ reached at $M=200M_\odot$.
Once this term is included, the new fitting coefficients that parametrize the sector of
$c_3$ away from the equal-mass, equal-spin limit read
$(p_1,p_2,p_3,p_4,p_5)=(917.59,-8754.35,20591.0,-78.95,83.40)$. For completeness, we evaluated
again the EOB/NR $\bar{F}$ with this new fit. The result is displayed in Fig.~\ref{fig:SXS_a02}.
It is remarkable to find that $\max(\bar{F})< 2.5\times 10^{-3}$ all over the SXS catalog.
It is also interesting to note that the two curves for $(3,+0.85,+0.85)$
and $(2,+0.85,+0.85)$ are essentially flat, which illustrates that all the difference
with the previous case was coming from the slightly inaccurate representation of the
spin-orbit coupling functions, now corrected by the improved representation of $c_3$.

\begin{figure}[t]
\center
\includegraphics[width=0.45\textwidth]{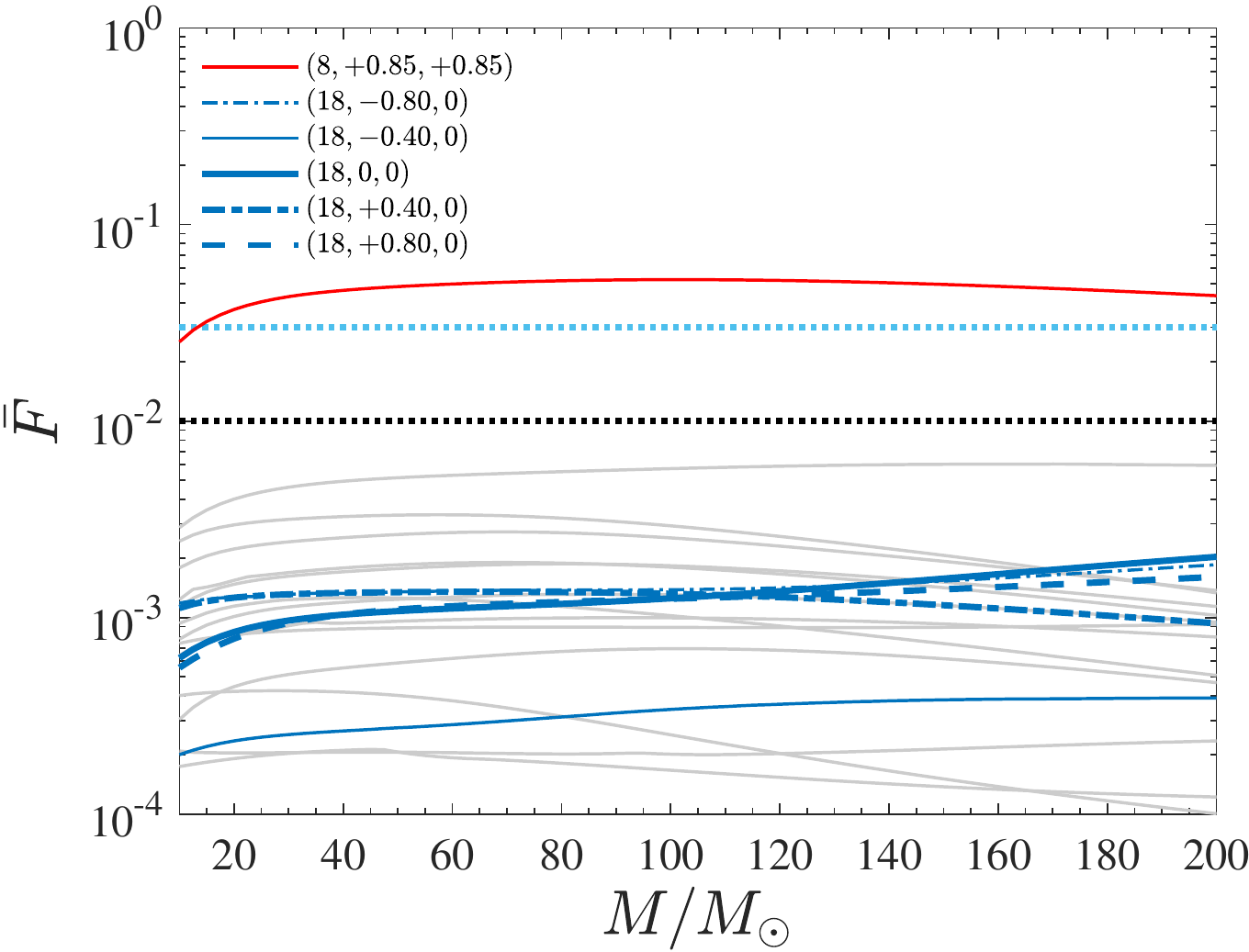}
\caption{\label{fig:BAM}Unfaithfulness comparison between \TEOBResumS{}
  and the set of \BAM{} waveforms mostly presented in
  Refs.~\cite{Husa:2015iqa,Khan:2015jqa,Keitel:2016krm}
  and listed in Table~\ref{tab:BAM} for completeness.
  The case $(8,+0.85,+0.85)$, where a new, high-resolution,
  \BAM{} waveform was produced explicitly for this work,
  is meaningfully above the $3\%$ limit and calls for
  an improvement of the model in that specific corner
  of the parameter space.}
\end{figure}

 \begin{table}[t]
   \caption{\label{tab:BAM}Portion of the parameter space
     covered by \BAM{} NR simulations.}
       \begin{center}
 \begin{ruledtabular}
   \begin{tabular}{llc}
     $\#$ & $(q,\chi_A,\chi_B)$ & $\hat{S}$ \\
     \hline
    1 & $(2,+0.75,+0.75)$ & 0.4167\\
    2 & $(2,+0.50,+0.50)$ & 0.2778\\
    3 & $(3,+0.50,+0.50)$ & 0.3125\\
    4 & $(4,+0.75,+0.75)$ & 0.51\\
    5 & $(4,+0.50,+0.50)$ & 0.34\\
    6 & $(4,+0.25,+0.25)$ & 0.17\\
    7 & $(4,0,0)$ & 0\\
    8 & $(4,-0.25,-0.25)$ & $-0.17$\\
    9 & $(4,-0.50,-0.50)$ & $-0.34$\\
    10 & $(4,-0.75,-0.75)$ & $-0.51$\\
    11 & $(8,+0.85,+0.85)$ & $0.6821$\\
    12 & $(8,+0.80,0)$ &   $0.6321$\\
    13 & $(8,-0.85,-0.85)$ & $-0.6821$ \\
    14 & $(10,0,0)$ & $0$\\
    15 & $(18,+0.80,0)$ & 0.7180\\
    16 & $(18,+0.40,0)$ & 0.3590\\
    17 & $(18, 0,0)$ & 0\\
    18 & $(18,-0.40,0)$ & $-0.3590$\\
    19 & $(18,-0.80,0)$ & $-0.7180$
 \end{tabular}
 \end{ruledtabular}
 \end{center}
 \end{table}

Let us turn now to discussing \TEOBResumS{}/\BAM{} comparisons, Fig.~\ref{fig:BAM}.
These waveforms cover a region of the parameter space,
for large mass ratios, that is not covered by SXS data (see Table~\ref{tab:BAM}).
Hence, we use them here as a probe of the phasing provided by \TEOBResumS{}.
In general, \BAM{} waveforms in the current database are shorter than the SXS
ones and have larger uncertainties. This is also the case for the $(8,+0.85,+0.85)$
configuration, that yields the largest NR/EOB disagreement, $\max(\bar{F})\simeq 5.2\%$,
which is above the usually acceptable level of $3\%$.
However, though this waveform is much longer
($\approx 18$ orbits) than the one previously used in~\cite{Nagar:2017jdw},
it was also obtained at higher resolution, so that its error assessment
is similar to those used for the \IMRPhenomD{} waveform
model~\cite{Khan:2015jqa,Keitel:2016krm}, with a mismatch error
of less than $10^{-3}$. The EOB/NR difference seen in
Fig.~\ref{fig:BAM}, originates then in the EOB model,
notably during the inspiral, and not in the NR data.
\begin{figure}[t]
\center
\includegraphics[width=0.4\textwidth]{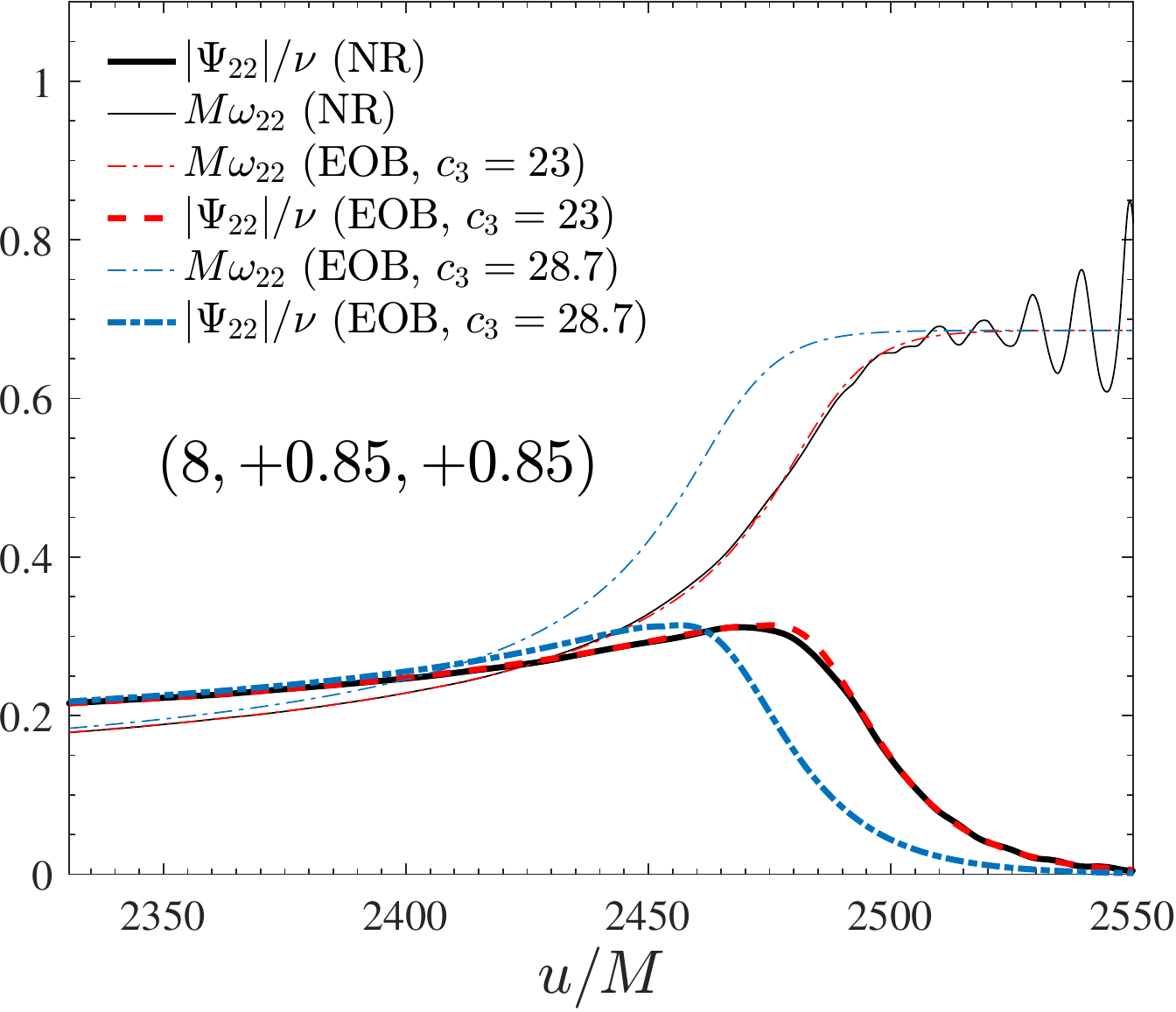}
\caption{\label{fig:c3eff}Effect of changing the value of the effective NNNLO spin-orbit parameter
  $c_3$ for the $(8,+0.85,+0.85)$ configuration. Time-domain evolution of frequency and amplitude.
  The $5.2\%$ value of $\bar{F}$ in Fig.~\ref{fig:BAM} comes entirely from the value $c_3=28.7$
  obtained by extrapolating from the SXS-based fit of Eq.~\eqref{Eq:c3}. A smaller value of the parameter,
  $c_3=23$, succeeds in getting a good EOB/NR agreement ($\bar{F}\simeq 1.3\times 10^{-3}$) .
  Despite this, both the NQC and post-merger sectors are correctly represented by the model
  because of the robust NR-informed fits.}
\end{figure}
To explicitly see that the origin of such EOB/NR discrepancy comes from the EOB-driven
inspiral dynamics and not from the ringdown part\footnote{This is the contrary of what was
stated in~\cite{Nagar:2017jdw}. The reason for this is that the \BAM{} waveform used
there was shorter than the one we are using now.}, we display in Fig.~\ref{fig:c3eff}
the waveform frequency and amplitude versus time. The figure compares three
datasets: (i) the \BAM{} data (black); (ii) the \TEOBResumS{} waveform with the value of
$c_3 \approx 28.7$ obtained from Eq.~\eqref{Eq:c3}
(blue, dash-dotted, lines) and $c_3=23$. Note that while the $c_3=28.7$ waveform was
obtained by iterating on NQCs parameter (i.e., the NQC correction is also added to the flux
for consistency with the waveform and then an iterative procedure is set until the 
values of $(a_1,a_2)$ are seen to converge~\cite{Nagar:2017jdw}), 
the $c_3=23$ one was not (see below). The waveforms
are aligned in the $(0.2,0.35)$ frequency interval region. The figure clearly illustrates
that the simple action of lowering $c_3$ (i.e. making the spin-orbit interaction less
attractive, see discussion in~\cite{Nagar:2017jdw}) is effective in getting the \TEOBResumS{}
waveform closer to the \BAM{} one: the waveform becomes longer and the frequency
behaviors get qualitatively more similar up to merger. Note also that the postmerger
part is perfectly consistent with the NR one. This is a remarkable indication of the 
robustness of our post-merger fits since the $(8,+0.85,+0.85)$ \BAM{} dataset was not used
in their construction. We mentioned above that the curves corresponding to $c_3=23$ 
were obtained without iterating on the amplitude NQC parameters $(a_1,a_2)$. 
The reason for this is that the value of the NQC parameters are rather large because
of the lack of robustness of the resummed waveform amplitude in this corner of the 
parameter space and they effectively tend to compensate the action of $c_3$,
that should be lowered further. The consequence of this is that, when  $c_3$ is chosen 
to be below 20, $(a_1,a_2)$ become so large that the iteration procedure 
is unable to converge. The use of the improved factorized and resummed waveform 
amplitudes of Refs.~\cite{Nagar:2016ayt,Messina:2018ghh}, that display a more robust and 
self-consistent behavior towards merger for high, positive spins is expected to solve 
this problem.

To summarize, the message of the analysis illustrated in
Fig.~\ref{fig:c3eff} goes as follows: (i) on the positive side, the figure 
illustrates that, even if we had not included $(8,+0.85,+0.85)$ data 
to obtaining the postmerger fit parameters, the resulting model is
rather accurate also for this choice of parameters; (ii) on the negative side, 
it also tells us that the dataset $(8,+0.85,+0.85)$ brings us new, genuine, 
NR information that is currently not incorporated in the model, but it should be 
in order to properly capture the correct phasing behavior in
this corner of the parameter space\footnote{We note in passing that \SEOBNRv4 also
used \BAM{} datasets with $(8,+0.85,+0.85)$, though different from the one we used here,
for its calibration.}. In principle, improving the model would be rather straightforward,
as it would just amount to adding a new value of $c_3$ in Table~\ref{tab:c3},
corresponding to an acceptable \BAM{}/EOB phasing up to merger, and redoing
the global fit. However, because of the aforementioned problems in obtaining
a consistent determination of the NQC parameters,  we shall postpone this to 
a forthcoming  study that will (partly) use the factorized and resummed waveform 
amplitudes of Ref.~\cite{Messina:2018ghh}.

\begin{figure}[t]
\center
\includegraphics[width=0.4\textwidth]{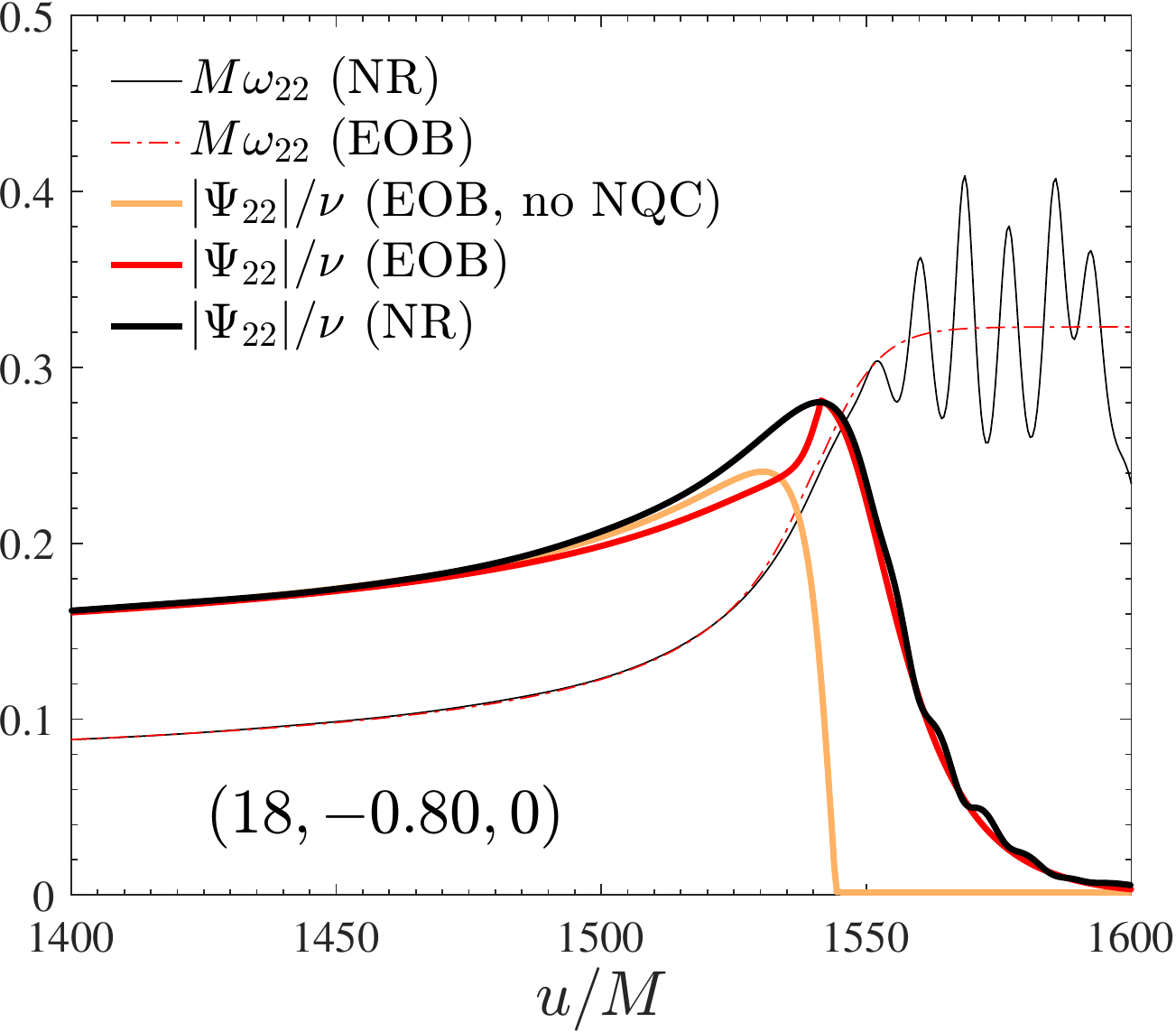}
\caption{\label{fig:q18_amp_freq}Frequency and amplitude comparison between \TEOBResumS{}
  and \BAM{} for $(18,-0.80,0)$. The full waveform amplitude develops a slightly unphysical
  feature due to the action of the NQC parameters. The frequency (as well as $\bar{F}$) is
  unaffected by this.}
\end{figure}
\begin{figure}[t]
\center
\includegraphics[width=0.4\textwidth]{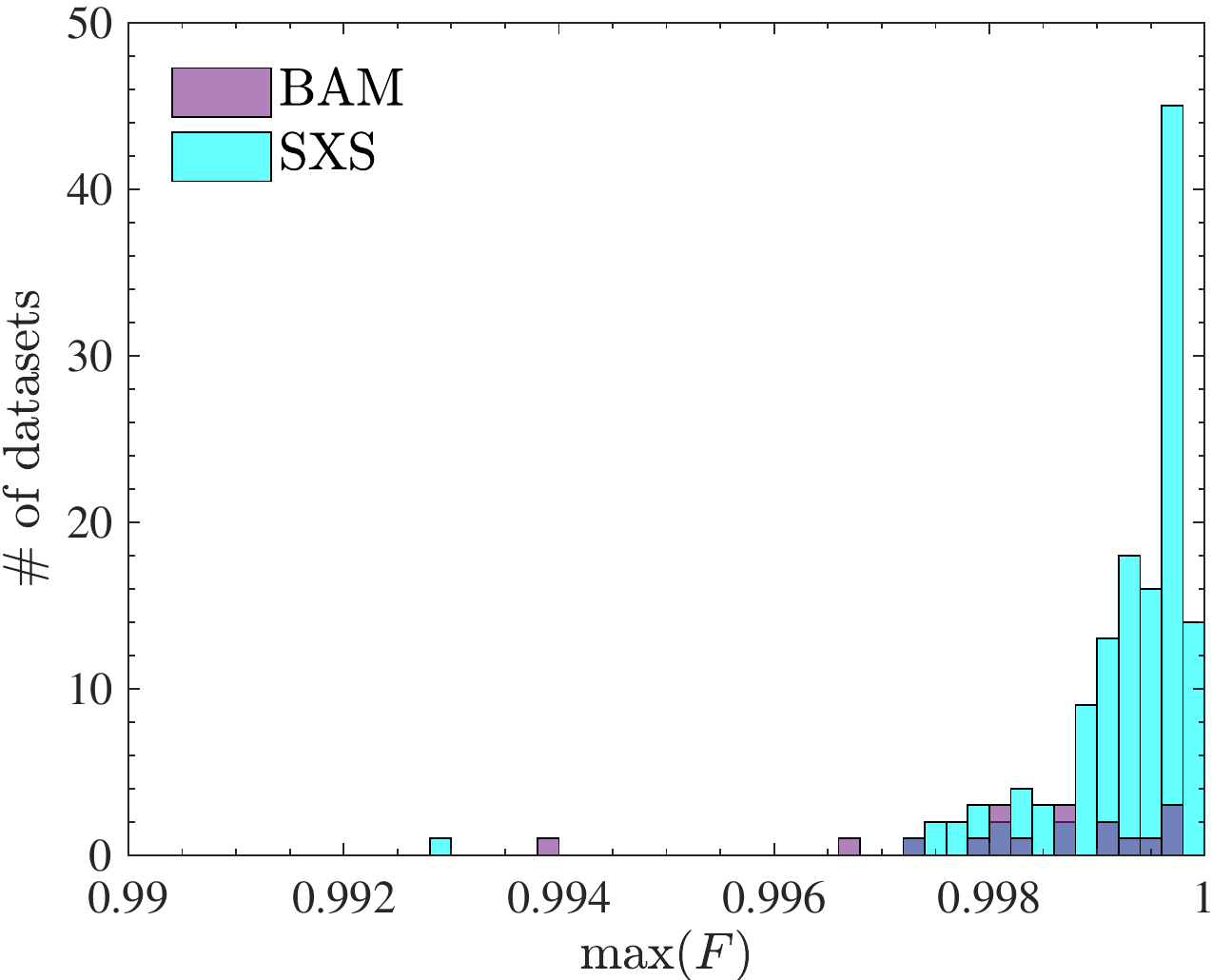}
\caption{\label{fig:histo}Global picture of the maximum value of the EOB/NR faithfulness $F$,
  Eq.~\eqref{eq:barF} over SXS and \BAM{} NR data. The only outlier above $3\%$, (8,+0.85,+0.85),
  is omitted from the figure.}
\end{figure}

Finally, Fig.~\ref{fig:q18_amp_freq} illustrates another difference between
\TEOBResumS{} and \BAM{} waveforms. The figure compares the analytical 
and numerical frequencies and amplitudes for $(18,-0.80,0)$. The waveforms
are aligned around merger. Although the frequencies are perfectly consistent,
the analytical amplitude (red line) shows a qualitatively incorrect behavior 
before merger. Although such feature in the amplitude might be interpreted
as due to an incorrect determination of the NQC corrections,
it is actually of dynamical origin. More precisely, it comes from the 
orbital frequency $\Omega$ crossing zero and then becoming negative 
due to a somewhat large values of the gyro-gravitomagnetic functions 
$(G_S,G_{S_*})$ for small values of the EOB radial separations. 
Since the spins are negative, the spin-orbit part of the  orbital frequency 
progressively compensates the orbital one, until dominating over it so
that $\Omega<0$ around merger time. We have tracked back the origin 
of this problem to the fact that, following Ref.~\cite{Damour:2014sva}, 
the argument of the functions $(\hat{G}_S,\hat{G}_{S_*})$ (see Eqs.~(36)-(37) of~\cite{Damour:2014sva})
where chosen, by construction, to be $1/r_c$, instead of of $1/r$, so as to 
effectively incorporate higher-order spin-orbit corrections.
Although it is not our intention to discuss this subject in more detail here, 
we have actually verified that going back to the standard $1/r$ dependence
of these functions is sufficiently to reduce and/or cure completely (as it is 
the case for the configuration $(11,-0.95,-0.50)$ discussed below) this
somewhat unphysical feature\footnote{Please note, however, that, likewise the
case of a test-particle plunging over a highly spinning black hole whose spin
is anti-aligned with the orbital angular momentum~\cite{Harms:2014dqa,Taracchini:2014zpa}, 
by continuity there might exist BBHs configurations where the orbital frequency 
is actually due to change its sign while approaching merger. 
This is however not the case of the $(18,-0.80,0)$ binary under consideration, 
since the positive-frequency QNMs branch is still more excited than the negative
frequency one. The contribution of this latter is not, however, negligible,
as illustrated by the large amplitude oscillation in the NR frequency
displayed in Fig.~\ref{fig:q18_amp_freq}.}. Although the behavior of the modulus in
Fig.~\ref{fig:q18_amp_freq} has not practical consequences, 
it is important to mention that similar features may occur systematically 
for binaries with large $q$ and large spins, anti-aligned with
the orbital angular momentum.  This statement will be recalled below when discussing 
the performance of the model outside the NR-covered region of the parameter space.
Finally, a global representation of the results of Figs.~\ref{fig:SXS}-\ref{fig:BAM}
is given in Fig.~\ref{fig:histo}, that displays the maximum value of the EOB/NR
{\it faithfulness} $F$, reached for each dataset varying the total mass $M$,
all over the SXS and \BAM{} waveform catalogs, only excluding the
$(8,+0.85,+0.85)$ outlier for readability.

\subsection{Waveform robustness outside the NR-covered region of parameter space}
\label{sec:outside_nr}
\begin{figure}[t]
\center
\includegraphics[width=0.4\textwidth]{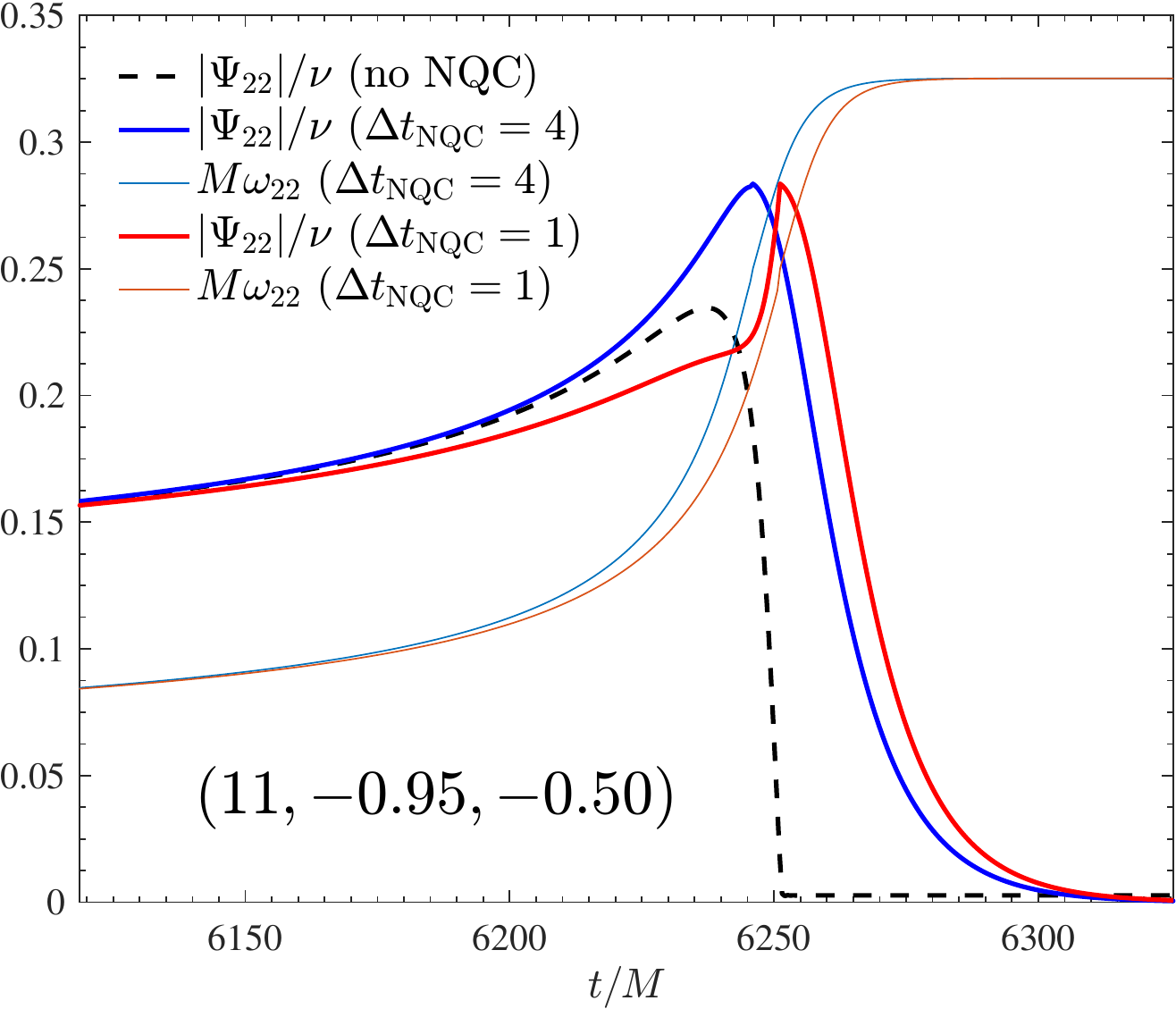}
\caption{\label{fig:Dt_bad_merger} Comparing the effect of using $\Delta t_{\rm NQC}=1$ and $\Delta t_{\rm NQC}=4$
  for $(q,\chi_1,\chi_2)=(11,-0.95,-0.50)$. The use of $\Delta t_{\rm NQC}=4$ makes the behavior of
  the waveform amplitude at merger consistent with the NR-fitted postmerger behavior.}
\end{figure}

\begin{figure}[t]
\center
\includegraphics[width=0.4\textwidth]{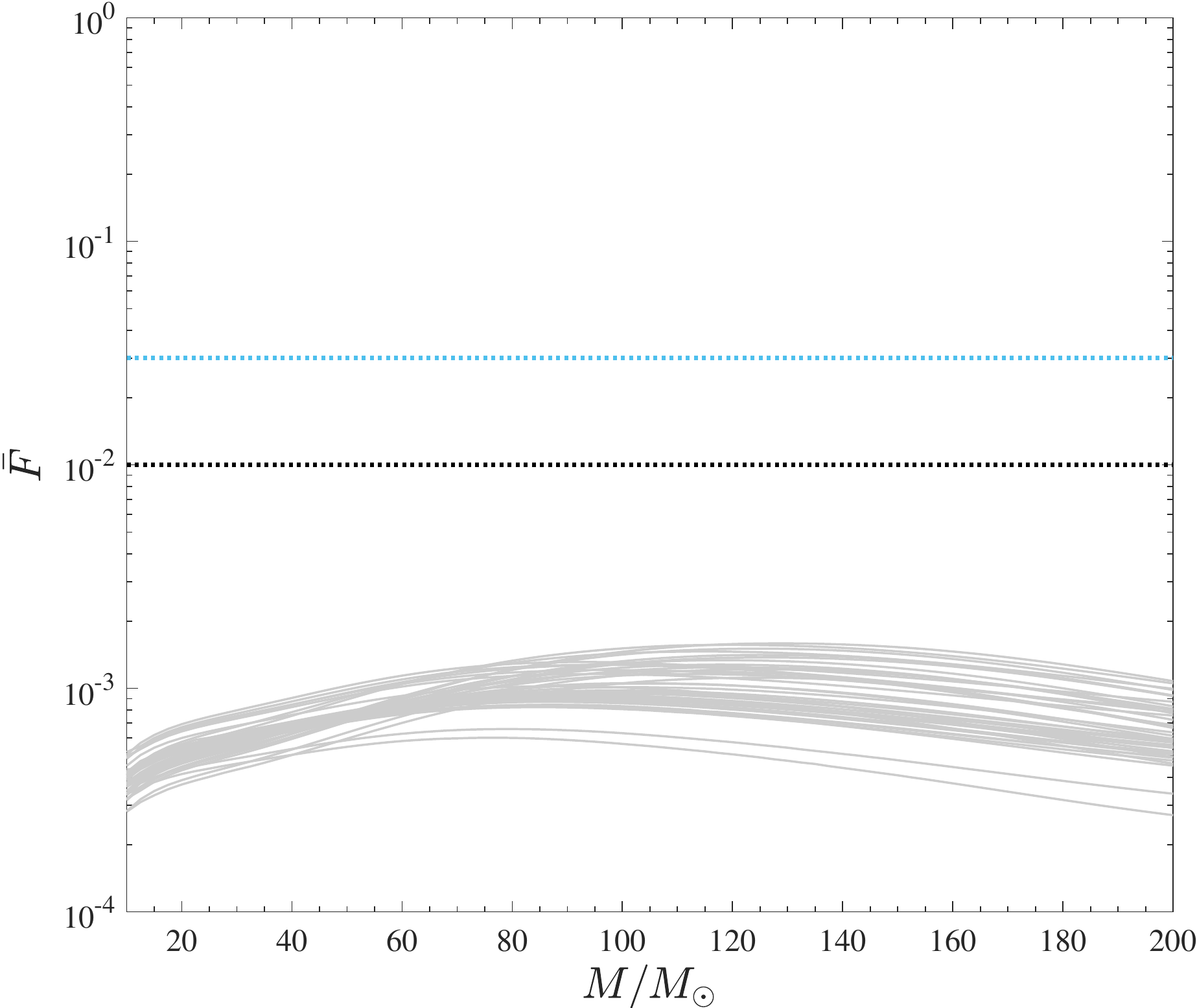}
\caption{\label{fig:Dt_nqc} Calculation of $\bar{F}$ between EOB waveforms with $\Delta t_{\rm NQC}=1$  
        and $\Delta t_{\rm NQC}=4$  at the boundary of the region of the 
        parameter space defined by Eqs.~\eqref{eq:m09}-\eqref{eq:m08}.  The consistency  between
        the two types of waveforms is excellent. }
\end{figure}

The model was tested to be robust in the most demanding corners of the parameter space, notably
for large mass ratios (though we limit ourselves to $q\leq 20$) and large values of the spin magnitudes.
In particular, no obvious problem was found for large mass ratios and when the spins are positive.
The absence of ill-defined behaviors in the waveform is mostly due to the use of robust fits 
across the whole parameter space and to the fact that the NQC corrections are able to effectively
reduce the residual inaccuracies in the EOB waveform. However, this comes at the price of
large NQC parameters (far from being order unity, as noted above for the specific case of
$(8,+0.85,+0.85)$) since they have to strongly correct a waveform in a regime where the radial
momenta are small. Large NQC parameters prevent the necessary iterative procedure of 
recomputing the flux from converging. We thus remove the NQC corrections to the flux, although
in this way it becomes mildly inconsistent with the waveform.

As anticipated above, when the mass ratio is moderately large ($q\geq 8$) and spins 
are equally large but anti-aligned with the angular momentum, the waveform amplitude 
may develop artifacts prompted by the underlying orbital frequency being small and 
eventually crossing zero (and thus strongly affecting the NQC amplitude correction factor) 
as we found for the  $(8,-0.80,0)$  configuration. 
\begin{figure}[t]
\center
\includegraphics[width=0.40\textwidth]{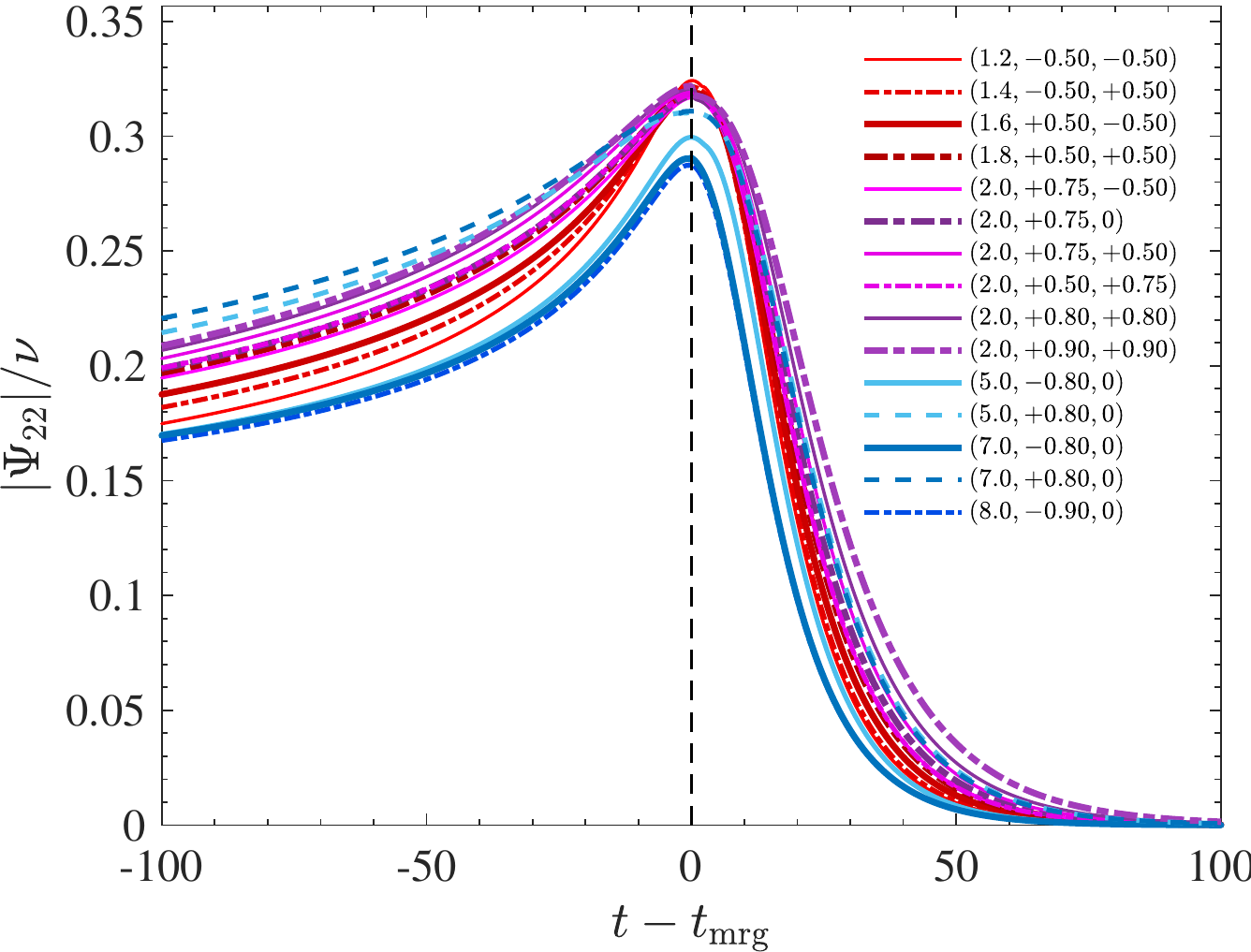}\\
\includegraphics[width=0.40\textwidth]{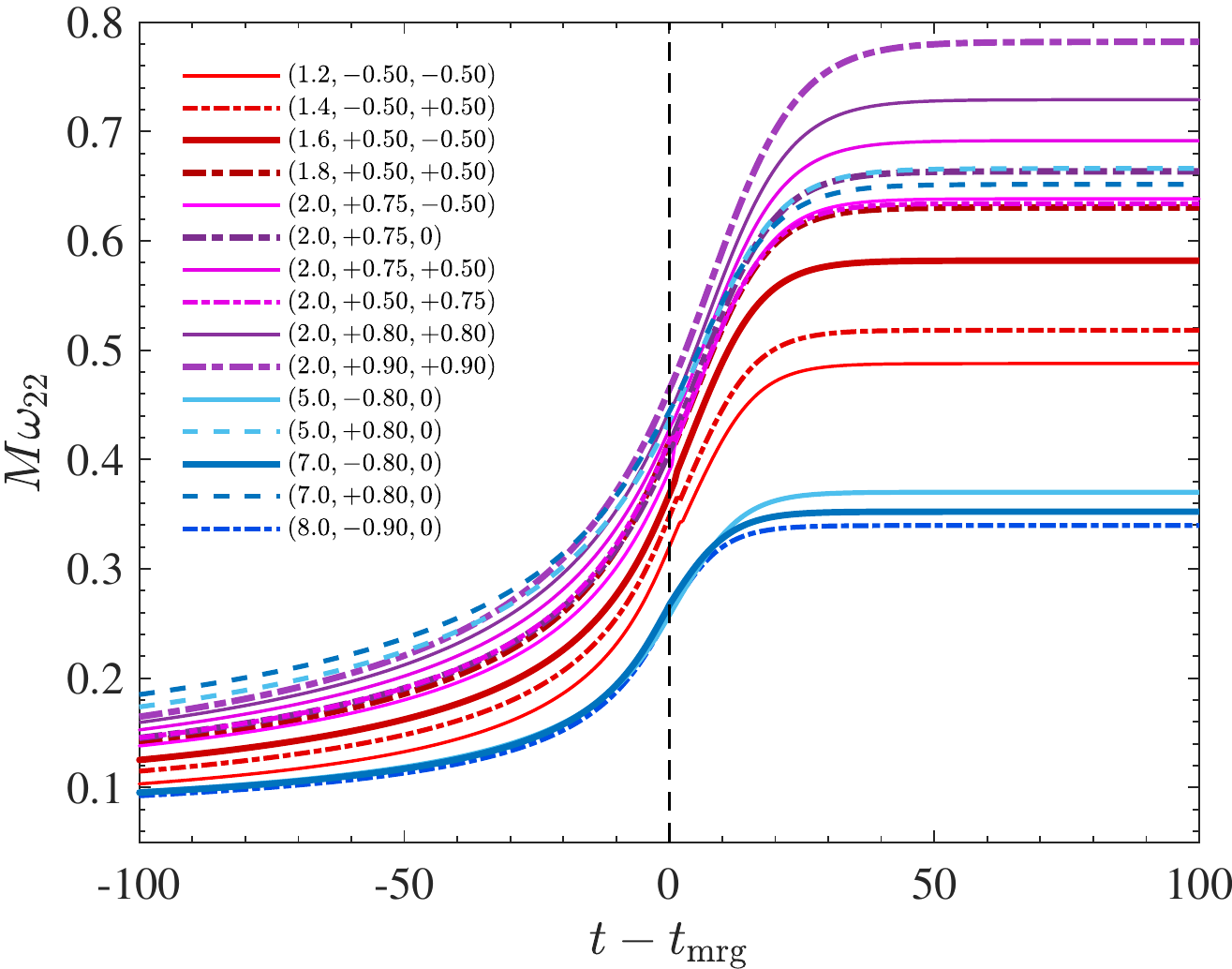}
\caption{\label{fig:bohe}Sanity check of EOB waveform modulus (top) and frequency 
(bottom) on the configurations considered in Table~I of Boh\'e~et~al.~\cite{Bohe:2016gbl}. 
Differently from what we do here, NR waveform data for these configurations
were used in~\cite{Bohe:2016gbl} to calibrate \SEOBNRv{4}.
The behavior of both functions look qualitatively and quantitatively consistent
and robust. Waveforms are time-shifted to be all aligned at merger time.}
\end{figure}
\begin{figure}[t]
\center
\includegraphics[width=0.4\textwidth]{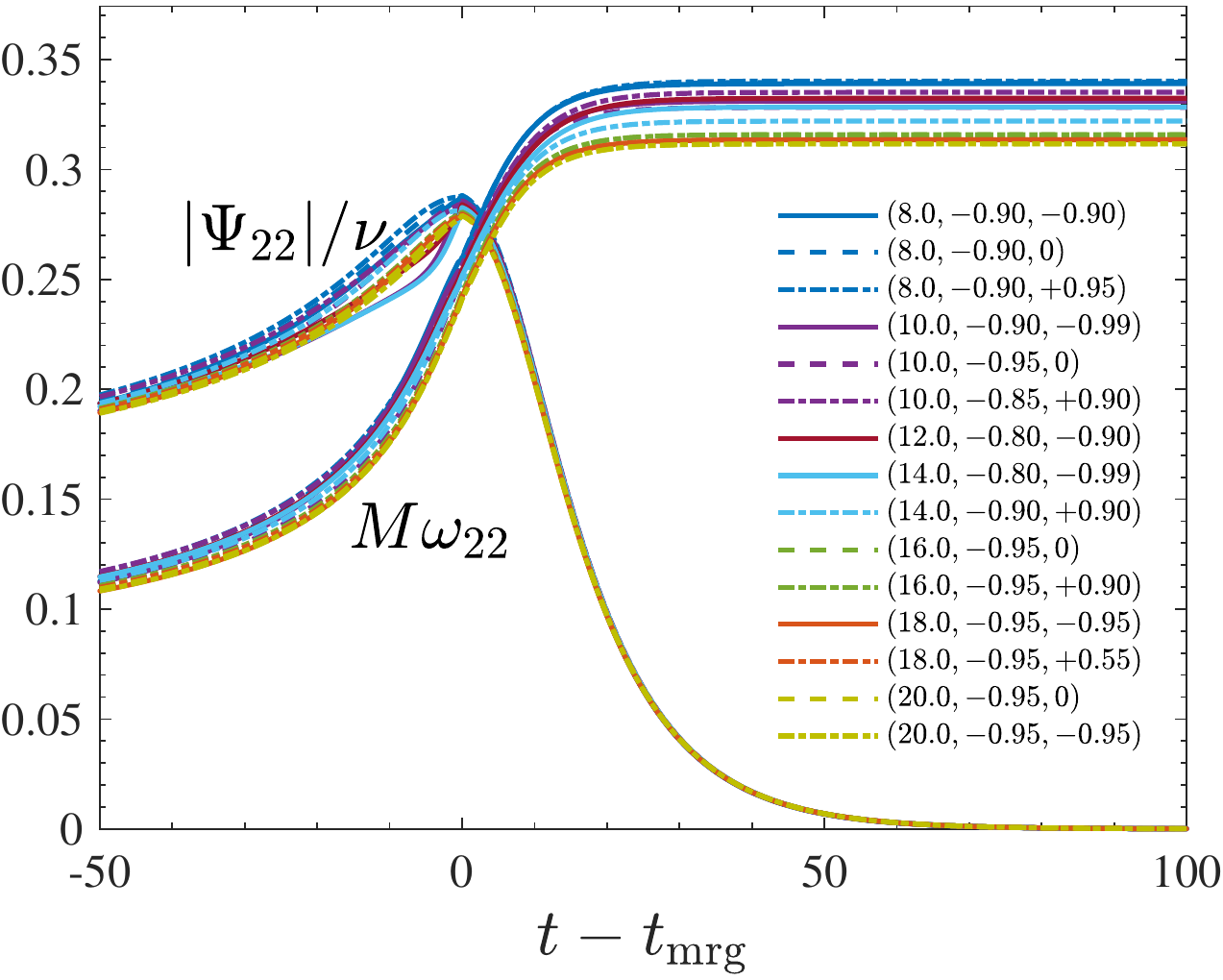}
\caption{\label{fig:NegSpin}Sanity check of EOB waveforms for large mass ratios
and large spins anti-aligned with the angular momentum. The good qualitative 
behavior of the waveform around merger is guaranteed by the value 
of $\Delta t_{\rm NQC}$ given by Eq.~\eqref{eq:m09}-\eqref{eq:m08}.}
\end{figure}
For example, Fig.~\ref{fig:Dt_bad_merger} illustrate the type of, qualitatively
incorrect, features that the waveform can develop towards merger due to the incorrect
action of the NQC factor. In the figure we show, with a red and an orange line, the amplitude 
and frequency for $(11,-0.95,-0.50)$ as generated by the model described above. 
The black dashed line is the {\it bare} EOB-waveform amplitude, without the NQC factor.
We have explicitly verified that $\Omega$ crosses zero also in this case. Although, as
we mentioned above, the theoretically correct way of solving this problem is to modify
the spin-orbit sector of \TEOBResumS{}, one finds that, if the standard value 
$\Delta t_{\rm NQC}=1$ is increased to $\Delta t_{\rm NQC}=4$, the weird behavior 
disappears and the inspiral EOB waveform amplitude can be connected smoothly to 
the postmerger part obtained via the global fit of the NR waveform data. 
The same kind of EOB/NR inconsistency also appears for configurations with even 
higher mass ratios and large, negative, spins. In some extreme situations, 
it can also affect the frequency.
We performed a thorough scan of the parameter space and we concluded that a pragmatical
approach to solve this problem is simply to impose $\Delta t_{\rm NQC}=4$  for a certain 
sample of configurations.
More precisely, we found that the ubiquitous $\Delta t_{\rm NQC}=1$ should be replaced
by  $\Delta t_{\rm NQC}=4$ when
\begin{align}
\label{eq:m09}
8 < q < 11&~\text{and}~\chi_A < -0.9 ,\\
\label{eq:m08}
11<q<19 &~\text{and}~\chi_A < -0.8. 
\end{align}
Note that, despite being independent of the value of $\chi_B$, such simplified
conditions allow to generate waveforms that present a sufficiently sane and smooth behavior
around the merger up to mass ratio $q=20$ and spins $\chi_A=\chi_B=\pm 0.95$.
Finally, the last question is about the magnitude of the uncertainty
that one introduces by choosing $\Delta t_{\rm NQC}=4$ instead of $\Delta t_{\rm NQC}=1$
at the boundary of the region of the parameter space defined by Eqs.~\eqref{eq:m09}-\eqref{eq:m08}. 
We evaluated this by (i) choosing several configurations at 
the interface, on the $(\nu,\chi_A)$ square, and by computing $\bar{F}$ between EOB 
waveforms with $\Delta t_{\rm NQC}=1$ and $\Delta t_{\rm NQC}=4$. We find values of $\bar{F}$
(see Fig.~\eqref{fig:Dt_nqc}) on average around $10^{-3}$, which means that having a discontinuous 
transition has in fact no practical consequences. 
Evidently, the radical solution to this problem will eventually be to change the argument
of the gyro-gravitomagnetic functions $(\hat{G}_S,\hat{G}_{S_*})$ as mentioned above.
In this respect, we have checked that doing so for the case $(11,-0.95,-0.50)$ of Fig.~\ref{fig:Dt_bad_merger}
allows one to (i) avoid the orbital frequency $\Omega$ crossing zero and (ii) consequently
recovering a qualitatively excellent modulus around merger simply keeping $\Delta t_{\rm NQC}=1$.
Since such an improved \TEOBResumS{} model will have also to rely on a different determination 
of $c_3$ to be consistent with all NR simulations, we postpone a detailed treatment to future work.

\begin{figure}[t]
\center
\includegraphics[width=0.4\textwidth]{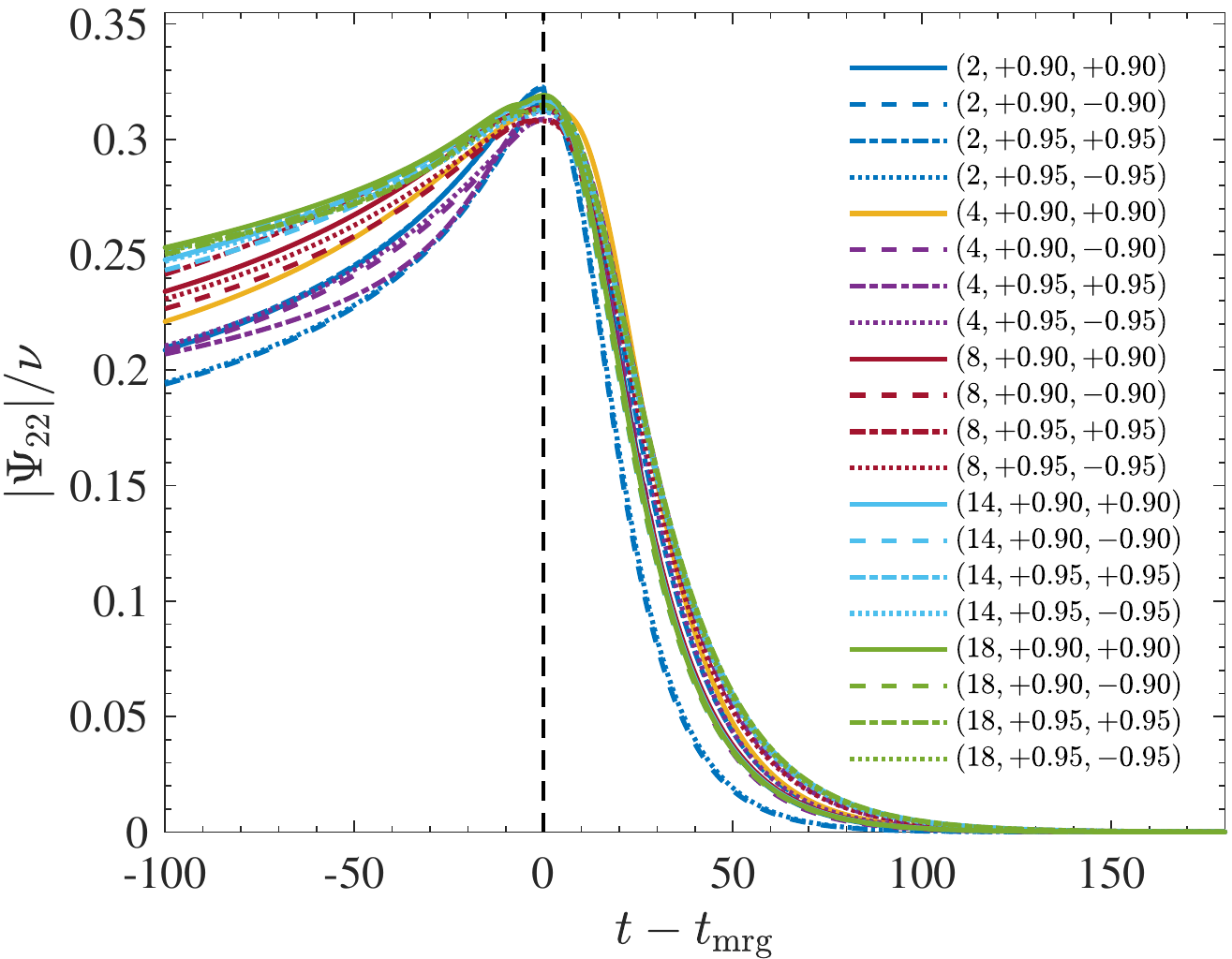}
\includegraphics[width=0.4\textwidth]{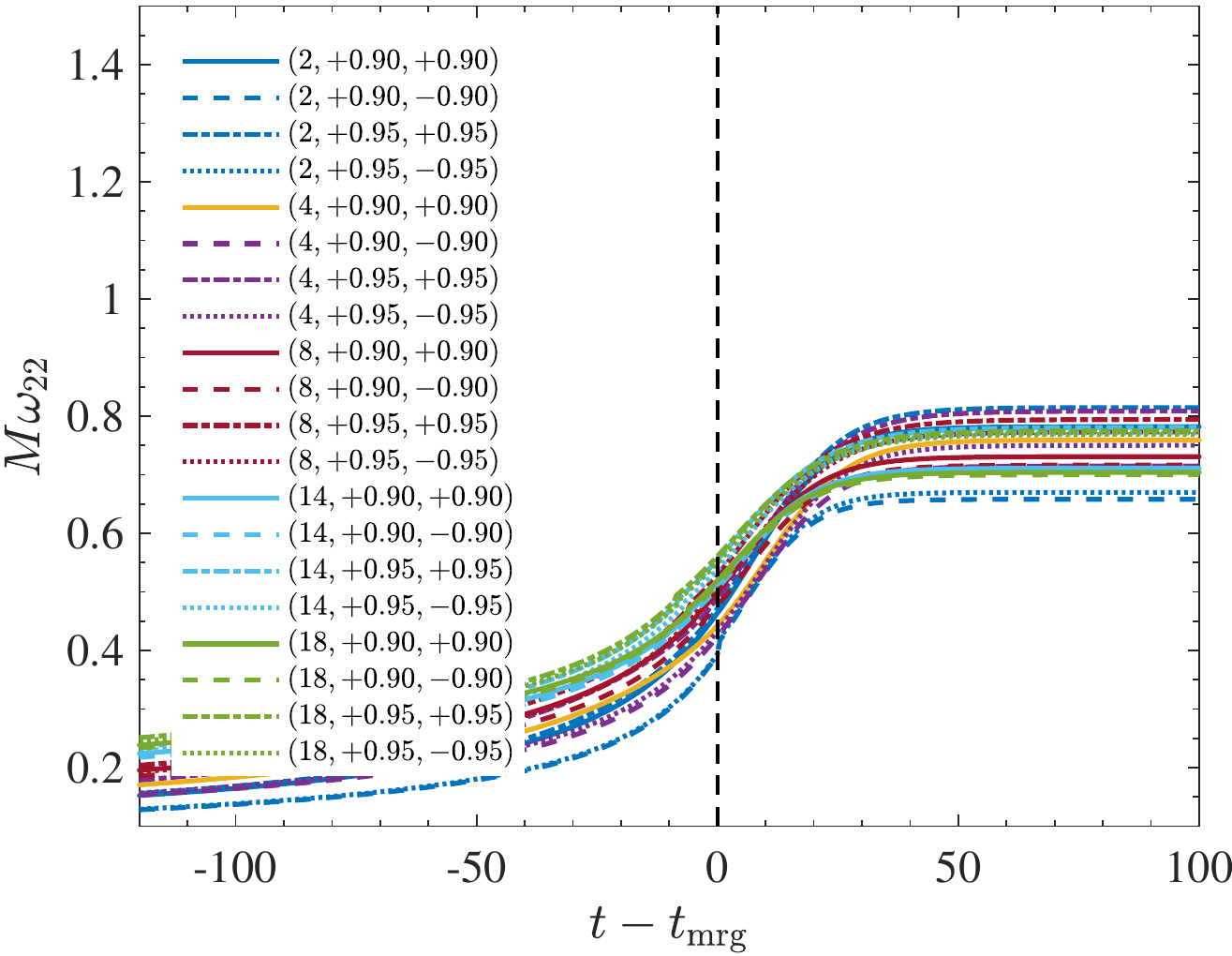}
\caption{\label{fig:HighSpin} Sanity check of EOB waveforms amplitude (top) and frequency (bottom) 
for several mass ratios and large spins aligned with the orbital angular momentum. The global 
consistency is highly satisfactory for both amplitude and frequency.}
\end{figure}

Finally, we test the robustness of the merger waveform provided by \TEOBResumS{} on
several specific configurations. In Fig.~\ref{fig:bohe} we cover that portion of the parameter
space listed in Table~I of Ref.~\cite{Bohe:2016gbl} (and notably covered by nonpublic 
SXS NR simulations). In addition, Figs.~\ref{fig:NegSpin}-\ref{fig:HighSpin} systematically
explore several configurations corresponding to the conditions given by Eqs.~\eqref{eq:m09}-\eqref{eq:m08}.
The figure stresses that neither the amplitude nor the frequency show any evident pathological
behavior around merger. This makes us confident that \TEOBResumS{} waveforms should
provide a reasonable approximation to the actual waveform for that region of the parameter
space. Evidently, like the case of (8,+0.85,+0.85) mentioned above, this does not a priori
guarantee that, had we at hand long NR simulations for such parameters,
we would get a phasing consistent with the numerical error, since modifications of $c_3$
might be needed. However, we think that constructing a waveform without evident pathologies
is already a good achievement seen the lack of NR-based complementary information
in these corners of the parameter space.

\section{Binary Neutron Stars}
\label{sec:bns}

General relativity predicts that the GW signal emitted by the quasi-circular
inspiral and plunge of BNSs is a chirp-like signal qualitatively similar to
that of a BBH system, but modified due to the presence of tidal effects.
At leading PN order, the latter arise because the gravitational field
of each star induces a multipolar deformation on the companion that
makes the binary interaction potential more attractive.
This means that,compared to the pure space-time BBH process,
the coalescence process is faster. Quadrupolar leading-order tidal interactions enter the dynamics 
at the 5th post-Newtonian order~\cite{Damour:1983a,Damour:1992qi,Racine:2004hs,Flanagan:2007ix,Damour:2009wj,Vines:2011ud}.
The impact on the phase evolution, however, is 
significant already at GW frequencies
$ f_{\mathrm{GW}}\gtrsim150\mathrm{Hz}$~\cite{Hinderer:2009ca}
and becomes the dominant effect towards the end of the inspiral~\cite{Bernuzzi:2014kca}. 
The magnitude of the tidal interaction is quantified
by a set of dimensionless tidal polarizability coefficients
for each star. The dominant one is usually addressed in the literature as ``tidal deformability'' and is defined as
\be
\label{eq:def_Lam2}
\Lambda_2 = \dfrac{2}{3}k_{2} \left(\dfrac{c^2}{G} \dfrac{R_*}{M_*}\right)^5,
\ee
where $k_{2}$ is the quadrupolar gravito-electric Love number
and $(R_*,M_*)$ are the NS areal radius and mass~\cite{Hinderer:2007mb,Damour:2009vw,Binnington:2009bb}.
The $\Lambda_\ell$ parameters are strongly dependent on the
NS internal structure; thus, their measurement provides a constraint
on the equation of state of cold degenerate matter at supranuclear
densities\footnote{Black holes are not deformed in this way;
  black hole static perturbations lead to
  $k_2=0$~\cite{Damour:2009va,Damour:2009wj,Binnington:2009bb,Gurlebeck:2015xpa}}. 
Reference~\cite{TheLIGOScientific:2017qsa} provided the first measure of
$\Lambda_2$ from GW data, setting upper limits and allowing
to disfavor some of the stiffest EOS models. 

\subsection{Main features}

Our starting point for describing the BNS evolution up to the merger is the model
discussed in Ref.~\cite{Bernuzzi:2014owa}, where the point-mass $A_0$ potential
(formerly denoted as $A(r)$) is augmented by a gravitational self-force (GSF)-informed
tidal contribution~\cite{Bini:2014zxa}. Following~\cite{Damour:2009wj},
the complete EOB potential is written as 
\be
\label{eq:A0_p_AT}
A = A_{0} + A_{T}^{(+)} \ ,
\ee
where
\be
\label{AT}
A_T^{(+)}(u;\nu)\equiv - \sum_{\ell=2}^{4} \left[ \kappa_{A}^{(\l)}
  u^{2\ell+2}\hat{A}^{(\ell^+)}_A + ({A}\leftrightarrow {B})\right] \ 
\ee 
models the gravito-electric sector of the interaction, with
$u\equiv 1/r$. In the expression above, the $\l=2,3,4$ tidal
coupling constants are defined as  
\begin{subequations}
\label{kappal}
\begin{align}
\kappa_A^{(\ell)} &= 2 \dfrac{X_B}{X_A} \left(\frac{X_A}{{\cal C}_A}\right)^{2\l+1} k^A_\l \ , \\
\kappa_B^{(\ell)} &= 2 \dfrac{X_A}{X_B} \left(\frac{X_B}{{\cal C}_B}\right)^{2\l+1} k^B_\l  \ ,
\end{align}
\end{subequations}
in which ${\cal C}_{A,B}=M_{A,B}/R_{A,B}$ are the compactness
of the two stars, $R_{A,B}$ their areal radii, while $k_\l^{A,B}$ are the
dimensionless relativistic Love
numbers~\cite{Damour:1983a,Hinderer:2007mb,Damour:2009vw,Binnington:2009bb,Hinderer:2009ca}.

At leading order, tidal interactions are
fully encoded in the total dimensionless 
{\it quadrupolar} tidal coupling constant
\be
\label{eq:def_k2T}
\kappa^T_2 \equiv \kappa^{(2)}_A + \kappa^{(2)}_B.
\ee
The above parameter is key to discovering and to interpreting EOS
quasi-universal relations for BNS merger
quantities~\cite{Bernuzzi:2014kca,Bernuzzi:2015rla,Zappa:2017xba}. In
GW experiments, however, one often measures 
separately $(\Lambda^A,\Lambda^B)$ and the masses~\cite{DelPozzo:2013ala,Agathos:2015uaa,TheLIGOScientific:2017qsa}.
The expression relating $\kappa_2^T$ to
$(\nu,\Lambda_2^A,\Lambda_2^B)$ can be easily obtained by inserting Eq.~\eqref{eq:def_Lam2} into Eq.~\eqref{eq:def_k2T}
and reads
\begin{align}
\kappa_2^T\left(\nu;\Lambda_2^A,\Lambda_2^B\right) &= \dfrac{3}{8}\nu\bigg[\left(\Lambda_2^A+\Lambda_2^B\right)\left(1+3 X_{AB}^2\right)\nonumber\\
           &+\left(\Lambda_2^A-\Lambda_2^B\right)X_{AB}\left(3+X_{AB}^2\right)\bigg]\ .
\end{align}

The relativistic correction factors $\hat{A}^{(\ell^+)}_A$
formally include all the high PN corrections to the
leading-order tidal interaction. The particular choice of 
$\hat{A}^{(\ell^+)}_A$ defines a particular \texttt{TEOB} model. For example,
the PN-expanded next-to-next-to-leading-order (NNLO) tidal model is
given by the, fractionally 2PN accurate, expression 
\be
\label{teob_nnlo}
\hat{A}_A^{(\ell^+ )}(u)= 1+ \alpha^{(\ell)}_1 u+
\alpha^{(\ell)}_2 u^2 \ \ \ {[\rm NNLO]} \ ,
\ee
with $\alpha^{(2),(3)}_{1,2}\neq 0$ computed analytically and
$\alpha^{(4)}_{1,2}=0$~\cite{Bini:2012gu}. This \TEOBNNLO{} model has
been compared against NR simulations in \cite{Bernuzzi:2012ci,Bernuzzi:2014owa}.
Significant deviations are observed during the last 2-3 orbits before
merger at dimensionless GW frequencies
$M\omega_{22}\gtrsim0.8$, that roughly correspond to the GW frequency
of the stars' contact.

The \TEOBResum{} model is defined from \TEOBNNLO{} by replacing
the $\ell=2$ term in \eqref{teob_nnlo} with the expression
\begin{align}
  \label{hatA2}
  \hat{A}^{(2^+)}_A(u) &= 1 + \dfrac{3u^2}{1-r_{\rm LR} u} 
  + \dfrac{X_A \tilde{A}_1^{(2^+) \rm 1SF}}{(1-r_{\rm LR} u)^{7/2}} 
    + \dfrac{X_A^2\tilde{A}_2^{(2^+) \rm 2SF}}{\left(1-r_{\rm
      LR}u\right)^{p}}, 
\end{align}
where $p=4$ and the functions $\tilde{A}_1^{(2^+)\rm 1SF}(u)$ and
$\tilde{A}_2^{(2^+)\rm 2SF}(u)$ are given 
in~\cite{Bini:2014zxa}, obtained by fitting to numerical data from \cite{Dolan:2014pja}.
The key idea of \TEOBResum{} is to use as pole location in Eq.~\eqref{hatA2}
the light ring $r_{\rm LR}(\nu;\kappa_A^{(\ell)})\,$ of the \TEOBNNLO{}
model, i.e., the location of the maximum of $A^{\rm
  NNLO}(r;\,\nu;\,\kappa_A^{(\ell)})/r^2$.
%
\TEOBResum{} is completed
with a resummed waveform \cite{Damour:2008gu} that includes the NLO tidal
contributions computed in~\cite{Damour:2009wj,Vines:2010ca,Damour:2012yf}. 
\TEOBResum{} is consistent with state-of-the-art NR simulations up to
merger~\cite{Bernuzzi:2014owa}. Consistently with
the BBH case, we here conventionally {\it define} the BNS merger as the peak of the 
$\ell=m=2$ amplitude of the strain waveform. The results
of~\cite{Bernuzzi:2014owa} span a sample of equation of states
(EOS) and consequently a large range of the tidal coupling
parameters. Such results were later confirmed by Hotokezaka et al.~\cite{Hotokezaka:2015xka,Hotokezaka:2016bzh}.
Similarly, Ref.~\cite{Dietrich:2017feu} showed that \TEOBResum{} is
consistent with an alternative tidal EOB model that does not incorporate GSF-driven
information but instead includes a way of accounting for the $f$-mode
oscillations of the NS excited during the orbital evolution~\cite{Hinderer:2016eia}.
A ROM version of \TEOBResum{} of Ref.~\cite{Bernuzzi:2014owa}
exists~\cite{Lackey:2016krb} and it is implemented in {\tt LAL}
under the name \TEOBResumROM.
In conclusion, despite a certain amount of approximations used to
build the model, we take the tidal EOB-model of
Ref.~\cite{Bernuzzi:2014owa} as our current best waveform
approximant for coalescing nonspinning BNS up to merger.
In the next Section, we use \TEOBResum{} as a starting point to
construct a BNS waveform model that puts together both tidal and spin effects.

\subsection{EOB formalism for self-spin term}

The spins of the two NSs (or in general of two deformable bodies) can be
easily incorporated in the formalism of Ref.~\cite{Damour:2014sva}.
Let us describe a two-step procedure starting from the case where 
the spin-spin terms are not present. This corresponds to posing the 
centrifugal radius $r_{c}=r$ in the framework of Ref.~\cite{Damour:2014sva},
i.e. Eq.~\eqref{eq:rc2_nlo} above.
In this case, moving from spinning BBHs to spinning BNSs is procedurally straightforward, since
the only trivial change is to replace the point-mass potential with the tidally augmented one.
The gyro-gravitomagnetic function $G_{S}$ and $G_{S_{*}}$ are the same as in the BBH case, 
and are resummed taking their Taylor-inverses as discussed in~\cite{Damour:2014sva}. 
A choice needs to be made for what concerns the NNNLO
effective parameter $c_{3}$, that for BBHs was tuned using NR data. 
Here we decide to simply fix it to zero. The reason behind this choice is that $c_{3}$ is an 
effective correction that depends on spin-square terms that are different in BBHs 
and BNSs and thus it is safer to drop it here. We have indeed  
explored the effect of keeping the BBH value of $c_3$ for
$\chi_A=\chi_B=0.1$ comparing with the BNS NR data corresponding to the SLy 
EOS and $1.35M_\odot+1.35M_\odot$. We find that such effect is not significant
  because it enters at high PN order in a frequency regime that is not really reached in
  a BNS system.

For what concerns spin-spin effects, it turns out that it is very easy to incorporate them
into the EOB model at {\it leading-order} (LO) also in the presence of matter objects like
NSs\footnote{Since the spin magnitude of each NS composing
the binary is expected to be small ($\chi \lesssim 0.1$), we may a priori expect this order
of approximation to be sufficient, although the corresponding Hamiltonian at NLO has been
obtained recently with different approaches~\cite{Levi:2014sba}.}.
When we talk of spin-spin interaction, let us recall that the PN-expanded Hamiltonian is made
by three terms: the mutual interaction term, $H_{S_{A}S_{B}}$, and the two self-spins ones 
$H_{S_{A}S_{A}}$ and $H_{S_{B}S_{B}}$. These two latter terms originate from the interaction
of the monopole $m_{B}$ with the spin-induced quadrupole moment of the spinning black
hole of mass $m_{A}$ and vice versa. For a NS, the same physical effect exists, but the spin-induced
quadrupole moment depends on the equation of state (EOS) by means of some, EOS-dependent, 
proportionality coefficient~\cite{Poisson:1997ha}. As we have seen above,  for BBHs,
Ref.~\cite{Damour:2014sva} introduced a prescription to incorporate into the EOB Hamiltonian
all three spin-spin couplings (at NLO)
in resummed form, by including them inside a suitable centrifugal radius $r_{c}$. This quantity
mimics, in the general, comparable-mass case, the same quantity that can be defined in the case  
of the Hamiltonian of a test particle  around a Kerr black hole. In this latter case, this takes 
into account the quadrupolar deformation of the hole due to the black hole rotation. 
For comparable-mass binaries, this may be thought as a way of incorporating the quadrupolar 
deformation of each black hole induced by its rotation. At LO, the definition of the
centrifugal radius of Eq.~\eqref{eq:rc2_nlo} simply reads
\be
\label{eq:rc2bbh}
r_c^2 = r^2 + \hat{a}^2_{0}\left(1 + \dfrac{2}{r}\right) .
\ee
where we recall that  the dimensionless effective Kerr spin is 
\be
\label{a0bbh}
\hat{a}_{0}=\tilde{a}_{A}+\tilde{a}_{B}
\ee
with $\tilde{a}_{A,B}=X_{A,B} \chi_{A,B}$. The use of these spin variables
is convenient for several reasons:
(i) the analytical expressions for spin-aligned binaries are nicely simplified and shorter 
compared to other standard notations~\footnote{Like, for example, the symmetric  
$\chi_{S}\equiv (\chi_{A}+\chi_{B})/2$ and antisymmetric  $\chi_{S}\equiv (\chi_{A}-\chi_{B})/2$
combinations of the dimensionless spins, or $S_{\ell}\equiv S_{A}+S_{B}$ and $\Sigma_{\ell}=X_{B}S_{B}-X_{A}S_{A}$ 
are typically used to express PN results.}; (ii) in the large mass ratio limit $M_{B}\ll M_{A}$,
one has that $\tilde{a}_{A}$ becomes the dimensionless spin of the massive black hole
of mass $M_{A}\approx M$, while $\tilde{a}_{B}$ just reduces to the usual spin-variable
of the particle $\sigma\equiv S_{B}/(M_{A}M_{B})$. 

Next-to-leading order spin-spin effects can be incorporated in a different fashion 
depending on whether the spins are generic or aligned with the orbital angular 
momentum. This is still ongoing work that needs further investigation~\cite{Balmelli:2015zsa}.
In the case of two NSs the recipe we propose
here to include spin-spin couplings at LO is just to replace the definition of the
effective spin $\hat{a}_{0}$ in Eq.~\eqref{eq:rc2bbh} by the following quadratic
form of $\tilde{a}_{A}$ and $\tilde{a}_{B}$
\be
\label{a0bns}
\hat{a}_{Q}^{2}= C_{QA}  \tilde{a}_A^2 + 2 \tilde{a}_A \tilde{a}_B + C_{QB} \tilde a_B^2
\ee
where $C_{QA}$ and $C_{QB}$ parametrize the quadrupolar deformation acquired by
each object due to its spin~\footnote{The notation $C_{Qi}$ we adopt here is mediated from
Ref.~\cite{Levi:2014sba} and we remind the reader that this quantity is identical
to the parameter $a$ in Poisson~\cite{Poisson:1997ha} and $C_{ES^2}$ of Ref.~\cite{Porto:2008tb}.
It is also the same parameter called $\kappa_{i}$ in Boh\'e et al.~\cite{Bohe:2015ana}.}.
For a black hole, $C_{Q}=1$ and in this case Eq.~\eqref{a0bns} coincides with
Eq.~\eqref{a0bbh}. For a NS (or any other ``exotic'' object different from a black hole,
like a boson star~\cite{Sennett:2017etc}) $C_{Q}\neq 1$ and needs to be computed starting
from a certain equation of state (see below). We can then follow Ref.~\cite{Damour:2014sva} 
and the EOB Hamiltonian will have precisely the same formal structure of the BBH case. 
In particular, the complete equatorial $A$ function entering $\hat{H}_{\rm orb}^{\rm eff}$ reads
\be
A(r,\nu,S_{i},\kappa_{i},C_{Qi})= \left[\dfrac{1 + 2 u_{c}}{1+2u}A_{\rm orb}(u_{c},\nu,\kappa_{i})\right]_{u_c(u,S_i,C_{Qi})},
\ee
where $u_c\equiv 1/r_c$ is obtained from Eq.~\eqref{eq:rc2bbh} and~\eqref{a0bns} and
we indicated explicitly the dependence on the various EOS-dependent parameters.
Note that $A_{\rm orb}$ is here depending explicitly on the tidal parameters $\kappa_{i}$,
because this is meant to be the sum of the point-mass $A$ function plus the tidal
part of the potential used in Ref.~\cite{Bernuzzi:2014owa} but everything is now 
taken as a function of $u_{c}$ instead of $u$.
One easily checks that, by PN-expanding the spin-dependent EOB Hamiltonian, 
as given by Eqs.~(23), (24) and (25) of~\cite{Damour:2014sva}, the LO spin-spin 
term coincides with the corresponding one of the ADM Hamiltonian given in 
Eqs.~(8.15) and (8.16) of~\cite{Levi:2014sba}, that in our notation just 
rereads as
\be
\hat{H}_{\rm ss_{\rm LO}}^{\rm ADM}=-\dfrac{1}{2 r_{\rm ADM}^3}\left\{C_{QA}\tilde{a}_A^2 + 2 \tilde{a}_A\tilde{a}_B+C_{QB}\tilde{a}_B^2\right\},
\ee
i.e $\hat{H}_{\rm ss}^{\rm ADM}=-\hat{a}_Q^2/(2r_{\rm ADM}^3)$ using Eq.~\eqref{a0bns}.
Since at this PN order the useful relation between the  ADM radial separation $r_{\rm ADM}$ 
and the EOB radial separation is just $r=r_{\rm ADM}$, it is immediate to verify the equivalence
of the two results.

Incorporating the full LO spin-spin interaction in the waveform, including monopole-quadrupole
terms, is similarly straightforward.  First, following Ref.~\cite{Damour:2014sva}, Eq.~(80) there, 
we recall that, for BBHs, this is done by including in the residual amplitude correction to the $(2,2)$ 
waveform a spin-dependent term of the form
\be
\rho_{22}^{\rm SS_{LO}}=c^{\rm SS}_{\rm LO}x^{2}=\dfrac{1}{2}\hat{a}^{2}_{0} x^{2}.
\ee
The monopole-quadrupole effect is then included by just replacing $\hat{a}_{0}^{2}$ 
by $\hat{a}_{Q}^{2}$ from Eq.~\eqref{a0bns}. One then verifies that, after PN-expanding
the resummed EOB flux, the corresponding LO spin-spin term coincides with the LO term 
for spin-aligned, circularized binaries, given in Eq.~(4.12) of Ref.~\cite{Bohe:2015ana}. 
Such Newton-normalized, spin-spin flux contributions, once rewritten using the 
$(\tilde{a}_{1},\tilde{a}_{2})$ spin variables, just gets simplified as
\be
\label{FssLO}
\hat{\cal F}_{\rm SS}^{\rm LO}=\Bigg\{\tilde{a}_{A}^{2}\left(\dfrac{1}{16}+2 C_{QA}\right)+\dfrac{31}{8}\tilde{a}_{A}\tilde{a}_{B}+\tilde{a}_{2}^{2}\left(\dfrac{1}{16}+2 C_{QB}\right)\Bigg\}x^{2},
\ee
so that the $A\leftrightarrow B$-symmetry is apparent\footnote{To obtain this
  result from Eq.~(4.12) of Ref.~\cite{Bohe:2015ana}
we recall the connection between the notations and spin variables: $\kappa_{i}=C_{Qi}$; $\kappa_{\pm}\equiv \kappa_{A}\pm \kappa_{B}$; $S_{\ell}=X_{A}\tilde{a}_{A}+X_{B}\tilde{a}_{B}$; $\Sigma_{\ell}=\tilde{a}_{B}-\tilde{a}_{A}$; $\delta=X_{A}-X_{B}=\sqrt{1-4\nu}$ and thus $X_{A}=(1+\sqrt{1-4\nu})/2$}.
This can be obtained by directly expanding the EOB-resummed  flux as defined in Ref.~\cite{Damour:2014sva}.
Actually, for this specific calculation, it is enough to consider the $(2,2)$ and $(2,1)$ waveform modes, the first
at LO in the spin-spin and spin-orbit interaction, while the latter only at LO in the spin-orbit interaction. 
The corresponding residual amplitudes, taken from Eqs.~(79), (84), (86), (89) and (90) of~\cite{Damour:2014sva}, read
\begin{align}
\rho_{22}      & = \rho_{22}^{\rm orb}+\rho_{22}^{\rm S},\\
X_{AB}f_{21} &=X_{AB}\left(\rho_{21}^{\rm orb}\right)^{2}+\tilde{f}_{21}^{\rm S},
\end{align}
where $\rho_{22}^{\rm S}$ is assumed here to incorporate only the LO spin-orbit
and spin-spin contribution
\begin{align}
\rho_{22}^{\rm S}&=c_{\rm SO}^{\rm LO} x^{3/2}+c_{\rm SS}^{\rm LO}x^{2}\nonumber\\
&=-\left[\dfrac{\hat{a}_{0}}{2}+\dfrac{X_{AB}}{6}(\tilde{a}_{A}-\tilde{a}_{B})\right]x^{3/2}+\dfrac{1}{2}\hat{a}_{Q}^{2} x^{2},\\
\tilde{f}_{21}^{\rm S}&=-\dfrac{3}{2}(\tilde{a}_{A}-\tilde{a}_{B}).
\end{align}
One verifies that, by keeping the orbital terms consistently, using these expressions in Eq.~(74) and (75)
of~\cite{Damour:2014sva}, one eventually obtains Eq.~\eqref{FssLO} above.
As a further check, we have also verified that the use of Eq.~\eqref{a0bns} is also fully consistent with the
calculation of the multipolar waveform amplitude $h_{22}$ that was done by S.~Marsat and A.~Boh\'e and
kindly shared with us before publication~\cite{Marsat_private}. 

At this stage, we have a complete analytical model that is able to blend, in a resummed (though
approximate) way spin and tidal effects. The model is complete once all the EOS dependent
information, schematically indicated by $\Lambda$ is given.
More precisely, the procedure is as follows: for a given choice of the EOS, one fixes the 
compactness  ${\cal C}$ (or the mass of the NS), which defines its equilibrium structure.
Then, following Ref.~\cite{Damour:2009vw} (see also Refs.~\cite{Hinderer:2007mb,Binnington:2009bb,Hinderer:2009ca}),
one computes the corresponding dimensionless Love numbers $(k_{2},k_{3},k_{4})$ as they
appear in the EOB potential. At this stage, the only missing piece is the EOS-dependent 
coefficient $C_{Q}$ for the two objects. Luckily, this can be obtained easily by taking advantage 
of the so-called I-Love-Q quasi-universal relations found by Yunes and Yagi~\cite{Yagi:2013bca,Yagi:2013awa}. 
In particular,  following Ref.~\cite{Yagi:2013awa}, defining $x\equiv \log(\Lambda_2)$
one has that, for each binary, the quadrupole coefficient $C_{Q}$ can be obtained as
 \begin{align}
\log \left(C_{Q}\right)&=0.194 + 0.0936x + 0.0474{x}^2\nonumber\\
&- 4.21\times10^{-3}{x}^3 + 1.23\times10^{-4}{x}^4 \ .
 \end{align}
 Since $C_{Q}$ is 1 for a BH but it is larger for a NS, depending on the EOS one is
 expecting a relevance of the monopole-quadrupole interaction terms. This was already
 pointed out by Poisson long ago~\cite{Poisson:1997ha} and more recently by Harry and Hinderer~\cite{Harry:2018hke}.

\subsection{Comparison with NR data}
 \begin{table*}[t]
   \caption{\label{tab:BNS_config}Equal-mass BNS configurations considered in this work.
     From left to right the column reports: the EOS, the gravitational mass of each star,
     the compactness, the quadrupolar dimensionless Love numbers, the leading-order tidal
     coupling constant $\kappa_2^T$, the corresponding value of the quadrupolar
     ``tidal deformability'' for each object, $\Lambda_2^{A,B}$, Eq.~\eqref{eq:def_Lam2},
     the dimensionless spin magnitude and the spin-induced quadrupole momenta $C_{QA,QB}$.}
   \begin{center}
 \begin{ruledtabular}
   \begin{tabular}{cclccccccccc}
     name &EOS & $M_{A,B}[M_\odot]$ & $C_{A,B}$ & $k_2^{A,B}$ & $\kappa_2^T$ &  $\Lambda^{A,B}_2$  & $\chi_{A,B}$  & $C_{QA,QB}$ \\
     \hline
     BAM:0095 &SLy      & 1.35             & $0.17$   & 0.093    & 73.51        & 392    &0.0  & 5.491 \\
     BAM:0039 &H4     & 1.37             & $0.149$  & 0.114    &  191.34      & 1020.5 &0.141& 7.396\\
     BAM:0064&MS1b   & 1.35             & $0.142$  & 0.134    &  289.67      & 1545   &0.0 & 8.396\\

 \end{tabular}
 \end{ruledtabular}
 \end{center}
 \end{table*}

\begin{figure*}[t]
\center
\includegraphics[width=0.45\textwidth]{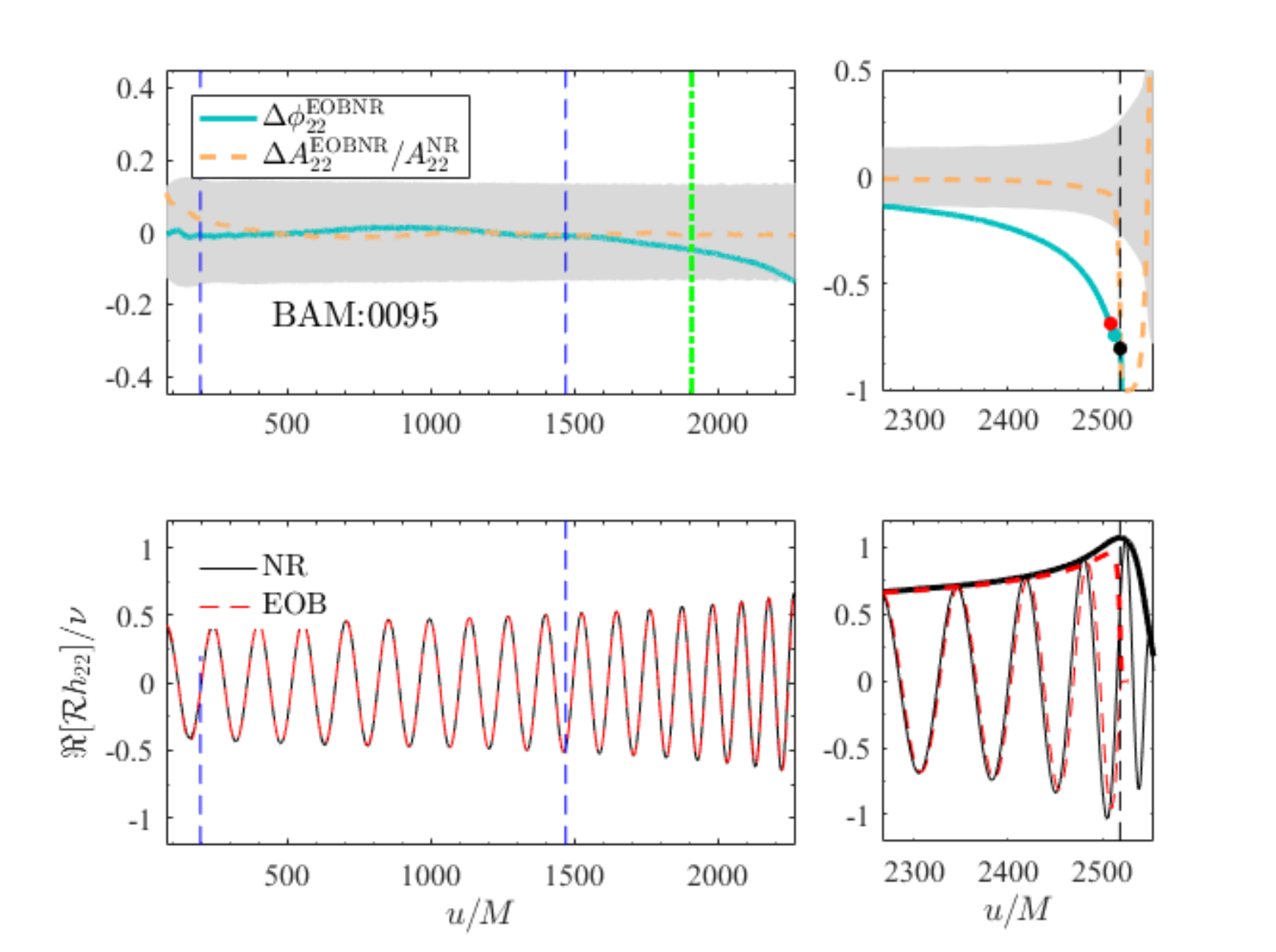}
\includegraphics[width=0.45\textwidth]{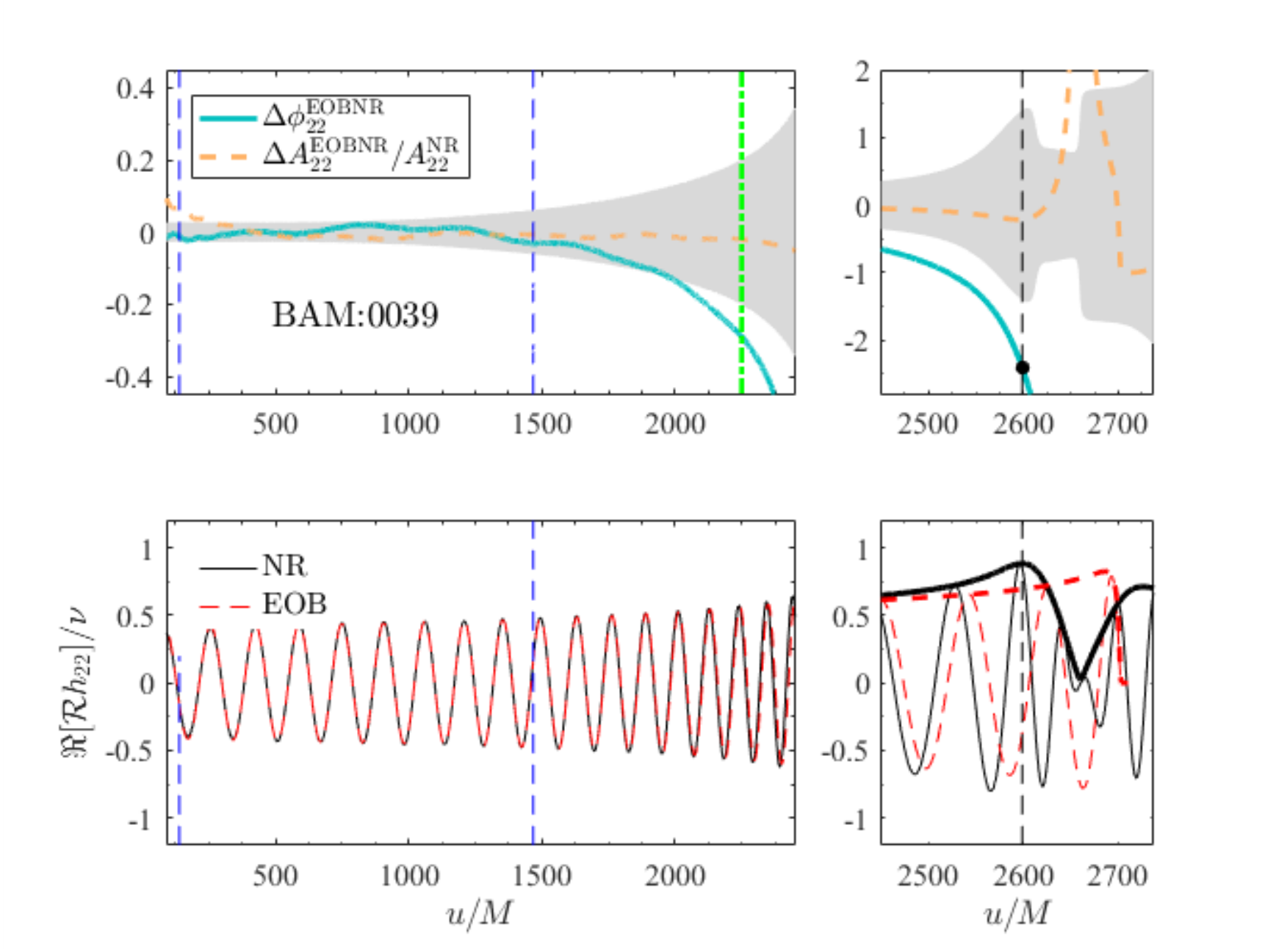}\\
\caption{\label{fig:bns_ms1b_sly}Phasing comparison between \BAM{}
  and \TEOBResumS{} waveforms for the SLy and Ms1b equal-mass BNS configurations
  of Table~\ref{tab:BNS_config}. The EOB and NR waveforms, once aligned during during
  the early inspiral (approximately over the first $1500M$ of evolution), are compatible,
  within the NR uncertainty (gray area in the figures) essentially up to the NR merger point,
  defined as the peak of the waveform amplitude $|h_{22}|$. Note however that the errors
  are {\it larger} for the MS1b configuration. The time marked by the vertical green
  line corresponds to 700Hz.}
\end{figure*}

\begin{figure}[t]
\includegraphics[width=0.45\textwidth]{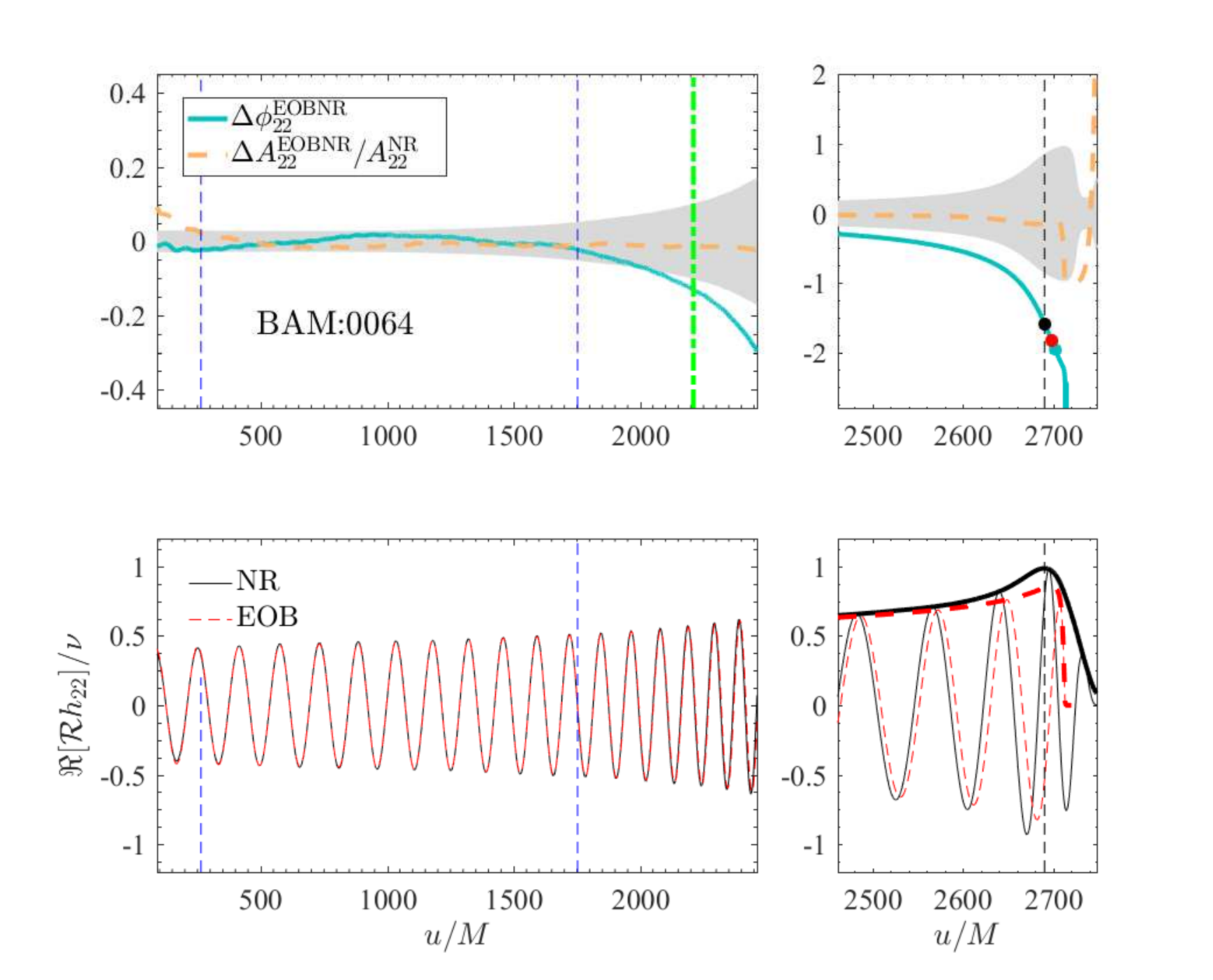}
\caption{\label{fig:bns_h4}Phasing comparison between \BAM{} and \TEOBResumS{}
  waveforms, effect of spin (H4 EOS, see Table~\ref{tab:BNS_config}).
  The figure refers to spinning binary with dimensionless spins $\chi_A=\chi_B\approx 0.14$.
  NR and EOB waveforms are still compatible, within the NR uncertainty (gray area in the figures),
  up to the NR merger point. The time marked by the vertical
  green line corresponds to 700Hz.}
\end{figure}
We verify the accuracy of \TEOBResumS{} against error-controlled NR
waveforms obtained from the evolution of spinning 
and eccentricity reduced initial data using multiple
resolutions. Initial data are constructed in the constant rotational velocity
 formalism using the \texttt{SGRID} code
 \cite{Tichy:2011gw,Tichy:2012rp}. The residual eccentricity of the
 initial data is
 reduced to typical values $e\sim10^{-3}-10^{-4}$ following the
 procedure described in \cite{Dietrich:2015pxa}.
 The main properties of the BNS configurations discussed in this work
 are listed in Table~\ref{tab:BNS_config}. The initial data are
 then evolved with \BAM{} \cite{Brugmann:2008zz,*Thierfelder:2011yi}
 using a high-order method 
 for the numerical fluxes of the general-relativistic hydrodynamics
 solver \cite{Bernuzzi:2016pie}. 

The \BAM{} waveforms employed here were produced and discussed in 
\cite{Dietrich:2017aum,Dietrich:2018upm}. We perform multiple
resolution runs, up to grid resolutions that allow us to make an unambiguous
assessment of convergence. We find a clear second order convergence in
many cases and build a consistent error budget following the convergence
tests~\cite{Bernuzzi:2016pie}. For this work we additionally checked
some of the waveforms by performing additional simulation with the
\THC{} code~\cite{Radice:2013xpa,Radice:2013hxh}.
The comparison with an independent code allows us to check some of the
systematics uncertainties that affect BNS simulations 
\cite{Bernuzzi:2012ci,Radice:2013xpa,Radice:2013hxh}.
We find that the two codes produce consistent waveforms.
Results are summarized in Appendix~\ref{sec:BAMvsTHC}. 

Figures~\ref{fig:bns_ms1b_sly} and~\ref{fig:bns_h4} illustrate EOB/NR phasing comparison.
The EOB waveforms are aligned, fixing a relative time and phase shift, to the NR
ones in the inspiral region marked by two vertical lines on the left panels that
correspond to the same frequency interval $(\omega_L,\omega_R)$ on both the EOB
and NR time series~\cite{Baiotti:2011am}. The alignment frequency intervals
are $(\omega_L,\omega_R)=(0.039,0.05)$ for BAM:0095; $(0.0365,0.045)$
for BAM:0039 and $(0.038,0.05)$ for BAM:0064.
The shaded areas in the top panels mark the NR phasing uncertainty as estimated in Appendix~\ref{sec:BAMvsTHC}.
For reference, the green, vertical line indicates the time at which the 700~Hz frequency is crossed.
The figure clearly illustrates that: (i) EOB and NR waveforms are {\it fully compatible}
up to our conventionally defined merger point, the peak of the $|h_{22}|$ waveform amplitude,
over the full range of values of $\kappa_2^T$ considered as well as for spins.
Interestingly, the leftmost panel of Fig.~\ref{fig:bns_ms1b_sly} also shows that
the EOB-NR phase difference towards merger is acceptably small ($< 1$~rad),
but also significantly {\it larger} than the NR uncertainty. This illustrates that,
for the first time, our NR simulations are finally mature to {\it inform} the
analytical model with some new, genuinely strong-field, information that can be
extracted from them.

The figures show that for the EOB dynamics, we typically underestimate the
effect of tides in the last orbit, since the phase of the NR data is
evolving faster (stronger tides). 
However, the opposite is true for BAM:0095.
This result is consistent with the ones of Ref.~\cite{Bernuzzi:2014owa} for
the same physical configuration (but different simulations, 
leftmost panel of Fig.~3) where one had already the indication that
for compact NS, tidal effects could be slightly overestimated with respect
to the corresponding NR description. 
Informing \TEOBResumS{} with the \BAM{} simulations
is outside the scope of the current work. However, we want to stress that this is
finally possible with our improved simulations.

\section{Contribution of self-spin terms to BNS inspiral}
\label{sec:MonQuad}

Now that we could show the consistency between the {\tt TEOBResumS} phasing
and state-of-the art NR simulations, let us investigate in more detail the
effect of spins on long BNS waveforms as predicted by our model.
First of all, let us recall that inspiralling BNS systems are not likely
to have significant spins. The fastest NS in a confirmed BNS system has
dimensionless spins $\sim 0.04$~\cite{Kramer:2009zza}. Another potential
BNS system has a NS with spin frequency of 239~Hz, corresponding to
dimensionless spin~0.2. The fastest-spinning, isolated, millisecond pulsar
observed so far has $\chi=0.04$. However, it is known that even a spin
of ~0.03 can lead to systematic biases in the estimated tidal parameters
if not incorporated in the waveform model~\cite{Favata:2013rwa,Cho:2017yaj}.
Those analysis are based on PN waveform models. A precise assessment of
these biases using \TEOBResumS{} is beyond the scope of the present work
and will hopefully be addressed in the future.
Since the most important theoretical novelty of \TEOBResumS{}
is the incorporation of self-spin effects in resummed form, our aim here is
to estimate their effect in terms of time-domain phasing up to
merger~\footnote{Note that it is currently not possible to reliably extract
  self-spin information from numerical simulations~\cite{Dietrich:2016lyp,Dietrich:2017aum}.},
notably contrasting the \TEOBResumS{} description with the standard PN one.

Before doing so, let us mention that LO, PN-expanded, self-spin terms~\cite{Poisson:1997ha}
in the TaylorF2~\cite{Damour:2000zb,Buonanno:2009zt} inspiral approximant have been used in parameter-estimation studies
by Agathos et al.~\cite{Agathos:2015uaa}, and, more recently, by Harry and Hinderer~\cite{Harry:2018hke}.
The LO  term (2PN accurate) to the frequency-domain phasing was originally computed by
Poisson~\cite{Poisson:1997ha}. Currently, EOS-dependent, self-spin information is computed
in PN theory up to 3.5PN order, so that one can have the corresponding 3.5PN accurate terms
in the TaylorF2 approximant. Let us explicitly review their computation.
Given the Fourier transform of the quadrupolar waveform as
\be
\tilde{h}_{22}\equiv \tilde{A}(f)e^{-\ii\Psi(f)},
\ee
the frequency domain phasing  of the TaylorF2 waveform approximant, that assumes the stationary
phase approximation, is obtained solving the integral given by Eq.~(3.5) of Ref.~\cite{Damour:2000zb},
\be
\label{eq:getTF2}
\Psi_f(t_f)=2\pi f t_{\rm ref}-\phi_{\rm ref}+2\int_{v_f}^{v_{\rm ref}}(v_f^3-v^3)\frac{E'(v)}{\mathcal{F}(v)}dv,
\ee
where the parameters $t_{\rm ref}$ and $\phi_{\rm ref}$ are gauge-dependent integration constants.
The $C_{Qi}$-dependent quadratic-in-spin energy and flux available in the literature at 3.5PN,
the maximum PN order actually known in this particular case, are given in Refs.~\cite{Marsat:2014xea}
and~\cite{Bohe:2015ana} respectively, where their notation $\kappa_{\pm}$ corresponds to
$\kappa_{+}=C_{QA}+C_{QB}$ and $\kappa_{-}\equiv C_{QA}-C_{QB}$.
It is important to stress that in Ref.~\cite{Levi:2014sba} a circularized spin-spin
$C_{Qi}$-dependent Hamiltonian, equivalent to the Multipolar post-Minkowskian (MPM)
result of Ref.~\cite{Marsat:2014xea} (see their Appendix~D), was computed via
effective field theory (EFT) techniques. From Eq.~\eqref{eq:getTF2}, by taking into
account all the orbital pieces at the consistent
PN order~\cite{Bernard:2016wrg,Damour:2016abl,Damour:2015isa,Damour:2014jta,Blanchet:2013haa},
one gets that the self-spin contribution is given by the sum of an LO term (2PN)~\cite{Poisson:1997ha},
an NLO term (3PN) and a LO tail\footnote{See Refs.~\cite{Blanchet:1987wq} and~\cite{Blanchet:1992br} for
  a physical insight to memory and tail effects in gravitational radiation.} term (3.5PN)
\be
\Psi_{\rm SS}^{\rm PN} = \Psi^{\rm PN,LO}_{\rm SS} + \Psi^{\rm PN, NLO}_{\rm SS} + \Psi^{\rm PN, tail}_{\rm SS}.
\ee
The LO tail term is computed here for the first time. It was obtained by expanding, at the corresponding
PN order, the EOB energy and flux adapting the procedure discussed in~\cite{Messina:2017yjg}.
These three terms explicitly read
\begin{align}
\label{eq:PsiPNqm}
\Psi^{\rm PN,LO}_{\rm SS}&= -\dfrac{75}{64\nu}\left(\tilde{a}_{A}^{2}C_{QA}+\tilde{a}_{B}^{2}C_{QB}\right)\left(\dfrac{\omega}{2}\right)^{-1/3},\\
\label{eq:PsiPNqmNLO}
\Psi^{\rm PN, NLO}_{\rm SS}&=\frac{1}{\nu}\biggl[\left(\frac{45}{16}\nu+\frac{15635}{896}\right)(C_{QA}\tilde{a}_A^2+C_{QB}\tilde{a}_B^2)\nonumber\\
  &+\frac{2215}{512}X_{AB}(C_{QA}\tilde{a}_A^2-C_{QB}\tilde{a}_B^2)\biggr]\left(\frac{\omega}{2}\right)^{1/3},\\
\label{eq:PsiPNqmNNLO}
\Psi^{\rm PN, tail}_{\rm SS}&= -\dfrac{75}{8\nu}\pi\left(\tilde{a}_{A}^{2}C_{QA}+\tilde{a}_{B}^{2}C_{QB}\right)\left(\dfrac{\omega}{2}\right)^{2/3},
\end{align}
where $\omega=2\pi M f$ denotes the circularized quadrupolar gravitational wave frequency.

To quantitatively investigate the differences between the PN-expanded and EOB-resummed
treatment of the self-spin contribution to the phase, it is convenient to use
the quantity $Q_{\omega}=\omega^{2}/\dot{\omega}$, where $\omega=\omega(t)$
is the time-domain quadrupolar gravitational wave frequency,  $\omega\equiv d\phi/dt$, 
where $\phi(t)\equiv \phi_{22}(t)$ is the phase of the time-domain
quadrupolar GW waveform $h_{22}(t)=A(t)e^{{\rm i}\phi_{22}(t)}$. 
This function has several properties that will be useful in
the present context. First, its inverse can be considered as an adiabatic
parameter $\epsilon_{\rm adiab}=1/Q_{\omega} = \dot{\omega}/\omega^2$ whose
magnitude controls the validity of the stationary phase approximation (SPA)
that is normally used to compute the frequency-domain phasing of PN approximants
during the quasi-adiabatic inspiral. Thus, the magnitude of $Q_\omega$ itself tells
us to which extent the SPA delivers a reliable approximation to the exact
Fourier transform of the complete inspiral waveform, that also incorporates
nonadiabatic effects. Let us recall~\cite{Damour:2012yf} that, as long as
the SPA holds, the phase of the Fourier transform of the time-domain quadrupolar
waveform $\Psi(f)$ is simply the Legendre transform of the quadrupolar
time-domain phase $\phi(t)$, that is
\be
\Psi(f) = 2\pi f t_{f}-\phi(t_{f})-\pi/4,
\ee
where $t_{f}$ is the solution of the equation  $\omega(t_{f})=2\pi f$.
Differentiating twice this equation one finds
\be
\label{getQomg}
\omega^{2}\dfrac{d^{2}\Psi(\omega)}{d\omega^{2}}=Q_{\omega}(\omega),
\ee
where we identify the time domain and frequency domain circular 
frequencies, i.e., $\omega_{f}=\omega(t)$.
Second, the integral of $Q_{\omega}$ per logarithmic frequency yields the phasing
accumulated by the evolution on a given frequency interval $(\omega_L,\omega_R)$,
that is
\be
\Delta \phi_(\omega_L,\omega_R)\equiv \int_{\omega_L}^{\omega_R}Q_{\omega} d\log\omega.
\ee
Additionally, since this function is free of the two ``shift ambiguities'' that
affect the GW phase (either in the time or frequency domain), it is perfectly
suited to compare in a simple way different waveform
models~\cite{Baiotti:2010xh,Baiotti:2010xh,Bernuzzi:2012ci,Damour:2012ky,Bernuzzi:2014owa}.
Then, the self-spin contribution to the PN-expanded $Q_\omega$ is given
by three terms
\be
Q_\omega^{\rm PN, SS} = Q_\omega^{{\rm SS}_{\rm PN,LO}}+ Q_\omega^{{\rm SS}_{\rm PN,NLO}} +  Q_\omega^{{\rm SS}_{\rm PN,tail}},
\ee
that are obtained from Eqs.~\eqref{eq:PsiPNqm}-\eqref{eq:PsiPNqmNNLO}
and read
\begin{align}
\label{eq:QomgMQpn}
Q_{\omega}^{{\rm SS}_{\rm PN,LO}}&=-\dfrac{25}{48\nu}\left(\tilde{a}_{A}^{2}C_{QA}+\tilde{a}_{B}^{2}C_{QB}\right)\left(\dfrac{\omega}{2}\right)^{-1/3},\\
\label{eq:QomgMQpnNLO}
Q_{\omega}^{\rm SS_{PN, NLO}}&=-\frac{1}{\nu}\biggl[\left(\frac{5}{8}\nu+\frac{15635}{4032}\right)(C_{QA}\tilde{a}_A^2+C_{QB}\tilde{a}_B^2)\nonumber\\
  &+\frac{2215}{2304}X_{AB}(C_{QA}\tilde{a}_A^2-C_{QB}\tilde{a}_B^2)\biggr]\left(\frac{\omega}{2}\right)^{1/3},\\
Q_{\omega}^{\rm SS_{PN, tail}}&=\dfrac{25}{12\nu}\pi\left(\tilde{a}_{A}^{2}C_{QA}+\tilde{a}_{B}^{2}C_{QB}\right)\left(\dfrac{\omega}{2}\right)^{2/3}.
\end{align}
\begin{figure}[t]
\center
\includegraphics[width=0.4\textwidth]{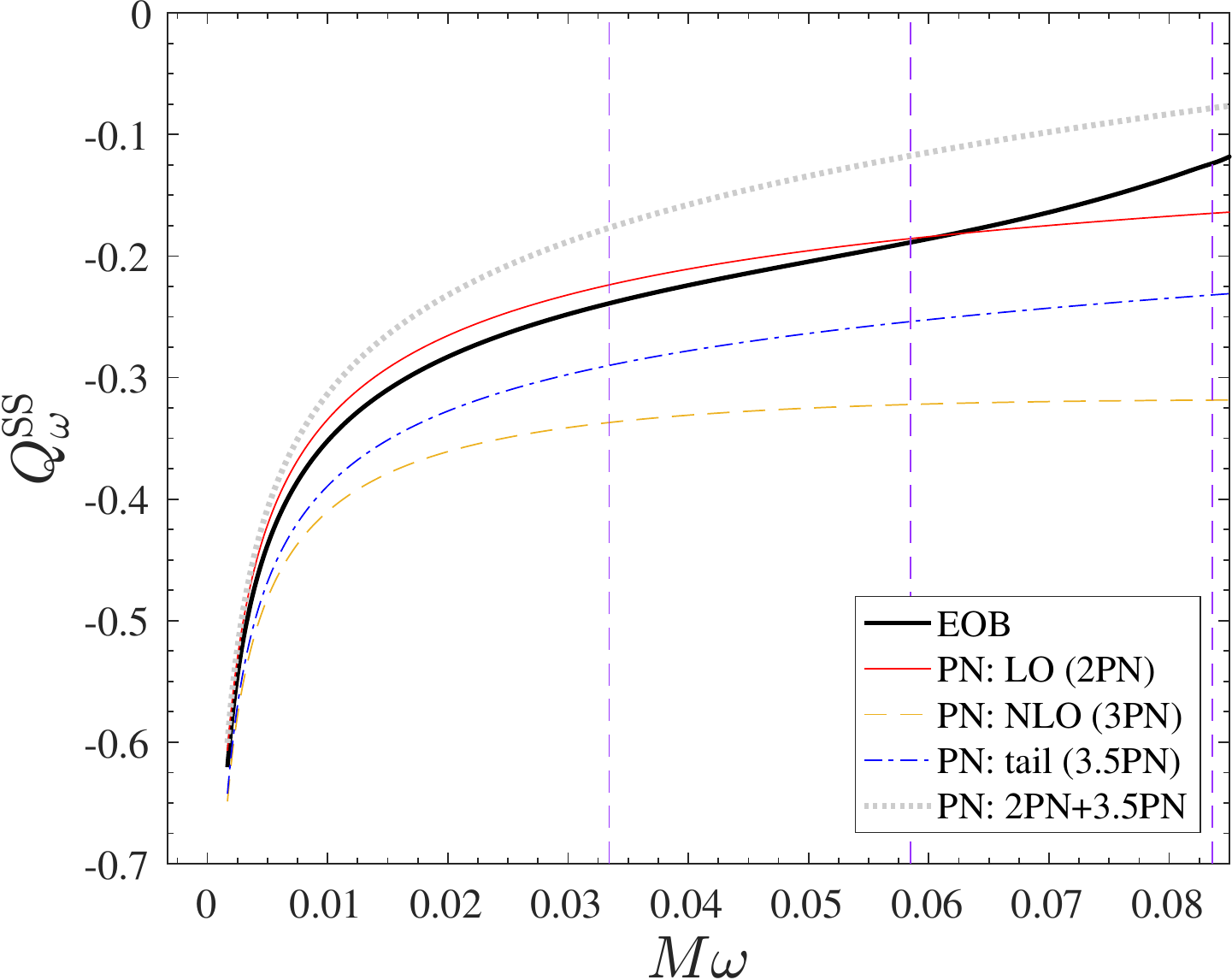}
\caption{\label{fig:QM} EOS-dependent self-spin effects on the phasing through the $Q_\omega^{\rm SS}$
  diagnostics. The figure contrasts the EOB description (incorporating LO dynamical and dissipative
  effects) with various PN approximations (see text) for the BAM:0095 tidal configuration
  with however $\chi_A=\chi_B=0.1$. The vertical lines mark respectively 400Hz, 700Hz and 1kHz.
  The EOB resummed description enhances the effect during most of the inspiral, though it
  reduces it towards merger. Consistency with all PN approximants is found in the low
  frequency regime (20Hz), though the PN regime is not yet reached there.}
\end{figure}

The corresponding function in \TEOBResumS, $Q_\omega^{\rm TEOBResumS,SS}$ is computed,
in the time domain, as follows. We perform two different runs, one with $C_{Qi}\neq0$
another with $C_{Qi}= 0$. In both cases we compute the time-domain $Q_\omega$ and
finally calculate
\be
Q_\omega^{\rm TEOBResumS,SS} \!\!= Q_\omega^{\rm TEOBResumS_{C_{Qi}\neq0}}\!\!\!-Q_\omega^{\rm TEOBResumS_{C_{Qi}=0}}.
\ee
Although the procedure is conceptually straightforward, since it only requires the computation of numerical derivatives of the time-domain phase $\phi(t)$, there are technical
subtleties in order to obtain a clean curve to be compared with the PN results.
First of all, any oscillation related to residual eccentricity coming from the initial
data, though negligible both in $\phi(t)$ or $\omega(t)$, will get amplified in $Q_\omega$
making the quantity useless. To avoid this drawback, the use of the 2PA initial data
of Ref.~\cite{Damour:2012ky}, discussed in detail in Appendix~\ref{sec:ID},
is absolutely crucial. Second, in order to explore the low-frequency regime one has to get
rid of the time-domain oversampling of the waveform, since it eventually generates high-frequency
(though low-amplitude) noise in the early frequency part of the curve. To this aim, the
raw time-domain phase $\phi(t)$ was suitably downsampled (and smoothed). Since the time-domain
output of \TEOBResumS{} is evenly sampled in time (but not in frequency) such procedure had
to be done separately on different time intervals of the complete signal
(e.g. starting from 20Hz) that are then joined together again. 

The outcome of this calculation is represented, as a black line, in Fig.~\ref{fig:QM}.
As case study, we selected the BAM:0095 configuration of Table~\ref{tab:BNS_config}
with $\chi_A=\chi_B=0.1$. To orient the reader, the vertical lines correspond to 400Hz,
700Hz and 1kHz. The figure illustrates two facts: (i) the EOB-resummed representation
of the self-spin phasing is consistent, as it should, with the PN description when
going to low-frequencies and (ii) it is stronger during most of the inspiral
(i.e.\ more attractive). More detailed analysis of the self-spin effects in
comparison with the various PN truncations displayed in the figure are discussed
in Sec.~VI of Ref.~\cite{ Dietrich:2018uni}, to which we address the interested reader.
One important information enclosed in the figure is  that the difference between the
EOB and NLO~(3PN) description of self-spin effects is nonnegligible.
It is likely that most of this difference comes from the bad behavior of
the PN-expanded NLO term. Note in fact that $Q_\omega^{\rm SS_{\rm PN,NLO}}$
has a quite large coefficient, $15635/4032\simeq 4$, (see Eq.~\eqref{eq:QomgMQpnNLO}),
that, e.g. at $M\omega\sim 0.04$, eventually yields a contribution that is comparable
to the LO one in the PN series. For this reason, we are prone to think that the
EOB description of self-spin effects, even if it is based only on the (limited)
LO self-spin term, is more robust and trustable than the straightforward PN-expanded
one. Clearly, to finally settle this question we will need to incorporate in the EOB
formalism, through a suitable $C_{Qi}$-dependent expression of the $\delta\hat{a}^2$
given in Eq.~\eqref{eq:deltaa2}, EOS-dependent self-spin effects at NLO.
This will be discussed extensively in a forthcoming study.

\section{Case study: Parameter estimation of GW150914}
\label{sec:pe}
\begin{table}
  \caption{Summary of the parameters that characterize GW150914 as found by \texttt{cpnest} and using \TEOBResumS{}
    as template waveform, compared with the values found by the LVC collaboration ~\cite{TheLIGOScientific:2016wfe}.
    We report the median value as well as the $90\%$ credible interval.
    For the magnitude of the dimensionless spins $|\chi_A|$ and $|\chi_B|$ we also report the $90\%$ upper bound.
    Note that we use the notation $\chi_{\rm eff}\equiv \hat{a}_0$ for the effective spin,
    as introduced in Eq.~\eqref{eq:hata0}.}
\begin{ruledtabular}
\begin{tabular}{p{0.66\linewidth} p{0.16\linewidth} p{0.16\linewidth}}
 & \!\!\!\!\TEOBResumS{} & \hspace{0.3cm}LVC \\
\hline
\vspace{0.005cm}Detector-frame total mass $M/\Msun$ & \vspace{0.005cm}$73.6_{-5.2}^{+5.7}$ & \vspace{0.005cm}$70.6_{-4.5}^{+4.6}$ \\
Detector-frame chirp mass $\mathcal{M}/\Msun$ & $31.8_{-2.4}^{+2.6}$ & $30.4_{-1.9}^{+2.1}$ \\
Detector-frame remnant mass $M_{f}/\Msun$ & $70.0_{-4.6}^{+5.0}$ & $67.4_{-4.0}^{+4.1}$\\
Magnitude of remnant spin $\hat{a}_{f}$ & $0.71_{-0.07}^{+0.05}$ & ${0.67}_{-0.07}^{+0.05}$ \\
Detector-frame primary mass $M_A/\Msun$ & $40.2_{-3.7}^{+5.1}$ & $38.9_{-4.3}^{+5.6}$ \\
Detector-frame secondary mass $M_B/\Msun$ & $33.5_{-5.5}^{+4.0}$ & $31.6_{-4.7}^{+4.2}$  \\
Mass ratio $M_B/M_A$ & $0.8_{-0.2}^{+0.1}$ & $0.82_{-0.17}^{+0.20}$  \\
Orbital component of primary spin $\chi_{A}$& $0.2_{-0.8}^{+0.6}$ & $0.32_{-0.29}^{+0.49}$  \\
Orbital component of secondary spin $\chi_{B}$& $0.0_{-0.8}^{+0.9}$ & $0.44_{-0.40}^{+0.50}$ \\
Effective aligned spin $\chi_{\mathrm{eff}}$ & $0.1_{-0.2}^{+0.1}$ & $-0.07_{-0.17}^{+0.16}$\\
Magnitude of primary spin $|\chi_{A}|$ & $\leq 0.7$ & $\leq 0.69$ \\
Magnitude of secondary spin $|\chi_{B}|$ & $\leq 0.9$ & $\leq 0.89$  \\
Luminosity distance $d_\mathrm{L}/\mathrm{Mpc}$ & $479_{-235}^{+188}$ & $410_{-180}^{+160}$  \\
\end{tabular}
\label{tab:pe-summary}
\end{ruledtabular}
\end{table}

\begin{figure}[t]
\center
\includegraphics[width=0.5\textwidth]{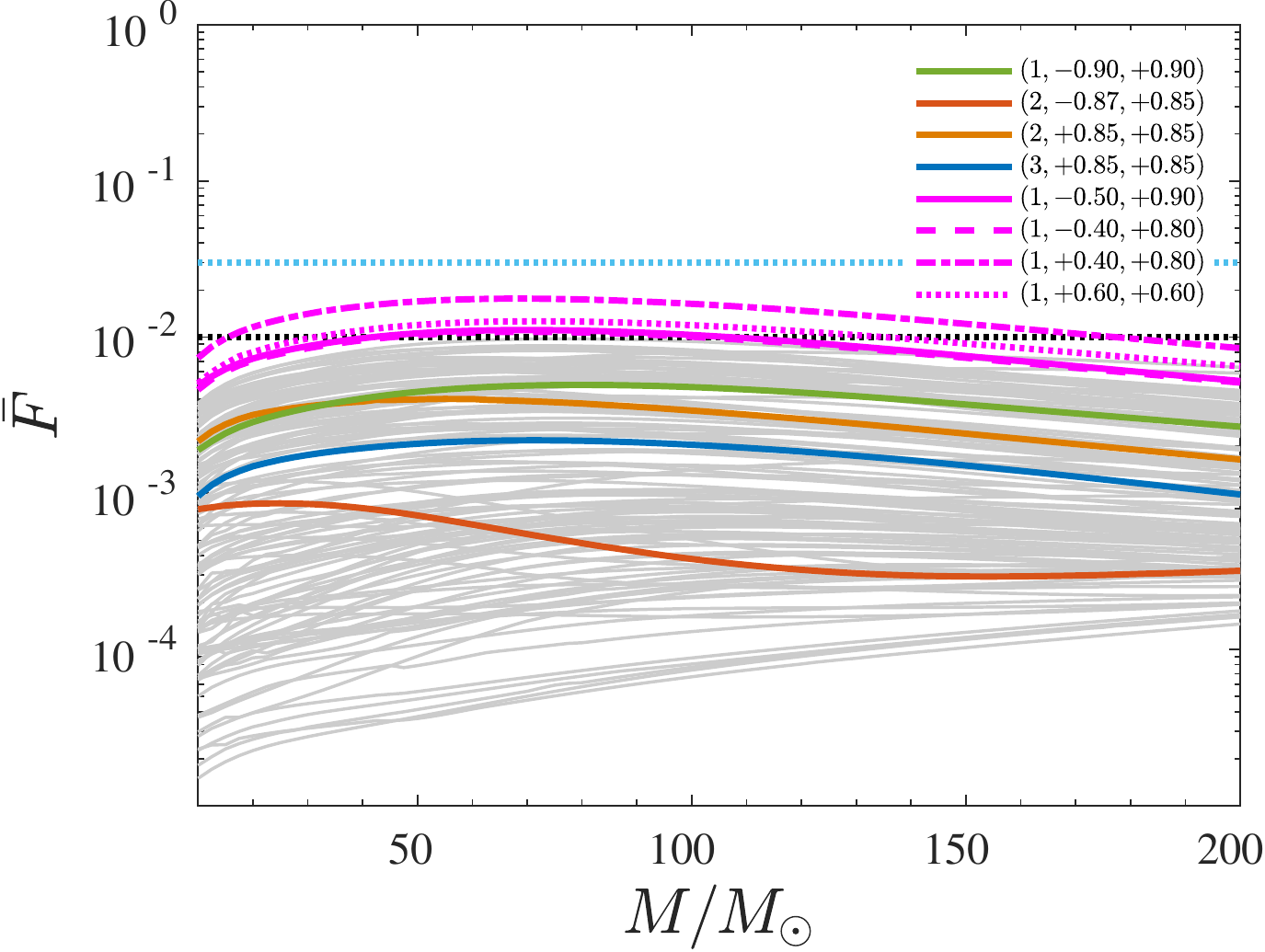}
\caption{\label{fig:sxs_noiter} Unfaithfulness comparison between \TEOBResumS{} and SXS waveforms
  obtained {\it without} iterating on the amplitude NQC parameters $(a_1,a_2)$, see Eq.~\eqref{eq:nqc_factor}.
  The performace of the model, where the parameters $(a_6^c,c_3)$ were NR-tuned {\it with}
  the iterative determination of $(a_1,a_2)$ (see Sec.~\ref{sec:mainfeats}), is slightly
  worsened with respect to Fig.~\ref{fig:SXS}, although it is still compatible with
  the $1\%$ limit. Such simplified version of \TEOBResumS{} is used for the parameter
  estimation of GW150914, with results reported in Table~\ref{tab:pe-summary}.}
\end{figure}

\begin{figure*}[t]
\center
\includegraphics[width=\textwidth]{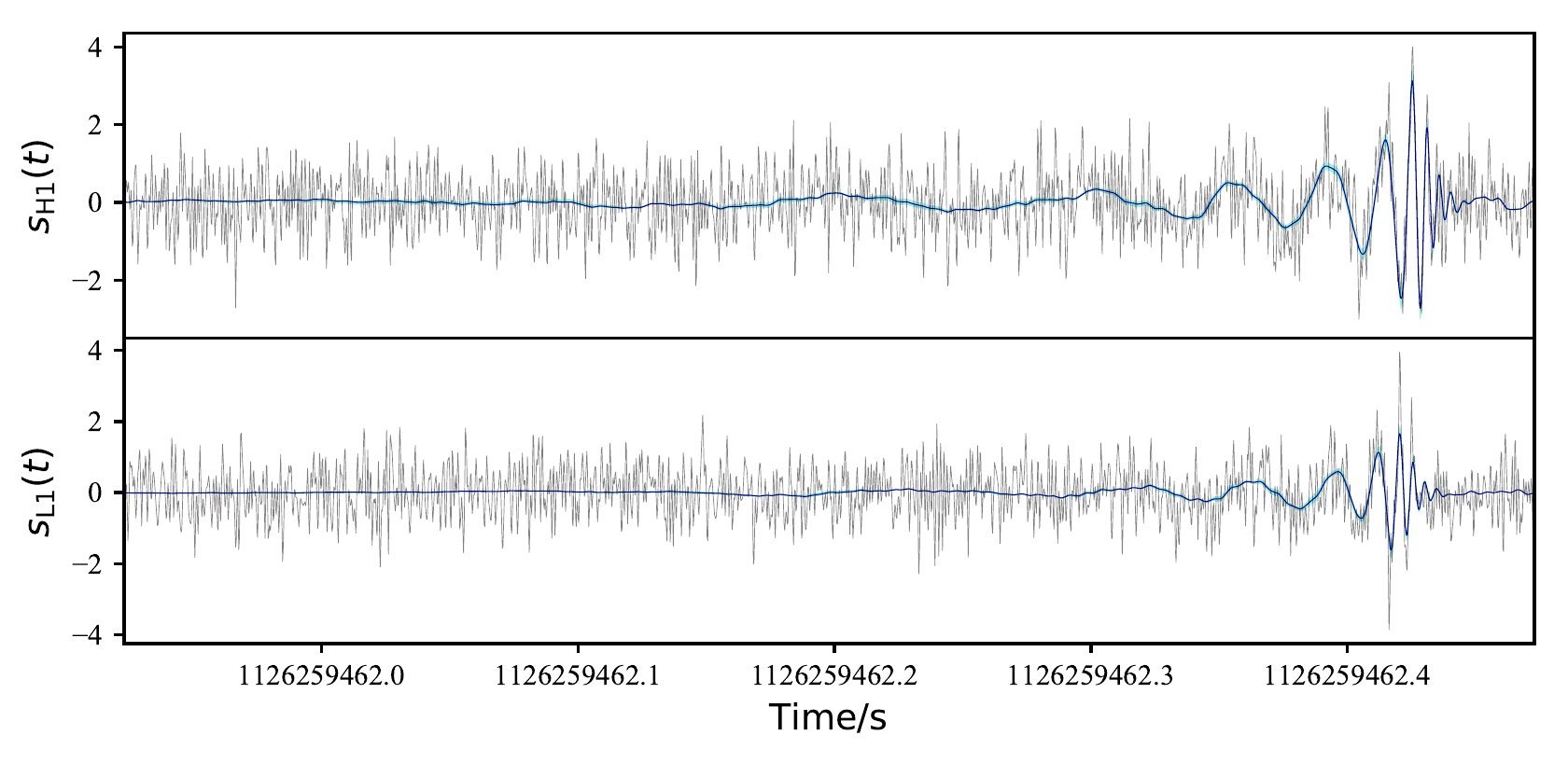}
\caption{\label{fig:wf} Reconstructed whitened GW waveforms in the Hanford (top panel)
  and in the Livingston (bottom panel) detectors. The solid lines indicate the median
  recovered waveforms. The cyan bands indicate the 90\% credible regions as recovered by
  our analysis. As a comparison we also overlay the whitened raw strain for the two detectors.}
\end{figure*}

\begin{figure}[t]
\center
\includegraphics[width=0.5\textwidth]{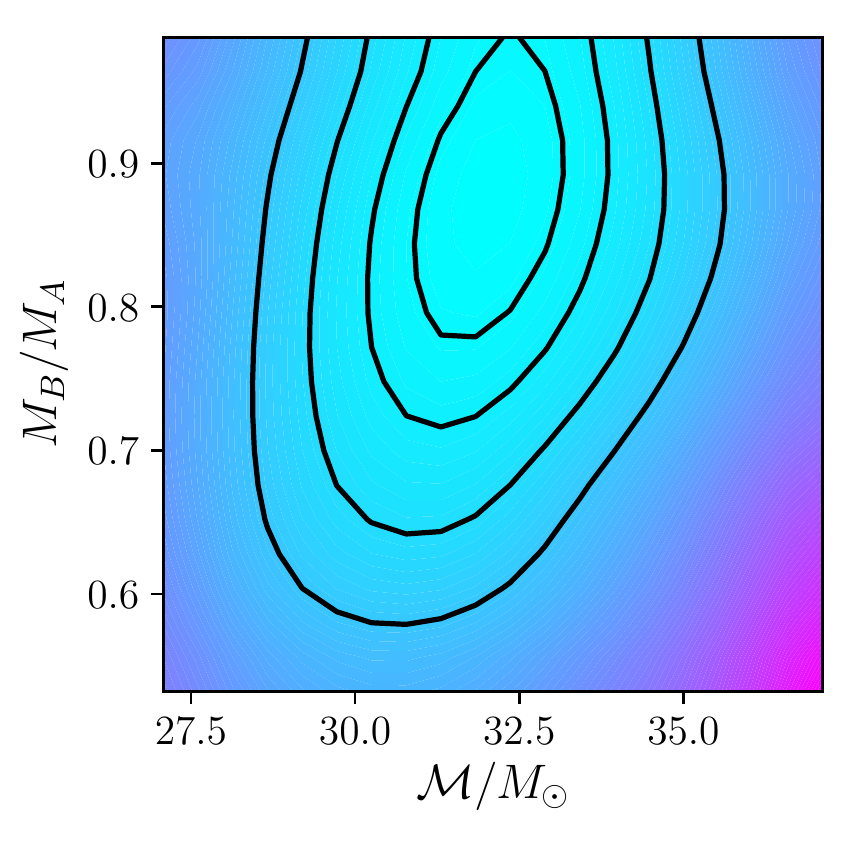}
\caption{\label{fig:mcq} Two-dimensional posterior distribution for $\mathcal{M}$ and $M_B/M_A$ for GW150914
  as inferred using \texttt{cpnest} and \TEOBResumS{}. The contours indicate the regions
  enclosing $90\%$, $75\%$, $50\%$ and $25\%$ 
of the probability.}
\end{figure}
We test the performance and faithfulness of our waveform model in a realistic setting 
by performing a parameter estimation study on the 4096 seconds of publicly available data for GW150914~\cite{LOSC}.
To do so efficiently, we do not iterate on the NQC parameters, so that the generation
time of each waveform from 20~Hz is $\sim 40$~ms using the \CC{} version of \TEOBResumS{}
discussed in Appendix~\ref{sec:code}. This worsens a bit the SXS/\TEOBResumS{} unfaithfulness,
as we illustrate in Fig.~\ref{fig:sxs_noiter}, though the model is still compatible with
the $\max{\bar{F}}\approx 1\%$ limit and below the $3\%$ threshold. The largest value
of $\bar{F}$ is in fact $\max{\bar{F}}\approx 0.018$, that is obtained for $(1,+0.40,+0.80)$.
We define $\theta$ as the vector of physical parameters necessary to fully characterize the
gravitational wave signal. 
For \TEOBResumS{}  and binary black hole systems, these are the component masses $(M_A,M_B)$,
their dimensionless spin components $(\chi_A,\chi_B)$ along the direction of the orbital angular
momentum, the three-dimensional coordinates in the Universe -- sky position angles and
luminosity distance --, polarization and inclination angles, and finally time and phase
of arrival at the LIGO sites.  We operate within the context of Bayesian inference;
given $k$ time series of $k$ detectors' data $d$, we construct the posterior distribution over the parameters $\theta$ as
\begin{align}
p(\theta|d_1,\ldots,d_k, H,I) = p(\theta|H,I)\frac{p(d_1,\ldots,d_k|\theta,H,I)}{p(d_1,\ldots,d_k| H,I)}\,
\end{align}
where we defined our gravitational wave model -- \TEOBResumS{} -- as $H$ and $I$ represents all ``background''
information which is relevant for the inference problem\footnote{For instance, the assumption of stationary
  Gaussian detector noise is hidden in the definition of $I$.}. For our choice of prior distribution
$p(\theta|H,I)$, we refer the reader to Ref.~\cite{TheLIGOScientific:2016wfe}.
Finally, we choose the likelihood $p(d_1,\ldots,d_k|\theta,H,I)$ to be the product of
$k$ wide sense stationary Gaussian noise distributions characterised entirely by their
power spectral density, which is estimated using the procedure outlined in Ref.~\cite{LOSC}.
We sample the posterior distribution for the physical parameters of GW150914 using the 
Python parallel nested sampling algorithm in \cite{cpnest}. The \texttt{cpnest} model
we wrote is available from the authors on request. In Table~\ref{tab:pe-summary} we summarize
our results by reporting median and $90\%$ credible intervals.
These numbers are to be compared with what reported in Table~I
in Ref.~\cite{TheLIGOScientific:2016wfe} and Table~I in Ref.~\cite{Abbott:2016izl}.
We also list them in the last column of Table~\ref{tab:pe-summary} for convenience.
As examples, we show the whitened reconstructed waveforms in Fig.~\ref{fig:wf} and 
the $\M$ and mass ratio posterior distribution in Fig.~\ref{fig:mcq}.
We find our posteriors to be consistent with what published by
the LIGO and Virgo collaborations, albeit our inference tends to prefer higher values 
for the mass parameters. However, no statistically significant difference is found. 
We find that \TEOBResumS{} is fit to perform parameter estimation studies and 
that on GW150914 it performs as well as mainstream waveform models.

\section{Selected comparisons with SEOBNRv4 and SEOBNRv4T}
\label{sec:vsSEOBNRv4}
To complement the above discussion, let us collect in this section a few selected
comparisons between \TEOBResumS{} and the only other existing state-of-the-art 
NR-informed EOB models \SEOBNRv4 and \SEOBNRv4{\tt T}~\cite{Bohe:2016gbl,Hinderer:2016eia,Steinhoff:2016rfi,Marsat-Vines-LIGO-tech-note}, that are currently being used on LIGO/Virgo data.
The tidal sector of the \SEOBNRv4{\tt T} model has been recently improved
so as to also include EOS-dependent self-spin terms in the Hamiltonian,
though in a form different from ours, and will be discussed in a forthcoming
publication. For the BBH case, our Fig.~\ref{fig:SXS},
when compared with Fig.~2 of~\cite{Bohe:2016gbl}, points out the excellent compatibility
between the two models at the level of unfaithfulness with the SXS catalog
of NR simulations, although the information (or calibration) of the model
was done in rather different ways. For \SEOBNRvq{} it relies on monitoring
a likelihood function that combines together the maximum EOB/NR faithfulness
and the difference between EOB and NR merger times (see Sec.~IVB of~\cite{Bohe:2016gbl}).
By contrast, the procedure of informing \TEOBResumS{} via NR simulations relies
on monitoring the EOB/NR phase differences and choosing (with a tuning by hand that
can be performed in little time without the need of a complicated computational
infrastructure, as explained in detail in~\cite{Nagar:2017jdw}) values of
parameters such that the accumulated phase difference at merger is within 
the SXS NR uncertainty obtained, as usual, by taking the phase difference between 
the two highest resolutions. This is possible within \TEOBResumS{} because of the
smaller number of dynamical parameters, i.e. $(a_6^c,c_3)$, and the rather ``rigid''
structure that connects the peak of the (pure) orbital frequency with the NQC point
and the beginning of ringdown, Eq.~\eqref{eq:tNQC}.
\begin{figure}[t]
\includegraphics[width=0.45\textwidth]{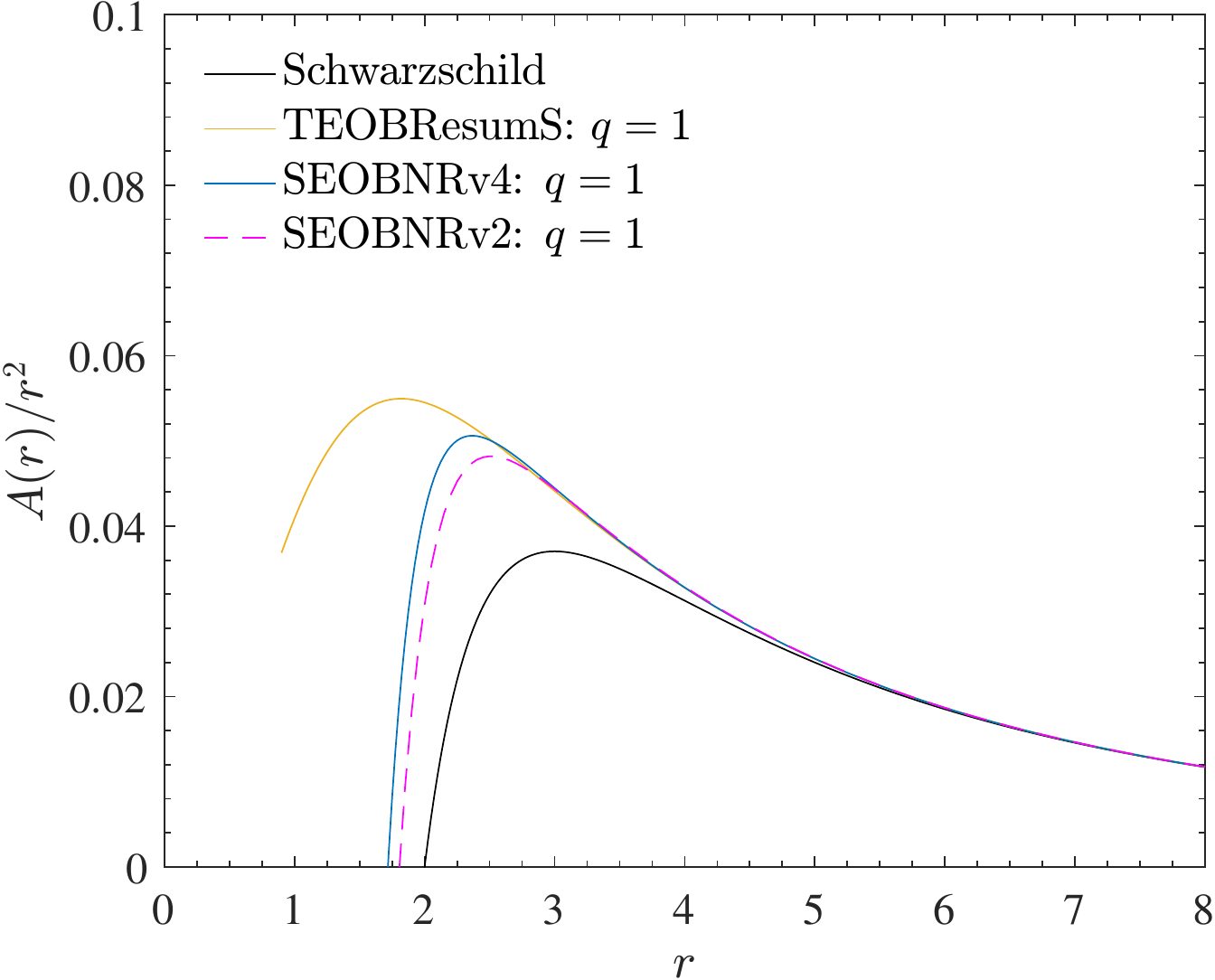}
\caption{\label{fig:Bq1} Comparison between two flavors of the \SEOBNRv* model and \TEOBResumS.
  The improved NR calibration incorporated in \SEOBNRv4{}~\cite{Bohe:2016gbl,Hinderer:2016eia,Steinhoff:2016rfi}
  pushed it closer to the \TEOBResumS{} curve than the \SEOBNRv2 one~\cite{Taracchini:2013rva}.}
\end{figure}

\begin{figure}[t]
\includegraphics[width=0.4\textwidth]{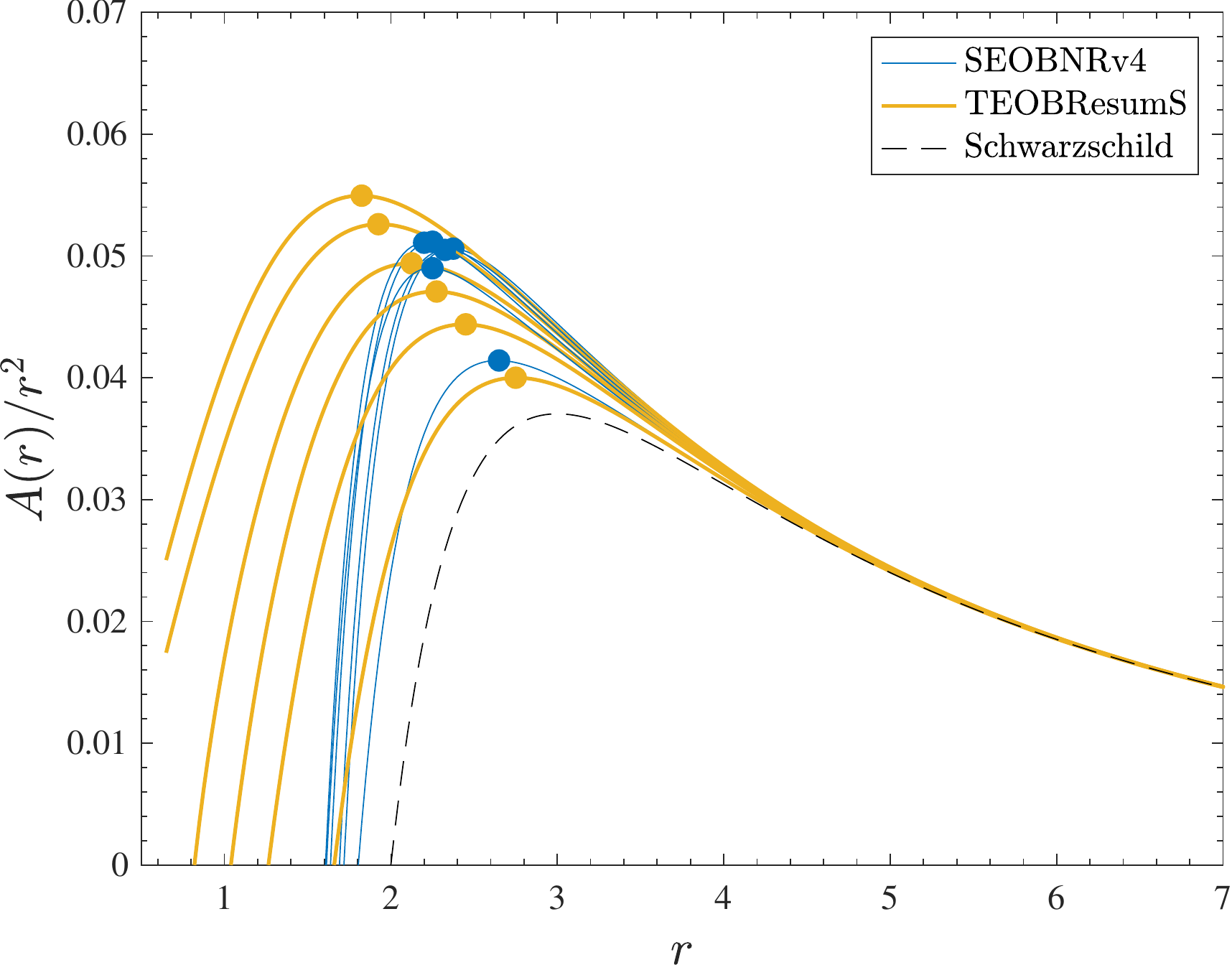}
\caption{\label{fig:Bq}EOB effective photon potential $A(r)/r^2$ for \SEOBNRv4
  and \TEOBResumS{} for mass ratios $q=(1,2,3,4,6,18)$.  The potentials are consistent,
  though different at the peak, also for medium mass ratios. The highest consistency is 
  found for $q=18$. The markers highlight the peaks of the functions, i.e. the locations of 
  the effective light-rings}
\end{figure}
\begin{figure}[t]
\includegraphics[width=0.4\textwidth]{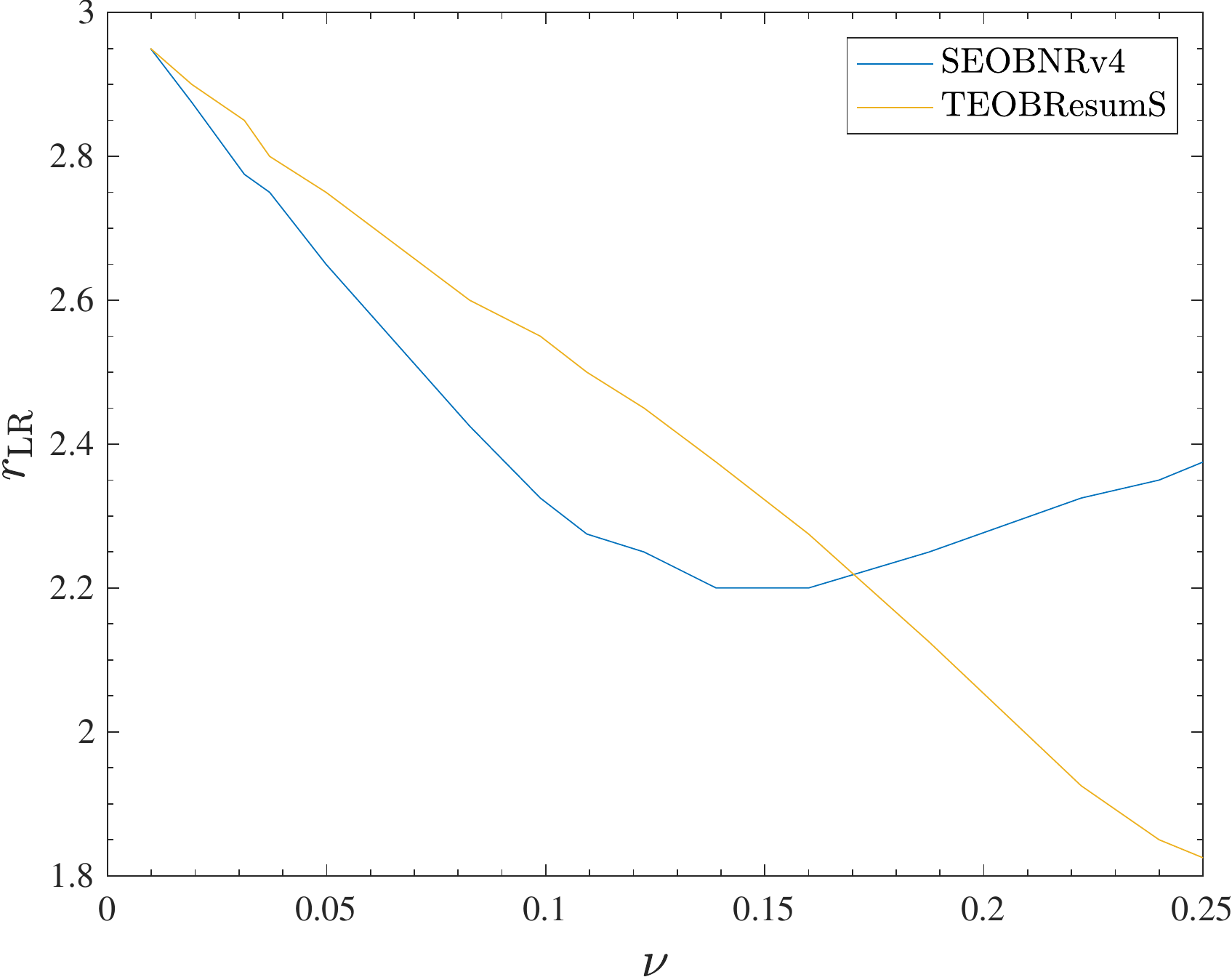}
\caption{\label{fig:LR_location} Dependence of the effective light-ring position, $r_{\rm LR}$,
i.e. the peak of $A(r)/r^2$ in Fig.~\ref{fig:Bq}, versus $\nu$. The behavior of the \TEOBResumS{} 
effective light-ring tends quasi-linearly to $r=3$, while the structure of the corresponding 
\SEOBNRv4 function is more complex. }
\end{figure}
Once this is done, and in particular once one has determined a global fit for $c_3$,
the EOB/NR unfaithfulness is computed as an additional cross check between waveforms.
Here we want to make the point that, even if the models look very compatible among 
themselves from the phasing and $\bar{F}$ point of view, they may actually hide  
different characteristics.  As a concrete example, we  focus on the (effective) photon potential 
function $A/r^2$, where $A$ is the EOB central interaction potential. In the test-particle
(Schwarzschild) limit, $A=1-2/r$ and $A/r^2$ peaks at the light ring $r=3$, 
which approximately coincides with (i) the peak of the orbital frequency; (ii) the peak 
of the Regge-Wheeler-Zerilli potential; (iii) the peak of the $\ell=m=2$ waveform 
amplitude~\cite{Damour:2007xr}. The location of the effective light ring (or the peak of the 
orbital frequency) is a crucial point in the EOB formalism, since, as in the test-particle limit,
it marks the beginning of the postmerger waveform part eventually
dominated by quasi-normal mode ringing.
We recall that \TEOBResumS{} and \SEOBNRv4 resum the $A$
potential in different ways: it is a (1,5) Pad\'e approximant
for \TEOBResumS{}, while it is a more complicated function
resummed by taking an overall logarithm for \SEOBNRv4~\cite{Barausse:2009xi}.
Moreover, while \TEOBResumS{} includes a 5PN-accurate logarithmic
term, \SEOBNRv4 only relies on 4PN-accurate analytic information.
In addition, both functions are NR-modified by a single, $\nu$-parametrized
function that is determined through EOB/NR phasing comparison.
This is the 5PN effective correction $a_6^c(\nu)$ mentioned above
for \TEOBResumS{} and the function $K_0(\nu)$ for
\SEOBNRv4. Explicitly, we are using $a_6^c(\nu)=3097.3\nu^2-1330.6\nu+81.38$
and $K_0=+267.788247\nu^3-126.686734\nu^2+10.257281\nu+1.733598$.
As a first comparison, we plot in Fig.~\ref{fig:Bq1}
the $q=1$ effective photon potential. Right to the point, 
the figure illustrates that the two potentials are nicely consistent among themselves, 
although the structure close to merger is different. The figure also includes 
the potential of the {\tt SEOBNRv2} model~\cite{Taracchini:2013rva}, 
a model that has been used on GW150914 and that was characterized
by $K_0 = 103.2\nu^3- 39.77\nu^2 - 1.804\nu + 1.712$.
Interestingly, the plot shows that the
\texttt{*v4} potential peak is closer to the \TEOBResumS{} one than
the \texttt{*v2} one. This finding deserves some mention for several reasons.
First, the \TEOBResumS{} nonspinning $A$ function behind the photon 
potential of Fig.~\ref{fig:Bq1} was NR-informed in Ref.~\cite{Nagar:2015xqa}  with the same 
nonspinning SXS NR simulations used for {\tt SEOBNRv2} (plus a $q=10$
dataset that became available after Ref.~\cite{Taracchini:2013rva}).
Second, {\tt SEOBNRv2} uses only linear-in-$\nu$ 4PN
information~\cite{Barausse:2011dq,Bini:2013zaa} while
{\tt SEOBNRv4} uses the full 4PN information~\cite{Damour:2015isa,Damour:2016abl},
as for \TEOBResumS{}. However, to our
understanding, the \SEOBNRvq{} potential was also calibrated
using more nonspinning NR simulations (notably with $q\gtrsim 1$)
than for \SEOBNRv2{} (see Ref.~\cite{Bohe:2016gbl}) and \TEOBResumS{}.
This suggests that the \TEOBResumS{} potential seems able
to naturally incorporate some amount of strong-field information
that needs to be extracted from NR when a SEOBNRv*-like~\cite{Barausse:2009xi}
potential is employed. These findings merit further investigation.

In  Fig.~\ref{fig:Bq} we display the same comparison
(though after omission of the \texttt{SEOBNRv2} curve)
for different mass ratios, $q=(1,2,3,4,6,18)$.
One sees that both \TEOBResumS{} and \SEOBNRv4 curves are  
smoothly and consistently connected to the Schwarzschild case.
This accomplishes the basic paradigm of the EOB formalism that
the dynamics of the two-body problem is 
a {\it continuous deformation} of the dynamics of a test-mass on a Schwarzschild 
black hole~\cite{Buonanno:1998gg,Buonanno:2000ef}, so that this limit should 
be properly incorporated by construction in the model and should be preserved by 
the addition of NR information. However, the way the Schwarzschild limit is 
reached is rather different in the two models. This is highlighted very well
by the markers in Fig.~\ref{fig:Bq}. These markers indicate the location of
the effective light-ring, $r_{\rm LR}$, that is shown, versus $\nu$,
in Fig.~\ref{fig:LR_location}. The figure highlights that, while
the $r_{\rm LR}(\nu)$ is approximately linear for \TEOBResumS{} (i.e. the Schwarzschild 
light-ring is reached  at constant speed in the space of the
nonspinning configurations parametrized by $\nu$) the behavior of the
corresponding quantity in \SEOBNRv4{} is more complicated, notably
it is not monotonic in $\nu$. This is not necessarily a problem from the practical 
point of view of generating NR-calibrated waveforms that are consistent with NR simulations.
However, from the theoretical point of view, this suggests a slight inconsistency
within the model, because the location of $r_{\rm LR}$ for $\nu=0.25$ is the same
as for $\nu\approx 0.09$. A priori, as it was pointed out in the foundation 
of the EOB model~\cite{Buonanno:2000ef,Damour:2000we}, one would expect that
the location of the LR is simply monotonically pushed to smaller radii
(i.e. higher frequency) due to the repulsive effect of the higher PN $\nu$-dependent
corrections that exist both at 2PN and at 3PN order. This is also suggested by NR simulations,
where one finds that the GW frequency at merger (that in the EOB formalism is connected
with the peak of the effective photon potential) is monotonically growing
with $\nu$ (see e.g. Fig.~3 of ~\cite{Healy:2017mvh}).
By contrast, \TEOBResumS{} seems to consistently incorporate this feature by construction,
even with the NR-informed function $a_6^c(\nu)$. However, one sees that $r_{\rm LR}(\nu)$
is a quasi-linear function, though not exactly a straight line. This suggests that
it would be interesting to investigate to which extent one can take it as a 
straight line (since it depends on $a_6^c$) and how this influences
the EOB/NR phasing performances. We hope to address these questions in future work.
As a last remark, we note that one can just plug the \SEOBNRv4 $A$ interaction
potential within the \TEOBResumS{} infrastructure and, without changing anything
else in the model, see whether or not the differences of Fig.~\ref{fig:Bq}
reflect on the waveform. It is easily found that, especially when $q>1$,
the dynamics yielded by the two NR-informed potentials are rather different
(and somehow not compatible), non-negligibly affecting the phasing. A detailed
comparison of these aspects is interesting, and will be possibly undertaken
in future work.

\begin{figure*}[t]
\includegraphics[width=0.45\textwidth]{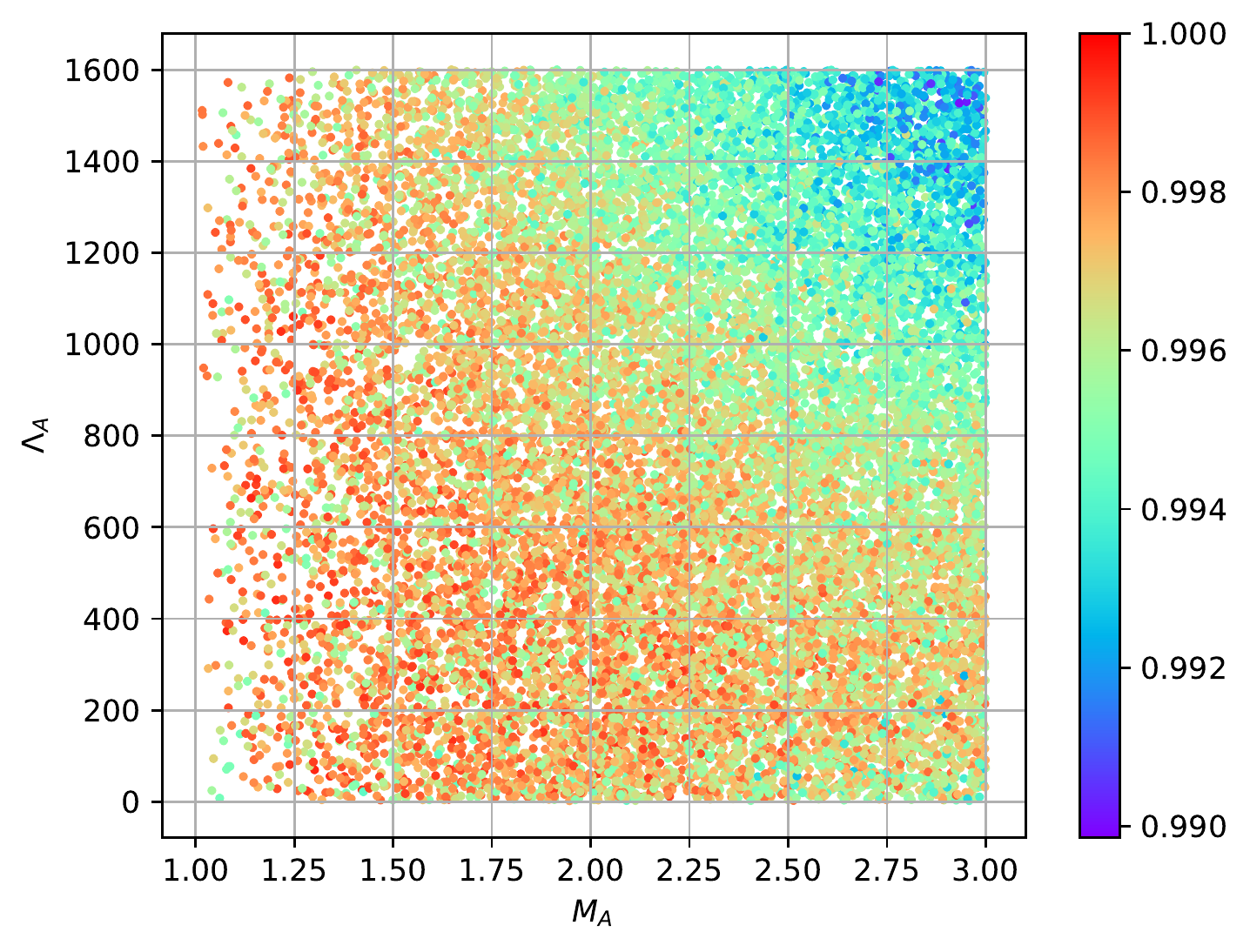}
\includegraphics[width=0.45\textwidth]{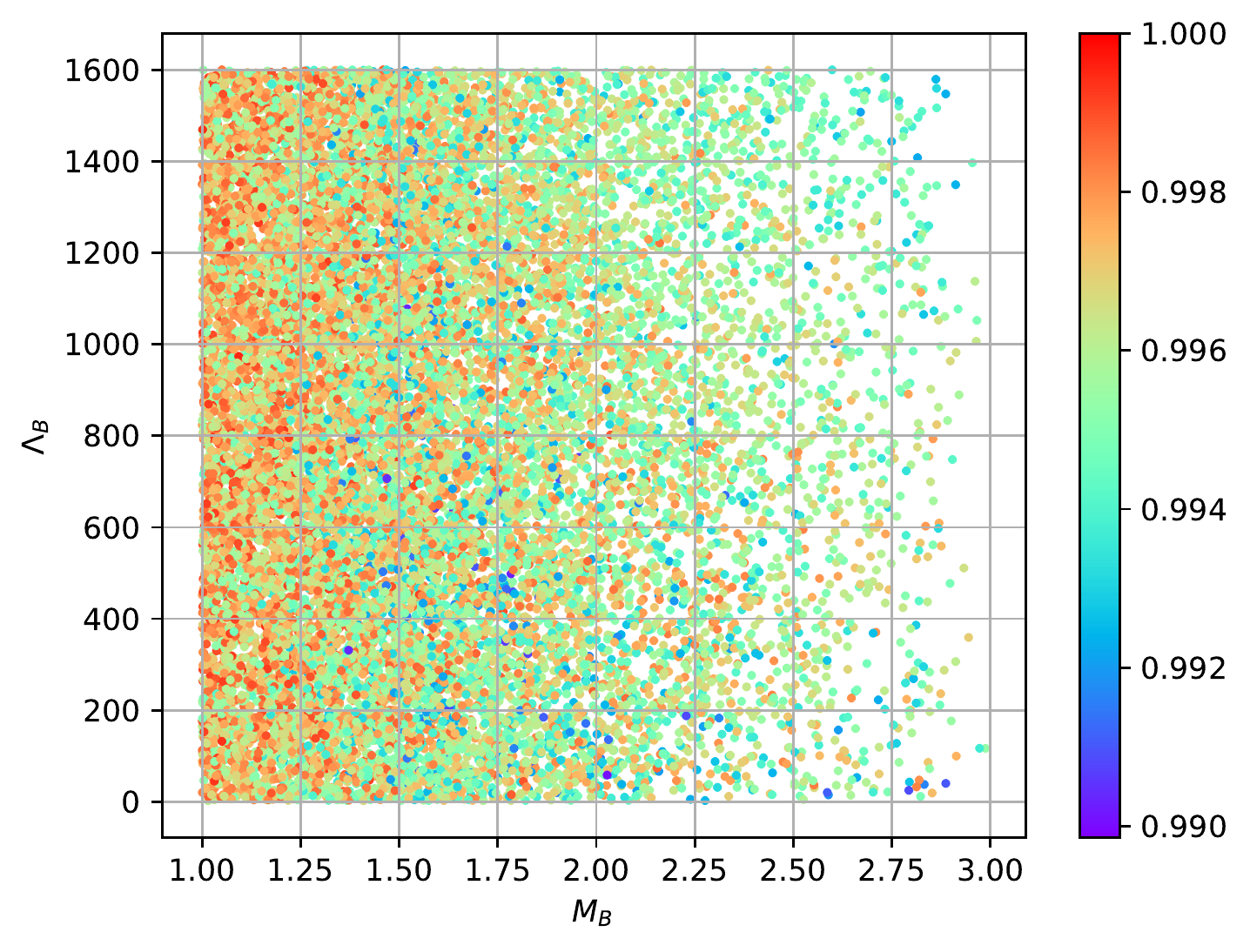}
\caption{\label{fig:Match_m_lam} The match computed between \SEOBNRv4{\tt T} and \TEOBResumS{}.
  The match values are color-coded. Based on 17300 randomly chosen points. The plot highlights 
  the high compatibility between the two models.}
\end{figure*}

As additional comparison between different EOB-based waveform models, we also computed the
faithfulness (or match) $F$ between \TEOBResumS{} and \SEOBNRv4{\tt T}, i.e., the tidal version
of \SEOBNRv4~\cite{Hinderer:2016eia,Steinhoff:2016rfi}. It has to be noticed that \SEOBNRv4{\tt T}
is conceptually different from \TEOBResumS{} in that the effects of enhancement of the tidal interaction
due to couplings with the internal oscillation $f$-mode of the stars is
incorporated in the model~\cite{Flanagan:2007ix,Steinhoff:2016rfi}. In addition, it also includes
EOS-dependent spin-spin terms, though not in the resummed form involving the
centrifugal radius~\cite{Marsat-Vines-LIGO-tech-note}. As above, the match here is the
overlap maximized over the time (time shift) and fiducial constant phase
\footnote{Note that, due to
  an incorrect flag, these results were obtained omitting, in \TEOBResumS{},
  the 3PN $\nu$-dependent, spin-independent, terms in $\rho_{31}$ and $\rho_{33}$
  as computed in Ref.~\cite{Faye:2014fra}. These terms were however correctly included
  to obtain all other results presented so far.}. The comparison was done in the part of
the parameter space that we expect to be astrophysically more relevant, namely,
we randomly draw parameters from the uniform distributions in the mass ratio $M_A/M_B \in [1,2]$,
the heaviest mass $M_A \in [1,3]M_\odot$, the spins (along orbital angular momentum) $\chi_{A,B} \in [-0.15, 0.15]$,
and the tidal parameters for each body $\Lambda_{A,B} \in  [2, 1600]$. Each waveform is computed
from a nominal initial frequency of 40~Hz. The most representative results are given in 
Fig.~\ref{fig:Match_m_lam} where we show the points drawn in the $(\Lambda_{A},M_{A})$ 
and $(\Lambda_{B},M_{B})$ planes. The match values, that are very high, are color-coded. 
The lowest match value found is 0.9898.

 \begin{table*}[t]
   \caption{\label{tab:ID}Initial conditions used to start the two EOB dynamics behind the
     waveforms of Fig.~\ref{fig:time_domain_worst} which yield the lowest match value 0.9898.
     The initial frequency was nominally fixed to be 40Hz in both models. From left to right
     we have: the name of the model; the initial relative separation; the corresponding value
     of the angular momentum; the corresponding value of the {\it circular} angular momentum
     and the value of the radial momentum. The initial values of the phase-space variables
     corresponding to 40Hz are slightly different in the two models. Due to the Newtonian 
     relation between frequency and radius that we use in \TEOBResumS{}, Eq.~\eqref{eq:Kepler},
     the consistency between initial configurations is recovered thanks to a slight modification
     in the initial nominal frequency of \TEOBResumS{} so that the values of $r_0/M$ coincide 
     up to the 5th decimal digit. See text for details.}
   \begin{center}
 \begin{ruledtabular}
   \begin{tabular}{cccccc}
     Model & $f_0$~[Hz] & $r_0$ & $p_\varphi$ &  $p_\varphi^{\rm circ}$ &$p_r$ \\
     \hline
     \TEOBResumS{}    &  40.000000 & 50.230212         & 7.3060375  & 7.3060378 & $-2.2938\times 10^{-5}$ \\
     \SEOBNRv4{\tt T} &  40.000000 & 50.296059         & 7.3105268  & 7.3105268  & $-2.2856\times 10^{-5}$  \\
     \hline
     \TEOBResumS{}    & 39.921474 & 50.296059  & 7.3105277  & 7.3105279 & $-2.2848\times 10^{-5}$ \\
 \end{tabular}
 \end{ruledtabular}
 \end{center}
 \end{table*}

\begin{figure}[t]
  \includegraphics[width=0.45\textwidth]{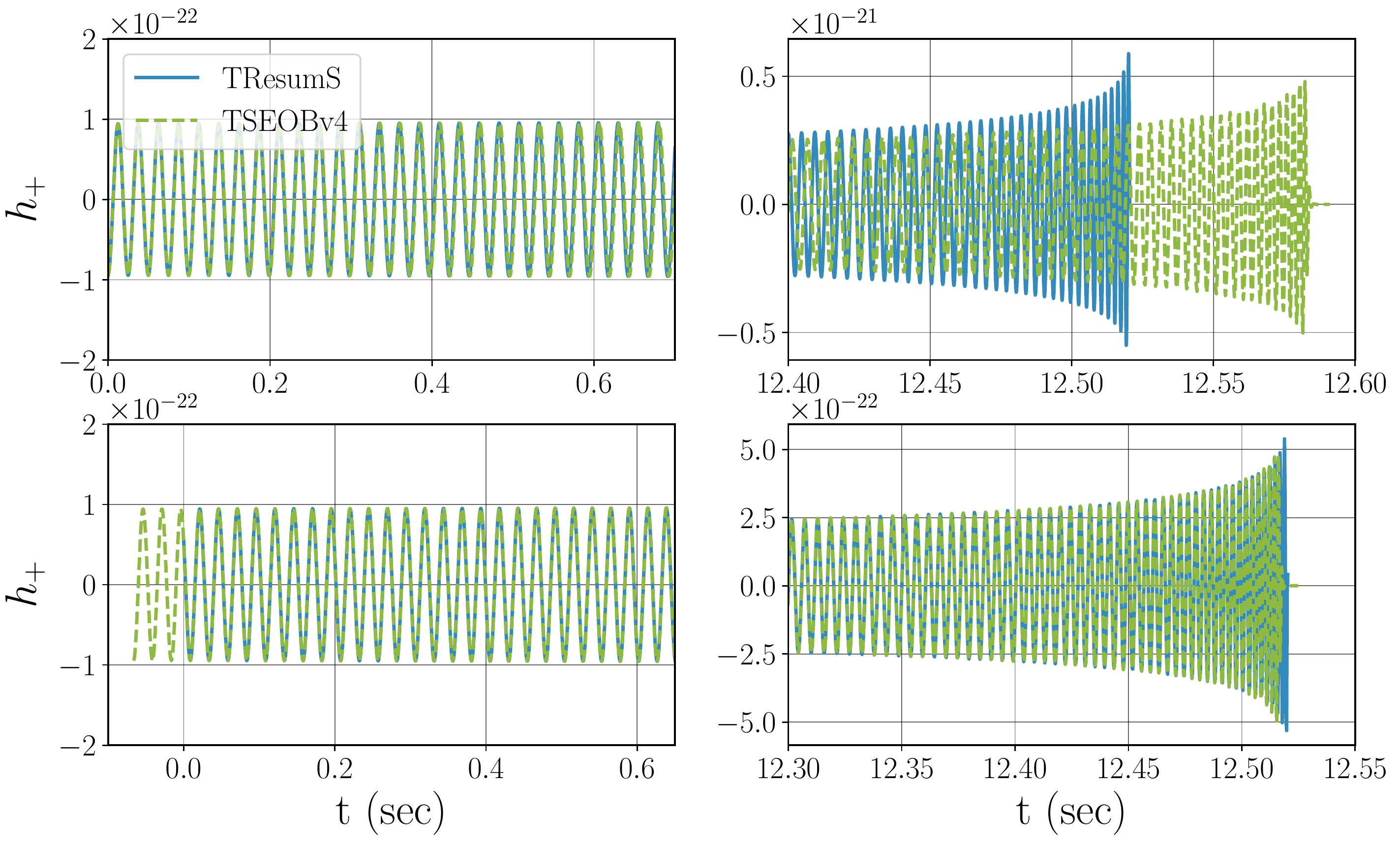}
    \includegraphics[width=0.45\textwidth]{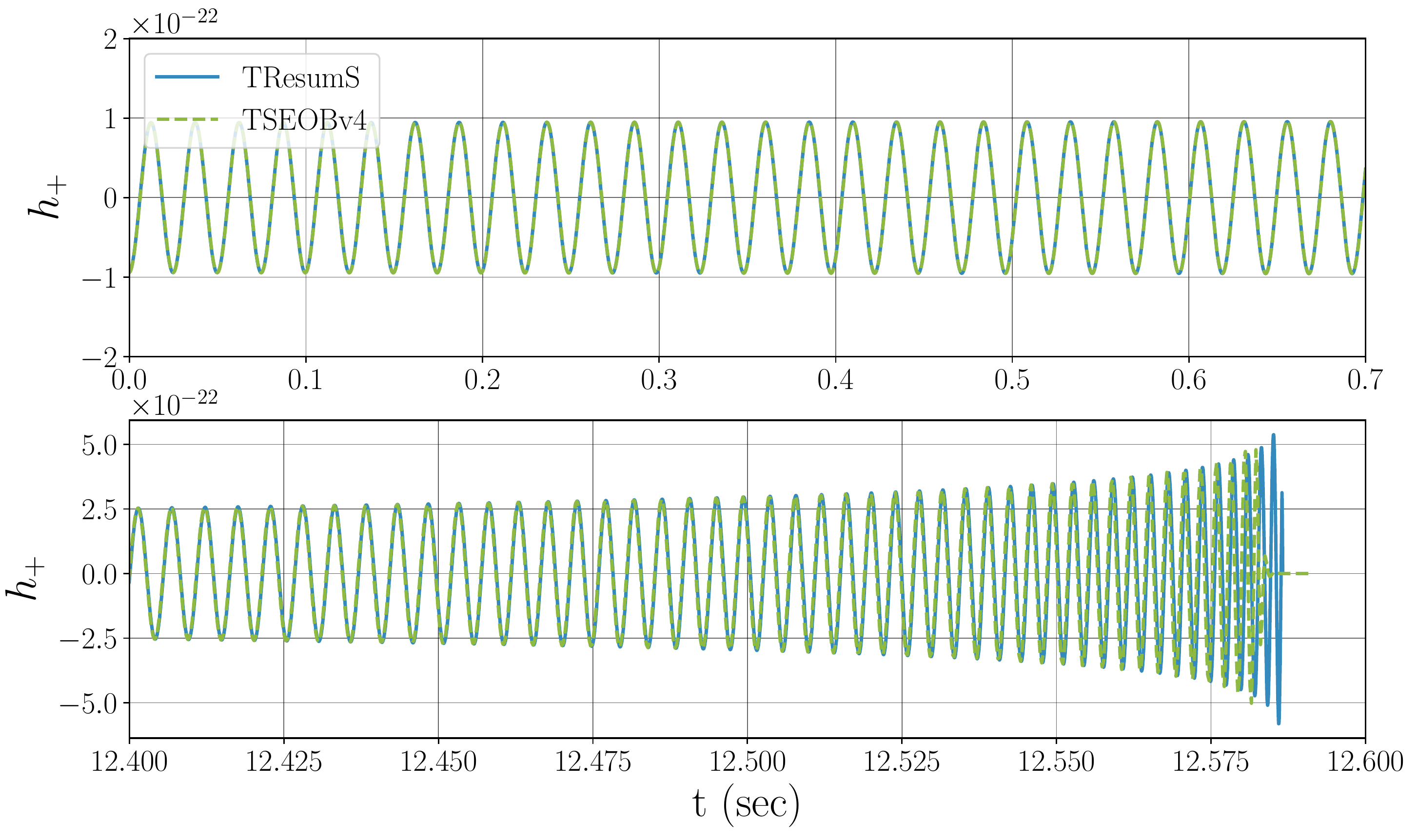}
\caption{\label{fig:time_domain_worst}Time-domain comparison between  \SEOBNRv4{\tt T} and \TEOBResumS{}
  for the case that delivers the lowest match, $F=0.9898$. Top panel: same initial nominal
  frequency, first two lines of Table~\ref{tab:ID}, the two waveforms are aligned by choosing
  a suitable relative time and phase shift. The first two rows of the plot show the waveforms
  before alignment, while the second ones after the alignment. Bottom panel: initial data
  for \TEOBResumS{} consistent with those of \SEOBNRv4{\tt T}, see second and third row
  of Table~\ref{tab:ID}. The two waveforms nicely agree directly, without the need
  of the additional alignment.}
\end{figure}

To better clarify the meaning of Fig.~\ref{fig:Match_m_lam} with complementary
information, we also depict in Fig.~\ref{fig:time_domain_worst} the direct time-domain 
comparison between the two waveforms corresponding to the lowest match value, $F=0.9898$.
The parameters of this binary are $M_A = 2.99173181168$,  $M_B = 1.54656708774$,
$\chi_A = -0.00403135733793$, $\chi_B = 0.104676230478$, 
$\Lambda_A^2= 1595.82370308$, $\Lambda_B^2 = 410.054257357$.
The corresponding values the spin-induced quadrupoles are 
$C_{QA}=8.47884798$ and $C_{QB}=5.56870361$.
The top panel of Fig.~\ref{fig:time_domain_worst} shows the two $h_+$ waveforms
without any relative time and phase shift. This is instead done in the bottom
panel, with these shifts dictated by the match calculation.
One notes that, although the initial GW frequency of the wave is chosen
to be 40~Hz for both models (and the waves seem to consistently start
in the same way) the initial conditions between the two models are different,
as highlighted in Table~\ref{tab:ID}. This difference comes from the
relation that connects the initial frequency $f_0$ to the initial radius
$r_0$. For \TEOBResumS{}, for simplicity, one is using the simple (though
approximate in this context) Newtonian Kepler's law
\be
\label{eq:Kepler}
r_0 = \left(\dfrac{\pi f MG}{c^3}\right)^{-2/3}.
\ee
On the contrary, \SEOBNRv4{\tt T} correctly recovers $r_0$ from Hamilton's
equations (see Eqs.~(4.8)-(4.9) of Ref.~\cite{Buonanno:2005xu}). The
difference in $r_0$ is then responsible for the difference in the
other phase-space variable, that is mostly behind the accumulated
time-domain relative dephasing between the two waveforms
highlighted in Fig.~\ref{fig:time_domain_worst}.
By contrast, what is not relevant for this case is the fact that,
while \TEOBResumS{} implements 2PA initial data~\cite{Damour:2007yf,Damour:2012ky}
(see also Appendix~\ref{sec:ID}) \SEOBNRv4{\tt T} only uses the post-adiabatic
(PA) approximation~\cite{Buonanno:2000ef}. Note that the effect of the 2PA
correction is very small at 40~Hz, since $p_\varphi^{\rm circ}$ is only changed
at the 7th decimal digit (see first row of Table~\ref{tab:ID}).
The last row of Table~\ref{tab:ID} illustrates that, if $f_0$ is slightly
changed so to compensate for the relativistic corrections that are not
included in Eq.~\eqref{eq:Kepler} and make \TEOBResumS{} start at the
same initial radius of \SEOBNRv4{\tt T}, the fractional difference
between the angular momenta is $\sim 10^{-7}$ and between the radial
momenta is $\sim 10^{-4}$. The \TEOBResumS{} waveform corresponding
to the last row of Table~\ref{tab:ID} is now largely more consistent
with the \SEOBNRv4{\tt T} {\it even without} time and phase alignment
(see bottom panel of Fig.~\ref{fig:time_domain_worst}).
The corresponding value of the match remains unchanged.

Since the \CC{} implementation of \TEOBResumS{} that was used in~\cite{Abbott:2018wiz}
was setting up the initial conditions using the simplified relation given by Eq.~\eqref{eq:Kepler}
above, we have decided not to modify it in the publicly available version of this code (see Appendix~\ref{sec:code} below).
By contrast, we are using a more correct relation between frequency and radius in
the corresponding C implementation of \TEOBResumS: the radius is obtained by
solving Eq.~\eqref{eq:Omg-Omgorb} for a given orbital frequency (assumed to be
half of the nominal initial gravitational wave frequency). In this way, we can
greatly improve the agreement with the corresponding \SEOBNRvqT{} initial conditions.
As an example, considering the case discussed above and detailed in
Table \ref{tab:ID}, the initial radius obtained in this way is found to be $r_0 = 50.296014$.

\section{Conclusions}
\label{end}

This paper has introduced and detailed \TEOBResumS{}, a state-of-the-art
effective-one-body model that generates time-domain gravitational waveforms
for nonprecessing, coalescing relativistic binaries. Our main results are as follows.
\begin{itemize}
\item[(i)] After correcting a minor coding error in the numerical implementation of the BBH sector 
of the model, we obtained a new determination of the NNNLO spin-orbit effective parameter $c_3$ 
with respect to Ref.~\cite{Nagar:2017jdw}. In addition, the merger and postmerger part was 
updated with respect of Ref.~\cite{Nagar:2017jdw} thanks to new effective fits that combine 
together NR information with test-particle results~\cite{Nagar:merger2018}. 
The parameter $c_3$ is determined by comparing EOB waveforms with 27, spin-dependent, 
NR waveforms from the SXS catalog. The model is then validated by computing the 
unfaithfulness (or mismatch) $\bar{F}$ over 135 NR waveforms from the SXS catalog 
obtained with the {\tt SpEC} code~and 19 NR waveforms from the \BAM{} code. 
Over the SXS catalog, $\max(\bar{F})\lesssim 2.5\times 10^{-3}$  except for a single 
outlier, $(3,+0.85,+0.85)$ where $\max(\bar{F})\lesssim 7.1\times 10^{-3}$. 
By incorporating more flexibility in the global fit for $c_3$,  notably allowing $c_3$ to 
depend {\it quadratically} on the individual spin variables also away from the 
equal-mass, equal-spin regime, one finds that $\max(\bar{F})\lesssim 2.5\times 10^{-3}$ {\it all over} 
the SXS waveform catalog. By contrast, $\bar{F}$ over the \BAM{} NR waveform 
is always well below the $1\%$ level except for the single outlier  $(8,+0.85,+0.85)$, 
that shoots up to $5.2\%$. We have identified the cause of this discrepancy to be 
the strength of the EOB-predicted spin-orbit interaction to be too small (i.e., resulting in
a dynamics plunging too fast with respect to the NR prediction) in that corner of the parameter space.
We have shown that the problem can be fixed by a new, NR-driven, choice for $c_3$. For simplicity, 
we have however decided not to provide a new fit of $c_3$ that also incorporates this strong-field 
information. This will be done in a forthcoming study that implements the factorized and resummed 
waveform amplitudes of  Refs.~\cite{Nagar:2016ayt,Messina:2018ghh} that are expected to be
more robust for large mass ratios and large, positive spins.
\item[(ii)] We comprehensively explored the behavior of \TEOBResumS{} waveform amplitude and 
frequencies outside the NR-covered portion of the parameter space. Thanks to the robustness of the 
merger and postmerger fits of Ref.~\cite{Nagar:merger2018}, the waveforms look sane and consistent 
among themselves even for large mass  ratios ($q\leq 20$) and high-spins $(\chi1=\chi_2=\pm 0.95)$. 
\item[(iii)] Building on previous work~\cite{Bernuzzi:2014owa}, the matter-dependent sector of \TEOBResumS{}
blends together, in resummed form, spin-orbit, spin-spin and tidal effects. Notably, the EOS-dependent
self-spin effects are also incorporated in the model (at leading order) in a similar fashion to the BBH case~\cite{Damour:2014sva}.
We showed that \TEOBResumS{} waveforms are compatible with state-of-the-art, long-end, error-controlled, 
NR simulations of coalescing, spinning BNSs for an illustrative choice of EOS.
\item[(iv)] We have produced selected comparisons with the EOB-based models \SEOBNRv4 and its tidal counterpart, 
\SEOBNRv4{\tt T}. In particular, for the case of spinning BNS, we computed the faithfulness (or match) between
\SEOBNRv4{\tt T} and \TEOBResumS, starting from 40Hz, with $M_A/M_B\in [1,2]$, the heaviest mass $M_A\in [1,3]M_\odot$,
dimensionless spins $\chi_{A,B}\in [-0.15,+0.15]$ and tidal parameters $\Lambda_{A,B}\in [2,1600]$.
We found excellent compatibility between the two models, with minimum match equal to 0.9898 for more than 17,000 events.
\item[(v)] Finally, we tested the performance of \TEOBResumS{} in a realistic setting by performing a parameter
estimation study on the publicly available data for GW150914. Our posteriors, listed in Table~\ref{tab:pe-summary}, 
are fully compatible with those inferred by the  LVC analysis of Refs.~\cite{TheLIGOScientific:2016wfe,Abbott:2016izl},
that are based on other NR-calibrated EOB waveform models.
\end{itemize}

While this paper was being finalized, a computationally efficient version
of \TEOBResumS{} based on the post-adiabatic approximation appeared~\cite{Nagar:2018gnk}.
In addition, \TEOBResumS{} is being used to test the RIFT algorithm to perform
Rapid parameter (RapidPE) inference of gravitational wave sources via Iterative
Fitting~\cite{Lange:2018pyp}. In particular, RapidPE results obtained
using \TEOBResumS{} on GW170817 data are reported in Ref.~\cite{Abbott:2018wiz}.

\begin{acknowledgments}
We acknowledge scientific discussion with J.~Lange, R.~O'Shaughnessy and M.~Rizzo,
that injected motivation to conclude this work.
We are grateful to A.~Buonanno, T.~Hinderer, S.~Marsat and J.~Vines for constructive
criticisms and clarifications concerning \SEOBNRvq{} and \SEOBNRvqT.
S.~B. and S.~A acknowledge support by the EU H2020 under ERC Starting Grant, no.~BinGraSp-714626, and
M.~H. and E.~F.-J. under ERC Consolidated Grant, no. 647839, as well as
Science and Technology Facilities Council (STFC) grant ST/L000962/1.
K.~W.~T. acknowledges support by the research programme of the Netherlands Organisation for Scientific Research (NWO).
D.~R. acknowledges support from a Frank and Peggy Taplin Membership at the
Institute for Advanced Study and the Max-Planck/Princeton Center (MPPC)
for Plasma Physics (NSF PHY-1523261). 
M.~A. acknowledges NWO Rubicon Grant No. RG86688.
P.~S. acknowledges NWO Veni grant no. 680-47-460. 
T.~Di. acknowledges support by the European Union's Horizon 
2020 research and innovation program under grant agreement No 749145, BNSmergers.
J.~A.~F. and P.~C.~D. acknowledge support by the Spanish MINECO
(AYA2015-66899-C2-1-P and RYC-2015-19074) and by the
Generalitat Valenciana (PROMETEOII-2014-069).
F.~M. thanks IHES for hospitality while this work was being developed.
P.~F. acknowledges support from the CARMIN Post-Doctoral Fellowship
programme (funding period 2014-2016) and would like to thank the IHES
for hospitality during the initial stages of this work in the spring of 2016.
Computations were performed on the supercomputers Bridges, Comet, and
Stampede (NSF XSEDE allocation TG-PHY160025), and on NSF/NCSA Blue Waters
 (NSF PRAC ACI-1440083 and NSF PRAC OAC-1811236), on the UK DiRAC Datacentric cluster,
 as part of the European PRACE petascale computing initiative on the clusters Hermit, Curie and SuperMUC,
 and on the BSC MareNostrum computer under PRACE and RES (Red Espa\~{n}ola de Supercomputaci\'{o}n) allocations.
A.~N. is very grateful towards his family for crucial help  to overcome especially 
difficult moments during the development of this work.
We are grateful to T.~Von der Liebe and A.~Charpentier  for pointing out 
the inconsistency related to the $\ell=5$, $m={\rm odd}$ flux modes.
\end{acknowledgments}

\appendix

\section{An extreme BBH configuration: $(8,-0.90,0)$}
\label{sec:8_m090_0}
After the work of the main body of this paper was finalized, we realized that
the SXS collaboration had publicly released one very interesting dataset,
SXS:BBH:1375~\cite{ian_hinder_2018_1215769} with $(8,-0.90,0)$.
This is interesting because it allows us to test
\TEOBResumS{} in the most difficult region of the
parameter space (i.e., when the spins are anti-aligned with
the orbital angular momentum) and, notably, it is 
marginally outside the portion covered by the \BAM{} simulations
of Table~\ref{tab:BAM} for $q=8$. In fact, it has $\hat{S}=-0.7111$,
to be compared with $\hat{S}=-0.6821$ corresponding to $(8,-0.85,-0.85)$.
\begin{figure}[t]
\center
\includegraphics[width=0.45\textwidth]{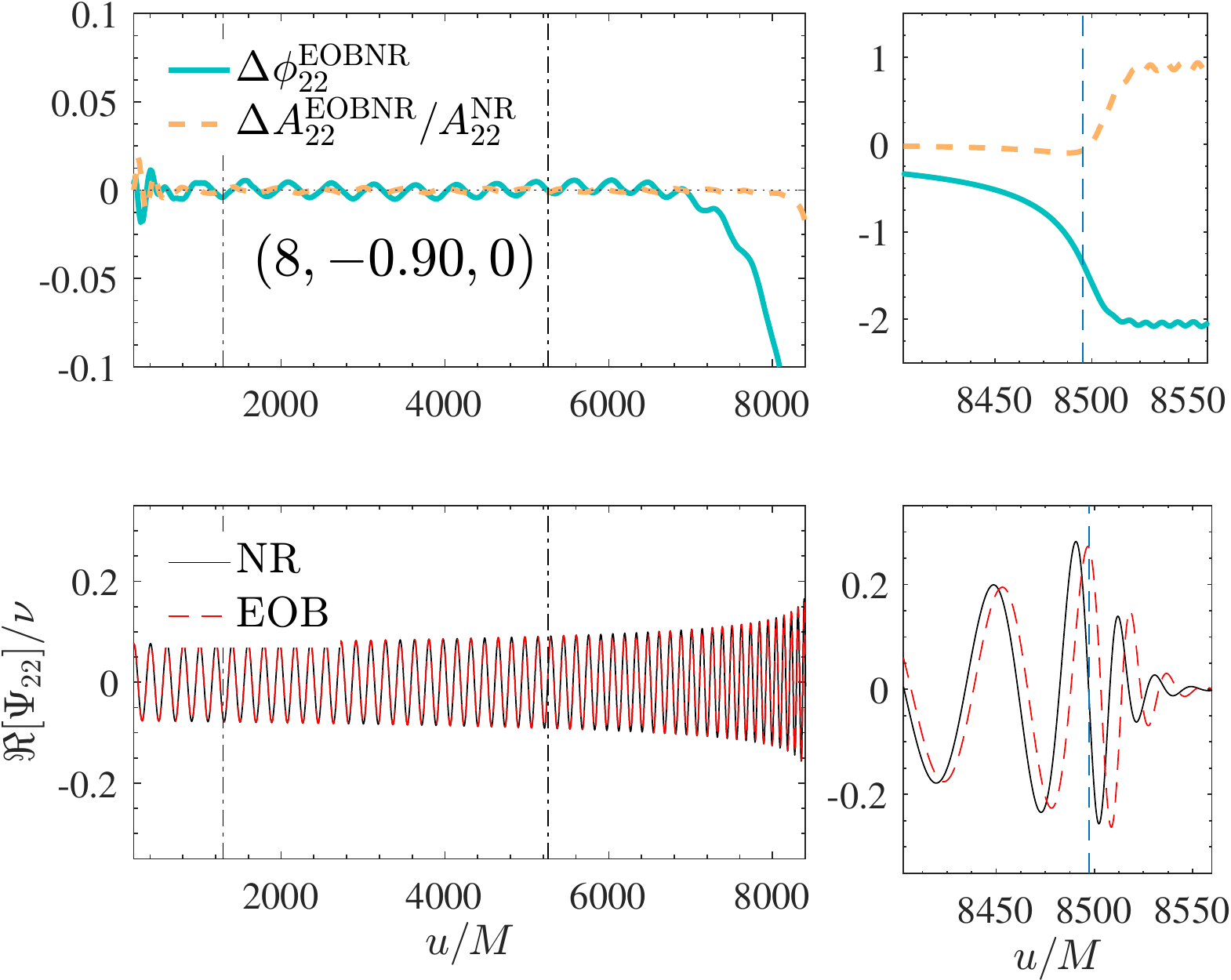}
\caption{\label{fig:8_m09}Phasing comparison between \TEOBResumS{} and
  SXS dataset SXS:BBH:1375. Alignment in the early inspiral (vertical lines).
  A EOB-NR phase difference of $-1.3$~rad is accumulated up to NR merger.}
\end{figure}
The phasing comparison is illustrated in Fig.~\ref{fig:8_m09}. We do
the following remarks. First, one sees that the phase difference (blue line)
oscillates around zero. This oscillation reflects the residual eccentricity
of the SXS waveform. Though it is rather small (i.e. $\sim 1.1\times 10^{-3}$)
it is visible because the \TEOBResumS{} waveform is started with essentially
eccentricity free initial data because of the 2PA approximation
(see Appendix~\ref{sec:ID} below). Second, the two waveforms dephase of about
1~rad up to the NR merger, with the \TEOBResumS{} plunging slightly slower
than the SXS one. The physical meaning of this plot is, for example,
that the spin-orbit coupling in \TEOBResumS{} is not strong enough.
In our current framework, this is understood as that the value of $c_3$
deduced by fitting the choices of Table~\ref{tab:c3} might be (slightly)
too large. Before pushing this reasoning further, let us focus on
Fig.~\ref{fig:8_m09_AmpFreq}, that illustrates the nice agreement
between the frequency and amplitude when the two waveforms are aligned
around merger, on a frequency interval (0.2,0.3).
\begin{figure}[t]
\center
\includegraphics[width=0.45\textwidth]{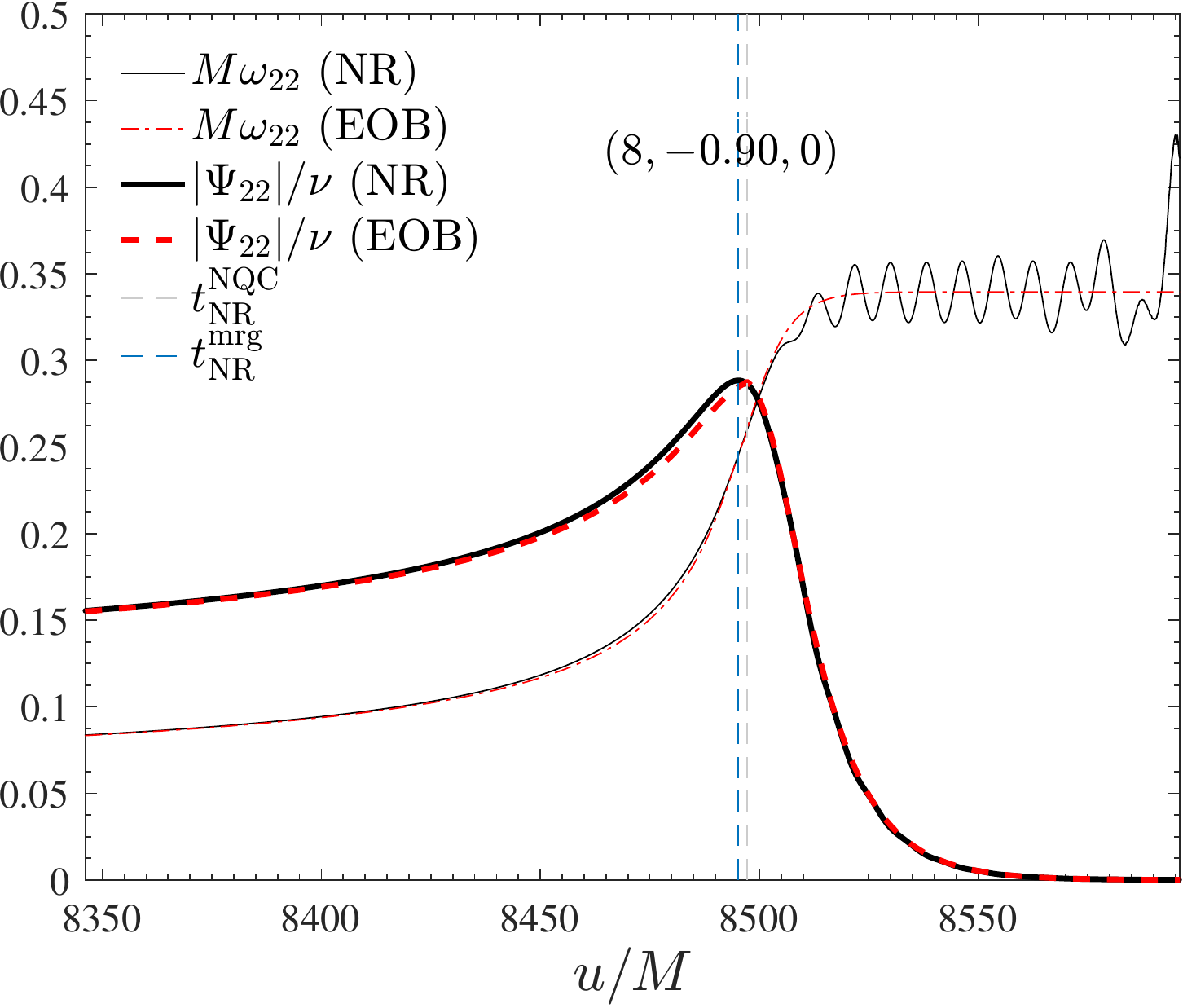}
\caption{\label{fig:8_m09_AmpFreq}Complement to Fig.~\ref{fig:8_m09}: excellent
  agreement between amplitude and frequency once the \TEOBResumS{} and SXS waveforms
  are aligned around merger.}
\end{figure}

\begin{figure}[t]
\center
\includegraphics[width=0.45\textwidth]{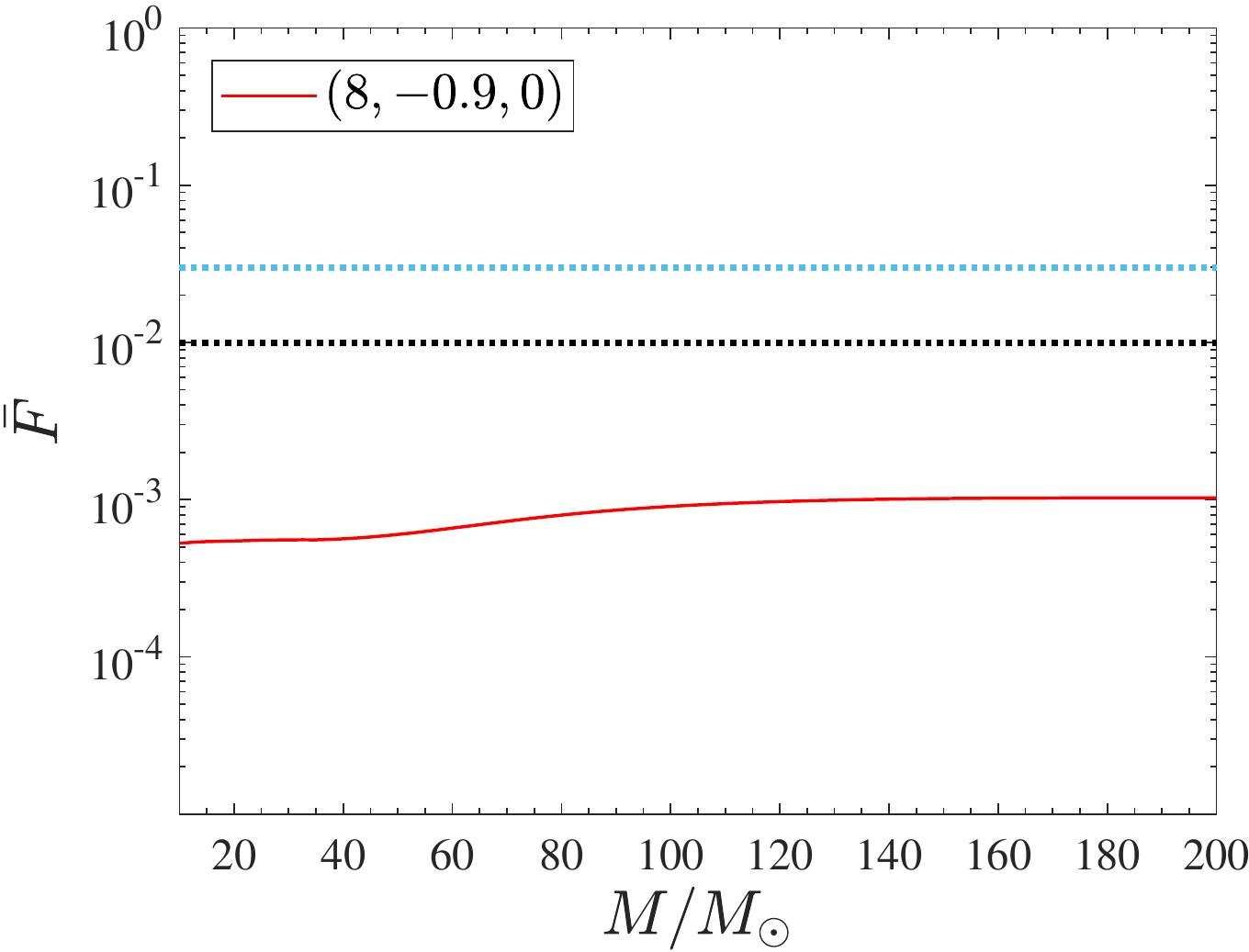}
\caption{\label{fig:barF_1375}Unfaithfulness calculation for the system of Fig.~\ref{fig:8_m09}. One finds that
$\max{(\bar F})=0.001027$}
\end{figure}
Note in passing that the oscillation in the frequency is physical
and is due to the beating between positive and negative frequency
quasi-normal-modes~\cite{Bernuzzi:2010ty}. This well-known feature
is currently not included in the EOB model.
As a last check, we computed, as usual, the EOB/NR unfaithfulness,
Fig.~\ref{fig:barF_1375}. One finds that $\max{(\bar F})=0.001027$.
This makes us conclude that, even if the time-domain analysis suggests
that the value of $c_3$ should be slightly reduced, we are not going to
do it now since the value of $\bar{F}$ is already one order of magnitude
smaller than the usual target of $0.01$.

\section{Black-hole -- Neutron-star binaries}
\label{sec:BHNS}

\begin{figure*}[t]
\center
\includegraphics[width=0.415\textwidth]{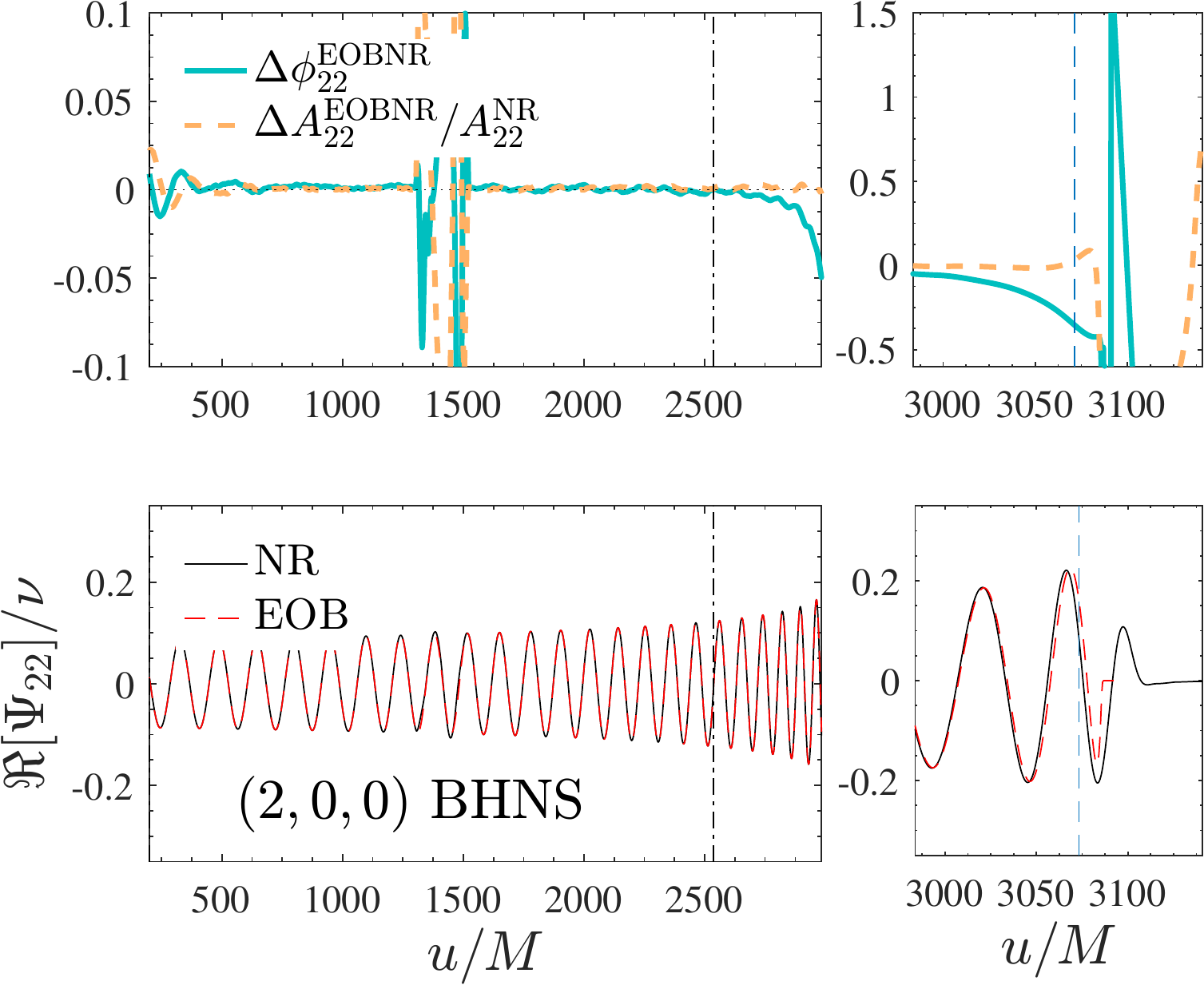}
\hspace{10 mm}
\includegraphics[width=0.4\textwidth]{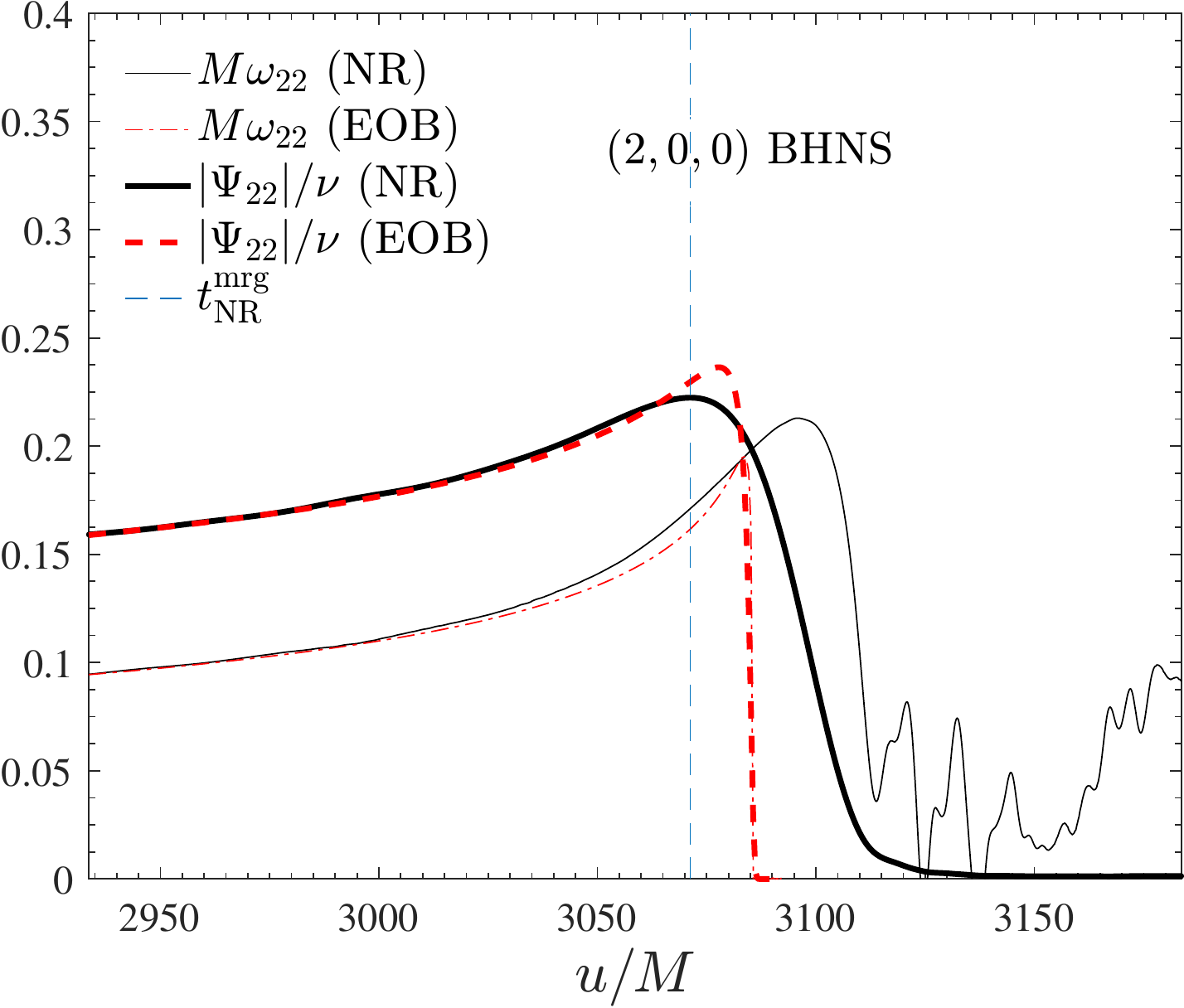}
\caption{\label{fig:bhns_q2} Phasing (left) and amplitude and frequency comparison (right)
  between \TEOBResumS{} and BHNS:0002 waveform for a BHNS merger with mass ratio $q=2$,
  with $M_B=1.4M_\odot$. Reference~\cite{Hinderer:2016eia} indicates that the accumulated
  phase errors to merger are about$\sim\pm0.5$~rad. The \TEOBResumS{} tidal waveform is
  well consistent with the NR one up to merger, even in the presence of tidal disruption.}
\end{figure*}
In this appendix we discuss the performances of \TEOBResumS{} for the
description of BH-NS waveforms. We stress that the model has not been 
developed for this type of waveforms and that this comparison is
preliminary to a forthcoming investigation. We focus on the two public
SXS datasets BHNS:0001 and BHNS:0002 that refer to a $q=2$ and $q=6$
nonspinning binaries where the NS is described by a $\Gamma=2$ polytropic
EOS with $K=101.45$ and $K=92.12$ respectively.
The dimensionless Love numbers are $k_{2,3,4}=(0.07524,0.0220429,0.0089129)$
and $k_{2,3,4}=(0.0658832,0.01873168,0.007341026)$ and the NS compactness
$C_B=0.144404$ and $C_B=0.1563007$.
The corresponding tidal parameters are $\Lambda_{2,3,4}=(470.8450,1095.9415,2511.5797)$
(BHNS:0001) and $\Lambda_{2,3,4}=(798.8698, 2244.6773, 6217.96765)$ (BHNS:0002).
The values of the tidal coupling constant
are $\kappa_2^T=0.50426$ for BHNS:0001 and $\kappa_2^T=19.725$ for BHNS:0002.
Given the very small value of $\kappa_2^T$ for BHNS:0001, and following the reasoning
of Ref.~\cite{Damour:2009wj} (see discussion related to Table~I), we expect 
that dataset to behave essentially like a BBH binary with the same mass ratio.

Let us focus first on the $q=2$ binary, BHNS:0002 Fig.~\ref{fig:bhns_q2}.
This binary dynamics is characterized by tidal disruption that suppresses
the ringdown oscillation after merger. The left panel of the figure illustrates
that  \TEOBResumS{} \textit{with tides and no NQC} captures well the waveform
up to merger, with a phase difference of $\sim -0.3$~rad there.
The ``glitch'' around $u/M\sim1300$ is in the Lev3 NR data (notably not in the Lev2 ones),
it is perhaps due to a re-gridding, but it is not relevant for our comparison.
The phase uncertainty at merger, estimated by just taking the
difference between Lev3 and Lev2 resolutions~\cite{SXS:catalog},
is of the order of 0.1~rad. This is of the order of the error budget
at merger estimated in Ref.~\cite{Hinderer:2016eia},
see Fig.~2 and Fig.~3 there, that is of the order of $\pm 0.5$~rad.
Hence, the BHNS waveform obtained with \TEOBResumS{} with tides is in
agreement with the NR data up to NR merger. Our result is comparable to
those presented in Ref.~\cite{Hinderer:2016eia}, but we stress here that
we do not use NQC-calibration and that the model only depends on the single
parameter $a_5^c(\nu)$ informed by BBH data; \TEOBResumS{} is not fed by
any strong-field information extracted from the BHNS:0002.

\begin{figure*}[t]
\center
\includegraphics[width=0.43\textwidth]{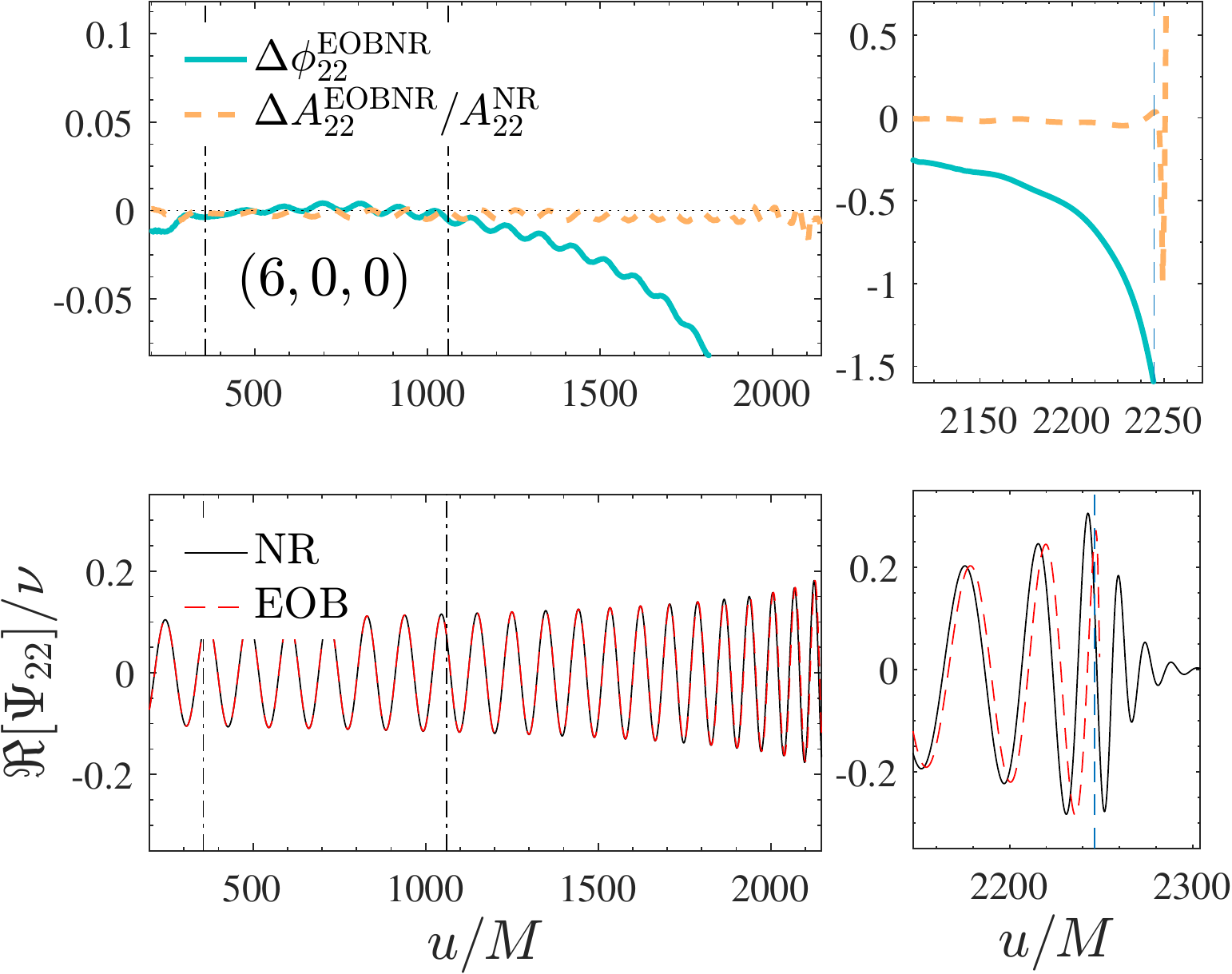}
\hspace{10 mm}
\includegraphics[width=0.4\textwidth]{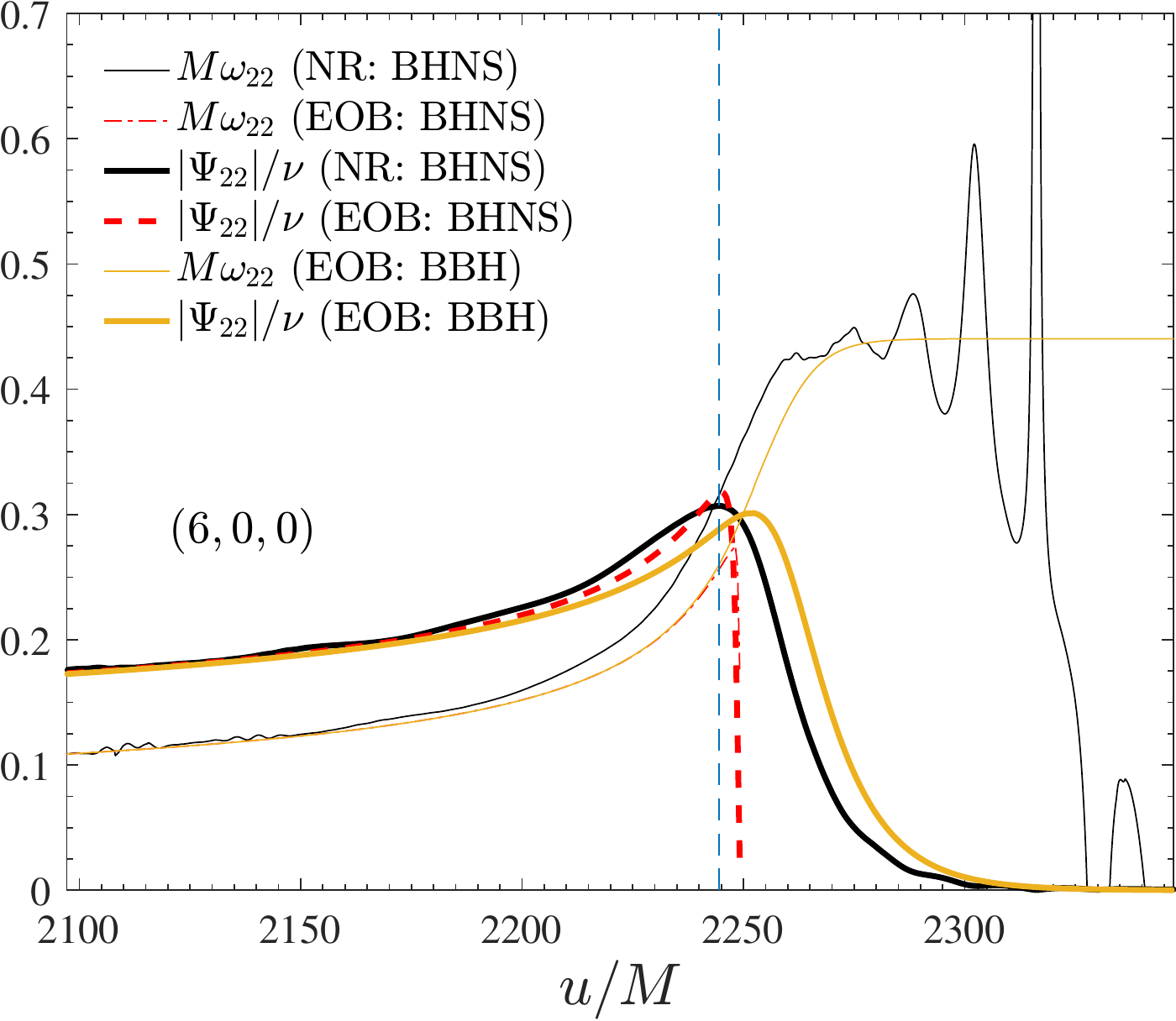}
\caption{\label{fig:bhns_q6}Left panel: \TEOBResumS{} and NR phasing for BHNS:0001, with $q=6$ and $M_B=1.4M_\odot$.
  A phase difference of $\simeq -1.6$ rad is accumulated up to merger. Right panel: frequency and amplitude plot.
  The orange line corresponds to the \TEOBResumS{} BBH (point-mass, no tides) waveform completed with NQC
  corrections and ringdown. Note that the frequency growth with tides (red-dashed) is almost indistinguishable
  from the corresponding curve without tides.}
\end{figure*}
\begin{figure}[t]
\center
\includegraphics[width=0.4\textwidth]{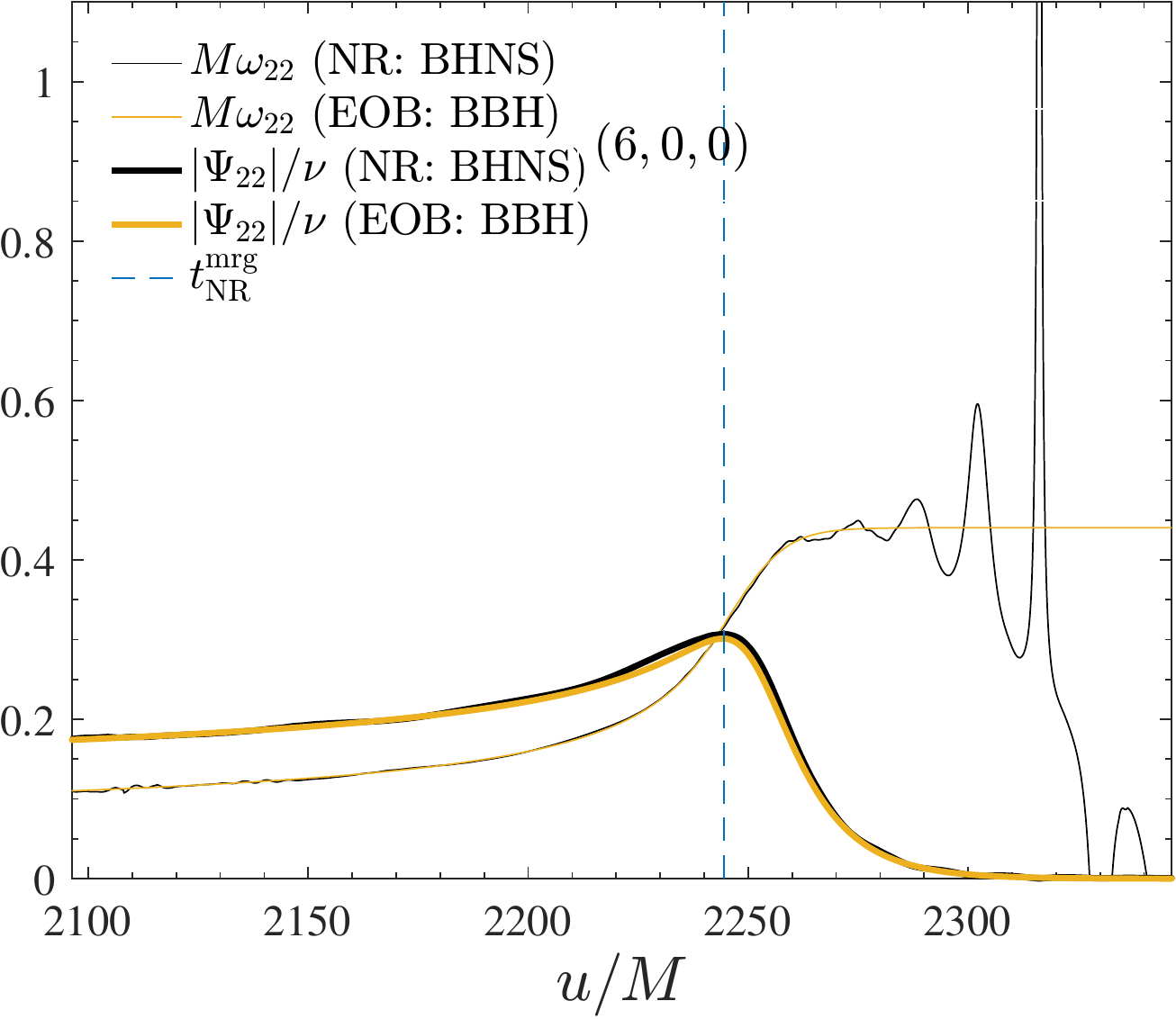}
\caption{\label{fig:bhns_mrg} Comparison between SXS BHNS waveform (black) with
  the \TEOBResumS{} BBH waveform (orange) {\it without} tides. Waveforms are aligned
  around merger. The suppression of the tidal interaction due to the effect of the
  large mass ratio is such that very little differences between the frequency
  and amplitude are seen.}
\end{figure}

Figure \ref{fig:bhns_q6} refers to the BHNS binary with larger mass ratio, $q=6$.
To our knowledge, this is the first time an EOB/NR comparison is done for this dataset,
as it was not included in Refs.~\cite{Hinderer:2016eia,Steinhoff:2016rfi}. 
The phasing analysis (left panel of Fig.~\ref{fig:bhns_q6}, alignment in the early inspiral)
is telling us that the EOB/NR phase difference is around $-1.6$~rad at NR merger.
The right panel illustrates that the \TEOBResumS{} tidal waveform (red lines) is sane,
notably with value of the merger amplitude very close to the NR one. On the same right
panel we also superpose the $q=6$ BBH \TEOBResumS{} amplitude and frequency (orange lines).
This waveform has {\it no tidal effects}, but it is completed by NQC and postmerger-ringdown.
Once the \TEOBResumS{} BBH  waveform is aligned to the SXS, see Fig.~\ref{fig:bhns_mrg} one
appreciates the high compatibility between the two waveforms during the plunge and merger,
consistently with the analytical understanding that a BHNS system with $\kappa_2^T=0.50426$
is {\it almost} a BBH binary. This brings also us to the conclusion that {\it most} of the
EOB/NR dephasing found in the phasing comparison of Fig.~\ref{fig:bhns_q6} is very likely
not physical, but of numerical origin. Due to the lack of different resolutions in the
SXS catalog (notably the Lev2 dataset was incomplete) we could not compute and estimate
of the numerical error on the BHNS:0001 waveform.

We conclude that the current design
of \TEOBResumS{} is very robust and does not lead to unphysical features in extreme
regions of the binary parameters. Hence, \TEOBResumS{} is a good starting points for
future BH-NS development. We also suggest that, lacking an accurate model for BH-NS,
\TEOBResumS{} can be used for the analysis of BH-NS by turning on tides in the
regime $1\leq q\lesssim 4-6$ while simply using the BBH waveform for larger mass-ratios.

\section{Post-post-adiabatic initial data for EOB dynamics}
\label{sec:ID}

In this Appendix we discuss in detail the post-post-adiabatic
prescription~\cite{Damour:2012ky} for generating initial data
for the EOB dynamics. This is a crucial input that allowed us,
among many things, to accurately compute the self-spin contribution
to the $Q_\omega$ diagnostic, $Q^{\rm SS}_\omega$, in Fig.~\ref{fig:QM}.
This was also crucial to properly extract the corresponding
tidal content of {\tt TEOBResumS} so to compare and contrast
it with the {\tt NRtidal} approximant~\cite{Dietrich:2017aum}
in Ref.~\cite{Dietrich:2018uni}.

Let us consider the EOB dynamics as described in Sec.~\ref{sec:mainfeats}.
To obtain circular orbits, we set $\hat{\F} = 0$ and we calculate the angular
momentum $j_0$ at a given radius $r_0$ solving the equation $\p \hat{H}_{\rm EOB} / \p r = 0$
for $p_\varphi$. The explicit expression of $j_0$ is obtained solving
\begin{align}
&\Bigg\{\Big[\Big(\frac{A}{r_c^2}\Big)'\Big]^2 - \frac{4A}{r_c^2} \Big[\tilde{G}'\Big]^2\Bigg\}j_0^4 + \nonumber \\
&+\Bigg\{2 A' \Big(\frac{A}{r_c^2}\Big)' - 4 A \Big[\tilde{G}'\Big]^2\Bigg\}j_0^2 + \Big[A'\Big]^2 = 0,
\end{align}
where $\tilde{G}\equiv G_S \hat{S} + G_{S_*}\hat{S}_*$. The idea behind the post-adiabatic (or post-circular)
approximation is to use the fact that, when the orbital separation is large, the gravitational
wave fluxes are small. We can then consider
\begin{equation}
\hat{\F}_\varphi = \bar{\F}_\varphi \ \vareps ,
\end{equation} 
where $\vareps$ is a, formal, small parameter. The quasi-circular inspiralling solution of
the EOB equations of motions can then be expanded as
\begin{align}
p_\varphi^2 =& \ j_0^2 \big(1+ k_2 \ \vareps^2 + \O[\vareps]^4 \big),\\
p_{r_*} =& \ \pi_1 \ \vareps+ \O[\vareps]^3.
\end{align} 
We now approximate $dp_\varphi / dt = (dp_\varphi / dr)(dr/dt) \sim (dj_0 / dr)(dr/dt)$,
in which we substitute Hamilton's equations. Keeping into account only the terms linear in $p_{r_*}$ in Eq.~\eqref{eq:drdt}, we get
\begin{equation}
\pi_1 \vareps = \hat{\F}_\varphi \Bigg[\frac{\nu \hat{H}_{\rm EOB} \hat{H}_{\rm eff}^{\rm orb} \Big(\frac{A}{B}\Big)^{-1/2}}{ \frac{dj_0}{dr} \Big\{1+\hat{H}_{\rm eff}^{\rm orb} \ j_0 \Big[\frac{\p \tilde{G}}{\p p_{r_*}}\Big]_1\Big\}}\Bigg]_0,
\end{equation}
where the subscript $0$ indicates that the term within brackets must be evaluated at $\vareps = 0$,
for example $[\hat{H}_{\rm eff}^{\rm orb}]_0 = \sqrt{A(1+ j_0^2 / r_c^2)}$. 
Here, $[\p \tilde{G}/\p p_{r_*}]_1$ denotes the coefficient of the term linear in $p_{r_*}$ of $\p \tilde{G}/\p p_{r_*}$.
Finally, $\pi_1 \vareps$ constitutes the post-adiabatic approximation and the first non-zero correction to $p_{r_*}$.
Using this result, we can calculate the post-post-adiabatic approximation (2PA) to $p_\varphi^2$. 
We thus solve $dp_{r_*} / dt \sim d(\pi_1\vareps) / dt$, in which we substitute Eq.~\eqref{eq:dprdt2} to the left hand side.
This is a quadratic equation in $p_\varphi$ given by 
\begin{align}
&\Big(\frac{A}{r_c^2}\Big)'p_\varphi^2 +2\big[\hat{H}_{\rm eff}^{\rm orb}\big]_0 \Big(G_S' \hat{S} + G_{S_*}'\hat{S}_*\Big) p_\varphi + A'+ \nonumber \\
&+ z_3 \big(\pi_1 \vareps \big)^4\Big(\frac{A}{r_c^2}\Big)'+ \Big(\frac{A}{B} \Big)^{-1/2} 2 \nu \big[\hat{H}_{\rm EOB} \hat{H}_{\rm eff}^{\rm orb}\big]_0 \frac{d[\pi_1\vareps]}{dt} = 0,
\end{align}
in which we approximated the Hamiltonians with their circular values and $p_{r_*}$ with $\pi_1 \vareps$.
Then, the derivative $d[\pi_1\vareps] / dt$ is numerically computed as $(d[\pi_1\vareps] / dj)(dj / dt) = (d[\pi_1\vareps] / dj)\ \F_\varphi$.
We could in principle keep going and calculate the post-post-post-adiabatic correction to $p_{r_*}$ 
by reiterating the same procedure using the computed $p_\varphi$ in place of the circular approximation $j_0$.

\section{Error budget and systematic uncertainties in NR BNS waveforms}
\label{sec:BAMvsTHC}

\begin{figure}[t]
\center
\includegraphics[width=.5\textwidth]{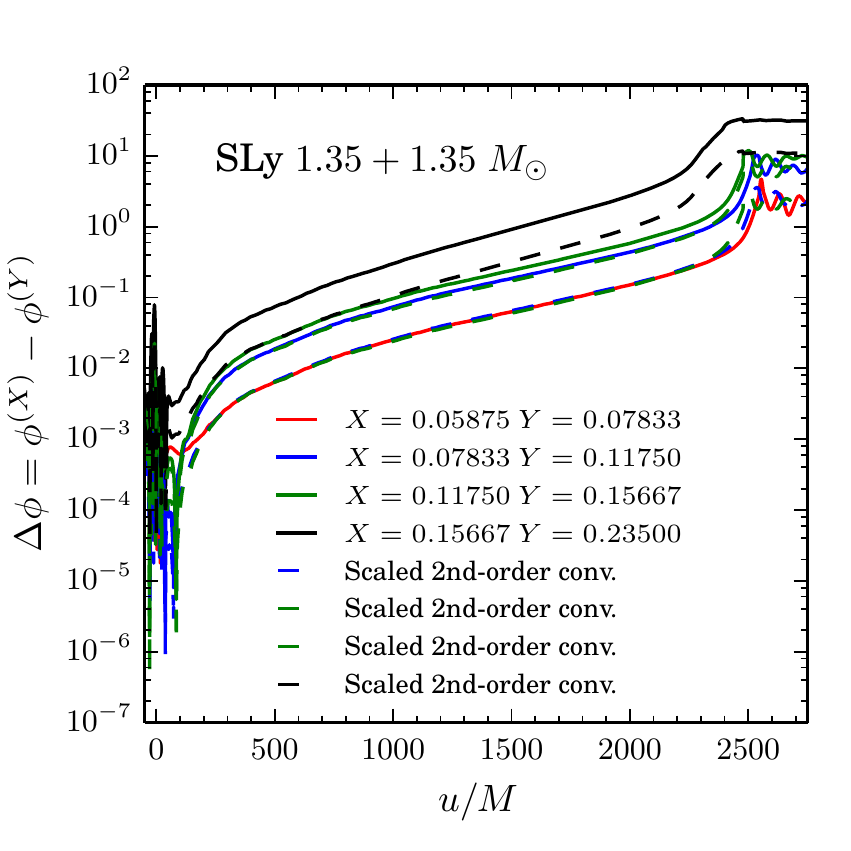}
\caption{\label{fig:BAM:conv} Self-convergence test for \BAM{}
  data. \BAM{} evolves eccentricity
  reduced initial data run with high-order methods. Second order
  convergence is observed except for the run at the lowest grid resolution.}
\end{figure}

The error budget of the \BAM{} BNS waveform is computed following the
method developed in~\cite{Bernuzzi:2011aq,Bernuzzi:2016pie}.
We perform simulations and convergence tests to identify the
resolutions at which the results are in the convergent
regime. Figure~\ref{fig:BAM:conv} shows an example of self-convergence
test for the GW phase in which differences between data sets at
different resolutions are plotted and rescaled by the factor relative
to second order convergence. The lowest resolution run, in which the NS are
covered with grid spacing $h=0.235$, does not give convergence
results. Higher resolutions are instead in the convergent regime as
indicated by the fact that dashed lines overlap with solid lines. We
thus choose the convergent data and perform a Richardson extrapolation
assuming second order convergence. The truncation error
$\delta\phi^{\rm h}_{22}$ is estimated as the difference between the
Richardson extrapolated phase and the highest resolution run. 

\begin{figure}[t]
\center
\includegraphics[width=.5\textwidth]{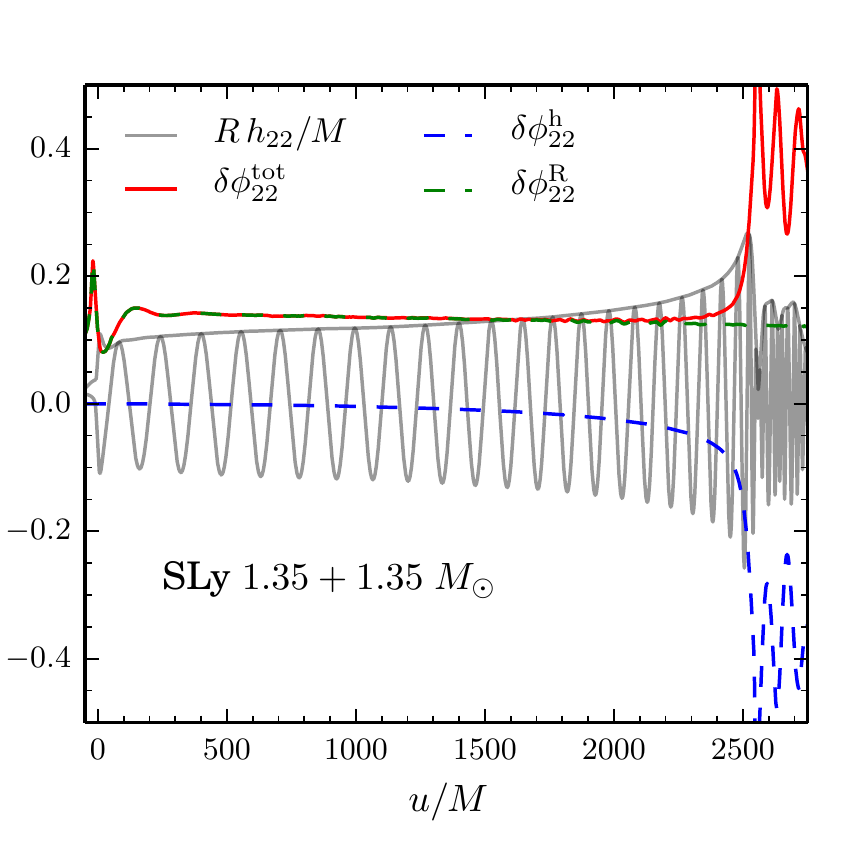}
\caption{\label{fig:BAM:errorbudget} Error budget for the waveform
  phase computed from the dataset of Fig.~\ref{fig:BAM:conv}.}
\end{figure}

Another source of uncertainty is that GW are extracted on spheres of finite
radius. To estimate such uncertainty, we pick waveforms from the highest
resolution run and from coordinate spheres with radii 
$r_{\rm obs}=600-1500~M_\odot$ and extrapolate to $r_{\rm obs}\to\infty$ with
low-order polynomials. The finite extraction error 
$\delta\phi^{\rm R}_{22}$ is estimate as the difference between the
extrapolated and the largest radius. Finally, the two independent
source of uncertainty are summed up in quadrature, 
\be
\delta\phi^{\rm tot}_{22} = \sqrt{(\delta\phi^{\rm
    h}_{22})^2+(\delta\phi^{\rm R}_{22})^2} \ .
\ee
As shown in Fig.~\ref{fig:BAM:errorbudget} the two error term
accumulate differently during the evolution and have opposite
signs. The finite extraction term $\delta\phi^{\rm R}_{22}$ dominates
the error budget up to the last orbits; close to merger the truncation
error term $\delta\phi^{\rm h}_{22}$ becomes the dominant one.

Beside truncation and finite radius 
uncertainties, BNS NR waveforms can be significantly affected by
systematic uncertainties related to the numerical treatment of
hydrodynamics
\cite{Bernuzzi:2012ci,Radice:2013xpa,Radice:2013hxh}. Here, we show
that our analysis is not affected by such 
systematics. We consider additional simulations with the independent
code \THC{}  \cite{Radice:2013hxh,Radice:2013xpa}.  
The \THC{} waveforms have been produced specifically for this work with the
goal of checking systematics uncertainties for the most challenging
case for the analytical model, i.e. the MS1b configuration. 
The numerical setup for the \THC{} runs 
is the same as in \cite{Radice:2016gym}, the employed 
resolutions are $h=(0.1,0.14,0.2,0.25)$. 
The \THC{} runs use a high-order scheme and typically 
show second - to - third order convergence with sufficiently high grid
resolutions.

\begin{figure}[t]
\center
\includegraphics[width=.5\textwidth]{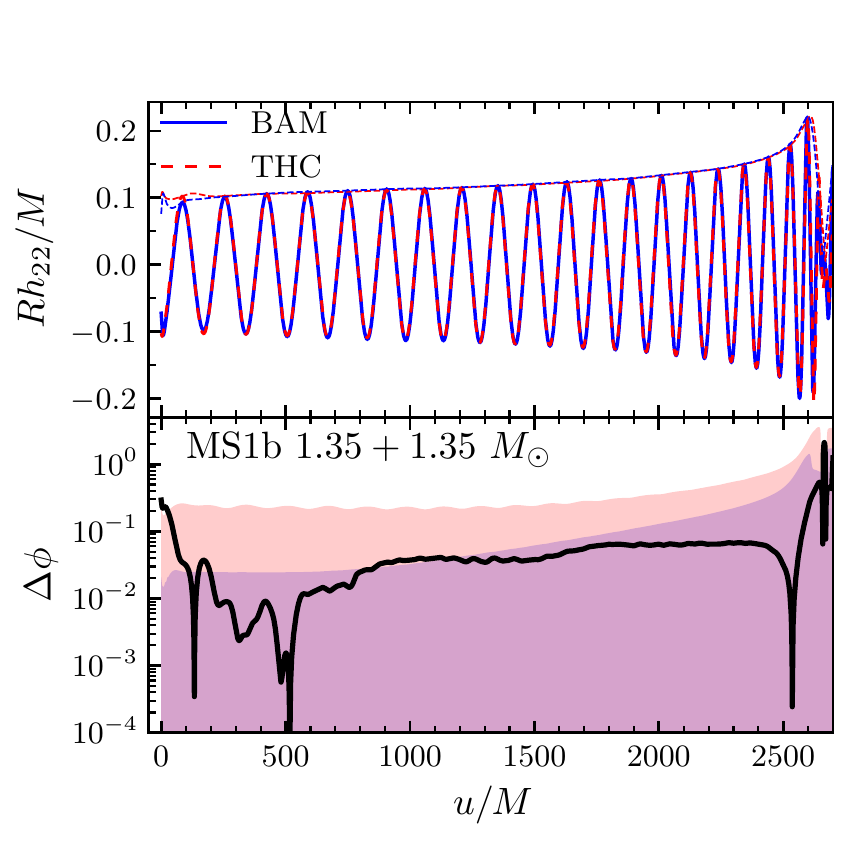}
\caption{\label{fig:BAM_THC} Systematic uncertainties in BNS numerical
  relativity inspiral-merger waveform. Waveforms from two independent and
  high-order codes are compared.}
\end{figure}

\BAM{} runs employ resolution
$h=(0.097,0.1455,0.194,0.291)$. 
The phase errors are estimated using Richardson extrapolation.
In Fig.~\ref{fig:BAM_THC} we compare such best waveforms
for the two codes. The waveform agree within the estimated
uncertainties. 

\section{TEOBResumS implementation details}
\label{sec:code}

An implementation of \TEOBResumS{} in C/\CC{} is publicly available at
\begin{center}
{\footnotesize \url{https://bitbucket.org/account/user/eob_ihes/projects/EOB}}
\end{center}
together with some of original \TEOBResum{} codes developed in {\tt Matlab}.
The inference results on GW150914 data of Sec.~\ref{sec:pe} and the match
results of Sec.~\ref{sec:vsSEOBNRv4} are obtained with this code.
An optimized implementation into the {\tt LALSuite}
library is currently in progress.

\subsection{General considerations}

The C/\CC{} implementation is straightforward. Main specific choices are
\begin{enumerate}
\item The equation of motion are written analytically and {\it not} using
  finite-difference derivatives of the Hamiltonian [Note this was an
    ``optimization'' of \SEOBNRv{2} as introduced in \cite{Devine:2016ovp}]. 
\item We use an eight-order adaptive-timestep Runge-Kutta order ODE
  solver as implemented in the GSL library. 
\item After the ODE solver has completed, both the solution of the Hamilton's
  equations as well as the waveform is sparsely and unevenly sampled in time
  (unless uniform timestep is requested). Since the waveform will eventually
  need to be Fourier transformed, it has to be uniformly sampled. We do it using
  cubic spline interpolant built within the GSL library. We note that, for our
  convenience, we interpolate on the evenly-spaced grid {\it both} the dynamics,
  that is the vector $(t,r,\varphi,p_{r_{*}},p_{\varphi})$, and the multipolar waveform. 
\end{enumerate}

We find that for long BNS waveforms starting at $10-20$~Hz, the computational
cost of the interpolation is almost as expensive as the solution of the ODE
system. Typical running times for BNS are of the order of 3.5~sec from 20Hz
using a time sampling of $4096$~Hz (see Table~\ref{tab:Benchmark} below).
Such performance is not yet competitive with  {\tt SEOBNRv2\_opt} of
Ref.~\cite{Devine:2016ovp} and thus cannot be directly used in parameter
estimation codes for long-inspiral signals (while it is sufficiently efficient
for short-inspiral ones like GW150914, as we showed in Sec.~\ref{sec:pe}).
This is not surprising, since no actual effort towards true optimization
was done at the moment and several unnecessarily repeated operations
are still present.
As an example of possible optimization, we discuss below an alternative 
implementation of the phase of the tail factor entering the resummed 
EOB waveform~\cite{Damour:2008gu}.

\subsection{Effective representation of the tail factor}
\label{subsec:tail}

This section describes the implementation of a specific term of the
multipolar waveform, the tail factor. We saw in the main text
that the circularized EOB multipolar waveform~\cite{Damour:2008gu}
is written in the following factorized form
\be
h_{\ell m} = h_{\ell m}^{\rm Newt} \hat{S}_{\rm eff}^{(\epsilon)} \hat{h}_{\ell m}^\text{tail}f_{\ell m}
\ee
where $h_{\ell m}^{\rm Newt}$ is the Newtonian prefactor, $\hat{S}_{\rm eff}^{(\epsilon)}$
is the effective source, $\epsilon$ is the parity of $\ell+m$, $f_{\ell m}$ is the
residual amplitude correction and $\hat{h}_{\ell m}^{\rm tail}$ is the tail
factor. The $f_{\ell m}$'s are given as Taylor series with $log$ functions
appearing, so that they are one of the most expensive part of the computation.
The tail factor $\hat h_{\ell m}^\text{tail}$ of the waveform is given by
\be
\hat{h}_{\ell m}^\text{tail}(y)=T_{\ell m}(y)e^{i\delta_{\ell m}(y)}
\ee
with
\be\label{Tlm}
T_{\ell m}(y) =\frac{\Gamma(\ell+1-2i\hat{\hat{k}})}{\Gamma(\ell+1)}e^{(\pi+2\ln(2k r_0)i)\hat{\hat{k}}}\,,
\ee
where
\begin{align}
k  &=m\Omega ,\\
\hat{\hat{k}} &=mGH_\text{EOB}\Omega,
\end{align}
with $r_0=2GM/\sqrt e$ and $y=(GH_\text{EOB}\Omega)^{2/3}$. Evaluating the full $T_{\ell m}$
as a complex number is computationally expensive as the $\Gamma$ functions need to be
evaluated separately for each multipole. 

To ease and speedup the implementation of those parts of the code where
the $\Gamma$ functions appear, it was chosen to work separately with its
argument and modulus, since only this latter is used for the computation
of the GW flux that drives the dynamics.
The squared modulus of the tail function, for each multipole, can be written as
\be
|T_{\ell m}(y)|^2=\frac{1}{(\ell!)^2}\frac{4\pi\hat{\hat k}}{1-e^{-4\pi \hat{\hat k}}}\prod^\ell_{s=1}\left(s^2+(2\hat{\hat k})^2\right) \ ,
\ee
and it can be thus computed via such simple formula.

Special routines for the $\Gamma$ function are needed only
for the {\it phase} of the tail term, for which a simple
formula like the one above does not exists. In fact, the
\TEOBResumS~C/\CC{} code uses instead the Lanczos approximation
to the complex $\Gamma$ function implemented in the GSL.

As an alternative, it is possible to construct a fast and
effective representation of the phase of $\Gamma$ as follows.
One starts from the following representation of the argument
\be
\arg\left[ \Gamma(x+iy)\right]=y\psi(x)+\sum_{n=0}^\infty\left[\frac{y}{x+n}-\arctan \frac{y}{x+n}\right]\,,
\ee
where $\psi(x)\equiv\Gamma'/\Gamma$ is the digamma function and the prime indicates the derivative
with respect to $x$. In Eq.~\eqref{Tlm}, the only complex $\Gamma$ is the one at the numerator,
where $y\equiv 2\hat{\hat{k}}$, $x=\ell+1$ and we can formally write its phase
in factorized form as
\be
\sigma_{\ell} (\hat{\hat{k}})\equiv \arg\left[\Gamma(\ell+1-2\ii\hat{\hat{k}})\right]=\sigma^{0}_{\ell}\hat{\sigma}'_{\ell},
\ee
where we defined $\sigma_{0}^{\ell}\equiv -2i\hat{\hat{k}}\psi(\ell+1)$.
The quantity $\sigma'_{\ell}$ can be represented by a polynomial in $\hat{\hat{k}}$
whose coefficients are fitted to the actual function $\sigma'_{\ell}(\hat{\hat{k}})$
evaluated on an interval large enough to include all possible frequencies spanned
by the binary evolution. For each multipole up to $\ell=8$ we find that
a fifth order polynomial of the form
$\sigma'_{\ell}=1 + n^\ell_{2}\hat{\hat{k}}^{2} + \dots +n^\ell_{5}\hat{\hat{k}}^{5}$
is able to give an accurate representation of the function, with fractional
difference typically of order $\lesssim 10^{-6}$.

\begin{figure}[t]
\center
\includegraphics[width=0.4\textwidth]{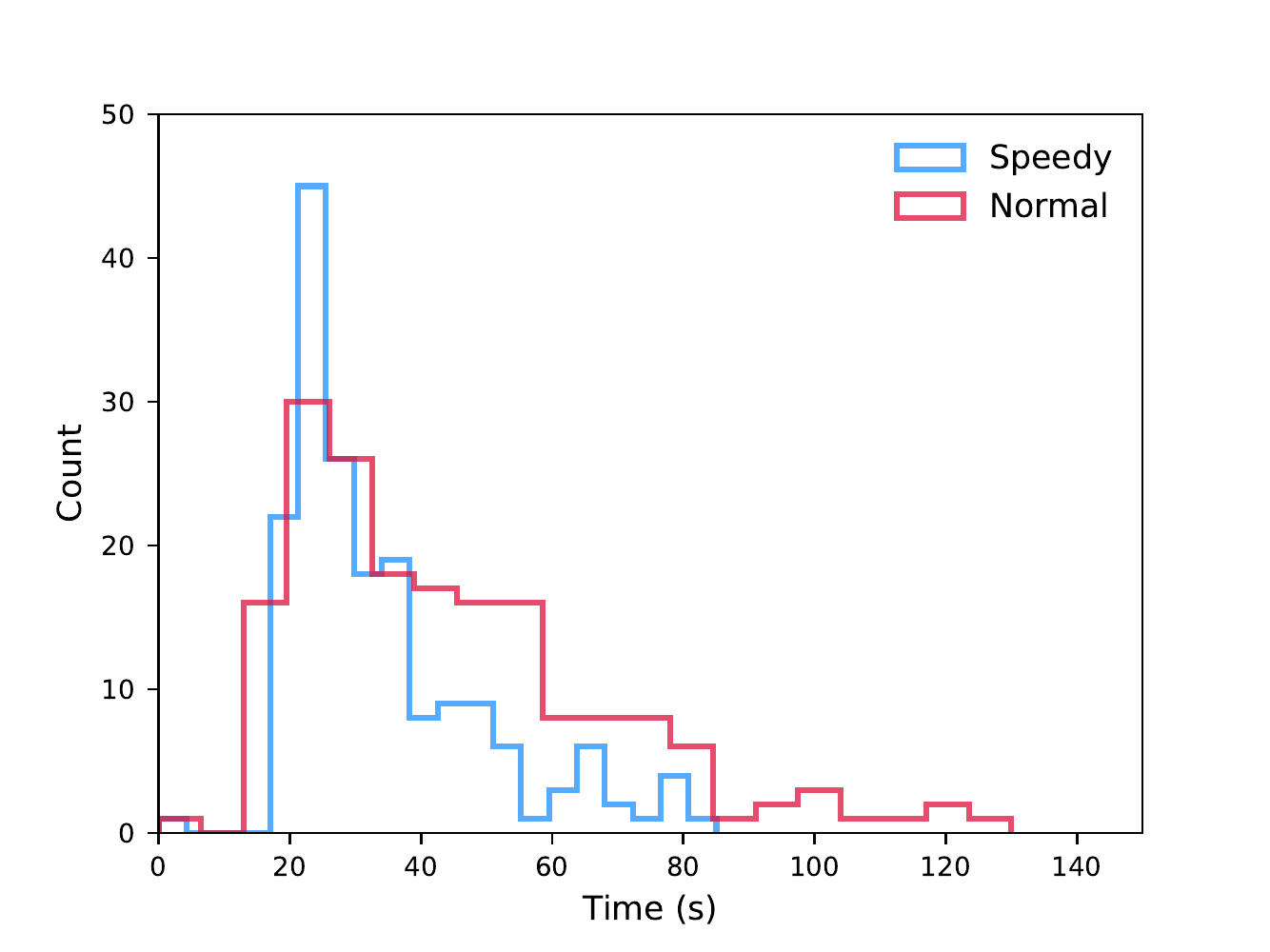}
\caption{\label{fig:speedy} Performance of the use of the "speedy" implementation
  of the phase of the $\Gamma$ function versus the standard Lanczos implementation
  of the full $\Gamma$ function. The plot refers to a fiducial $1.35M_\odot+1.35M_\odot$
  BNS system starting from 10 Hz with sampling rate of 16384Hz.
  $\Lambda^A_2=\Lambda^B_2$  varying from 1 to 1800.
  The use of the effective phase fit brings a non-negligible improvement to the performance of the code. The horizontal axis represents runtimes on a MacBook Pro with an Intel Core i7 (2.5GHz) and 16GB RAM}
\end{figure}
The improvement of performance brought by the effective representation
of the tail phase is illustrated in Fig.~\ref{fig:speedy}. The plot
refers to a fiducial $1.35M_\odot+1.35M_\odot$ BNS system starting from
10 Hz with sampling rate of 16384~Hz and $\Lambda^A_2=\Lambda^B_2$
varying from 1 to 1800. Note that to setup the \TEOBResumS{} run, from the initial
frequency one computes the initial separation using the Newtonian Kepler's constraint,
so that $r_0/M = \left(G\pi f M/c^3\right)^{-2/3}$. To convert from physical to dimensionless
units we use the value of the solar mass in time
units~\cite{Damour:1991rd}
$T_\odot=GM_\odot/c^3=4.925490947\times 10^{-6}$~sec.
The figure highlights that the use of the effective phase
fit brings a non-negligible improvement to the performance of the code.
 \begin{table*}[t]
   \caption{\label{tab:Benchmark}Runtime performance benchmark for \TEOBResumS{} over a set
     of standard configurations. All benchmarks were completed on a MacBook Pro with an Intel
     Core i7 (2.5GHz) and 16GB RAM. The code was compiled with the ${\rm g}{++}$ GNU compiler
     using O3 optimization. Typical performance for a BNS system from 10Hz is on the order
     of 45s and from 20Hz on the order of 6s. All benchmarks quoted are calculated by averaging
     over multiple waveform generation operations. Unless otherwise stated, we adopt the
     effective tail representation detailed in Appendix \ref{subsec:tail}. }
   \begin{center}
 \begin{ruledtabular}
   \begin{tabular}{cclccccccccc}
   System & EOS & $M_{A,B}[M_\odot]$ & $\chi_{A,B}$ & $\Lambda^{A,B}_2$ & $f_{\rm min} \left( \rm{Hz} \right)$ & Sample Rate $(\rm{Hz})$ & Effective Tail & Benchmark, gcc (s) \\
     \hline
     BNS & SLy    & $(1.35,1.35)$ & $(0,0)$   & $(392,392)$  & $10$ & $4096$ & Yes & $21.76$ \\
     BNS & SLy    & $(1.35,1.35)$ & $(0,0)$   & $(392,392)$  & $10$ & $8192$ & Yes & $41.66$ \\
     BNS & SLy    & $(1.35,1.35)$ & $(0,0)$   & $(392,392)$  & $10$ & $8192$ & No & $43.59$ \\
     BNS & SLy    & $(1.35,1.35)$ & $(0,0)$   & $(392,392)$  & $20$ & $4096$ & Yes & $3.44$ \\
     BNS & SLy    & $(1.35,1.35)$ & $(0,0)$   & $(392,392)$  & $20$ & $8192$ & Yes & $5.84$ \\
     BNS & MS1b    & $(1.35,1.35)$ & $(0.1,0.1)$   & $(1531,1531)$  & $10$ & $8192$ & Yes & $40.67$ \\
     NSBH & N/A & $(6,1.35)$ & $(0.4,0)$   & $(0,0)$  & $10$ & $4096$ & Yes & $17.28$ \\
     BBH & N/A    & $(10,10)$ & $(0.6,0.6)$   & $(0,0)$  & $10$ & $4096$ & Yes & $2.64$ \\
     BBH & N/A    & $(36,29)$ & $(0.5,-0.2)$   & $(0,0)$  & $10$ & $2048$ & Yes & $0.14$ \\
     BBH & N/A    & $(36,29)$ & $(0.5,-0.2)$   & $(0,0)$  & $10$ & $16384$ & Yes & $0.71$ \\

 \end{tabular}
 \end{ruledtabular}
 \end{center}
 \end{table*}
 Additional information is also listed in the second and third rows of Table~\ref{tab:Benchmark},
 where one evaluates the impact of the effective representation of the tail on a specific BNS case.
 The table also lists the performance of the \TEOBResumS{} \CC{} code for several standard binary configurations.

\section{NR-informed description of merger and postmerger}
\label{sec:postmerger_NQC}
The purpose of this Appendix is to collect all the fits used
by \TEOBResumS{} to describe the $\ell=m=2$ postmerger/ringdown
waveform part. We report results for the QNM quantities
(i.e. frequency and damping time) as well as for all other
parameters that enter the postmerger template of
Ref.~\cite{Damour:2014yha,Nagar:2016iwa,Nagar:2017jdw}.

These fits are an improvement with respect to those
of~\cite{Nagar:2016iwa,Nagar:2017jdw} in that:
(i) we use SXS data where the unphysical CoM drift is corrected~\cite{Nagar:2017jdw}
and (ii) we employ datasets that were not previously available.
While the performance in the nonspinning case remains practically
unchanged, the new information makes a difference when spins are
considered. The fits are informed by the 135 SXS waveforms,
5 BAM waveforms ($q=18$) and test-particle data. 
All fits depend on the symmetric mass ratio $\nu$ and on a
spin variable that is a suitable combination of the individual
spins of the two objects. The fits are built using a hierarchical approach:
(i) one first obtains results for the nonspinning sector using all available
nonspinning waveforms; (ii) then, the $\nu=1/4$ behavior is determined with
one-dimensional fits relying on the SXS $q=1$ waveforms; (iii) finally, all
remaining data is used to determine the rest of the coefficients.

\subsection{Postmerger and ringdown}
\label{sec:postmerger}
Let us start by discussing the new fits of the parameters
$(A_{22}^{\rm mrg},\omega_{22}^{\rm mrg},c_3^\phi,c_4^\phi,c_3^A,\omega_1^{22},\alpha_1^{22},\alpha_{21}^{22})$
entering the postmerger template as defined below.
Following Ref.~\cite{Nagar:2016iwa}, the QNM-rescaled ringdown waveform
is defined as $\bar{h}(\tau)\equiv e^{\sigma_1^{22}\tau + i\phi_0}h_{22}(\tau)/\nu$,
where $\tau= (t-t_{\rm mrg})/M_{\rm BH}$, $\sigma_1^{22}\equiv \alpha_1^{22}+i\omega_1^{22}$ is the 
(dimensionless, $M_{\rm BH}$ rescaled) complex frequency of 
the fundamental (positive frequency, $\omega_1>0$) QNM of the $\ell=m=2$
mode and $\phi_0$ the phase at merger.
The function $\bar{h}(\tau)$ is decomposed into phase and amplitude as
\begin{align}
\bar{h}(\tau )=\hat{A}_{\bar{h}}(\tau )e^{i\phi_{\bar{h}}(\tau )}.
\end{align}
The amplitude and phase are fitted using the following ans\"atze
\begin{align}
\label{eq:amp_temp}
\hat{A}_{\bar{h}}(\tau ) = &c_1^A \tanh \left(c_2^A\tau +c_3^A \right) +c_4^A,\\
\label{eq:phi_temp}
\phi_{\bar{h}}(\tau ) = & -c_1^{\phi} \mathrm{ln} \left( \frac{1+c_3^{\phi} e^{-c_2^{\phi} \tau } +c_4^\phi e^{-2c_2^\phi \tau }}{1+c_3^{\phi}+c_4^{\phi}}\right).
\end{align}
Following Ref.~\cite{Damour:2014yha}, only three of the eight coefficients
$(c_3^A,c_3^\phi,c_4^\phi)$ are independent and need to be fitted, while the
others are related to these three via several physically motivated constraints
as discussed in~\cite{Damour:2014yha}. In practice one has
\begin{align}
c_2^A = & \frac{1}{2}\alpha_{21}^{22},\\
c_4^A = & \hat{A}_{22}^{\rm mrg} - c_1^A \tanh \left(c_3^A\right),\\
c_1^A = & \hat{A}_{22}^{\rm mrg} \alpha_1^{22} \frac{\cosh^2\left(c_3^A\right)}{c_2^A},\\
c_1^{\phi} = & \Delta\omega \frac{1+c_3^{\phi}+c_4^{\phi}}{c_2^{\phi}\left(c_3^{\phi}+2c_4^{\phi}\right)},\\
c_2^{\phi} = & \alpha_{21}^{22},
\end{align}
with $\Delta\omega \equiv \omega^{22}_1 - M_{BH}\omega^{\rm mrg}_{22}$ and 
$\alpha_{21}^{22}\equiv \alpha_{2}^{22} - \alpha_{1}^{22}$.

\subsubsection{Waveform amplitude and frequency at merger}

The fits of the waveform amplitude and frequency at merger
time, $t_{\rm mrg}\equiv t_{22}^{\rm peak}$, are obtained as follows.
First, we found it useful to write the $\nu$-scaled merger amplitude as
\begin{align}
\label{eq:amp_temp1}
\hat{A}_{22}^{\rm mrg}&=\hat{A}_{\rm orb}^{\rm mrg}\hat{A}_{\rm LO}^{\rm SO} \hat{\hat{A}}_{\rm S}^{\rm mrg}.
\end{align}
In this equation, $\hat{A}^{\rm mrg}_{\rm orb}$ is the nonspinning (or orbital) contribution,
$\hat{A}_{\rm LO}^{\rm SO}$ takes into account, in an heuristic way, the leading-order
spin-orbit dependence (see below) and  $\hat{\hat{A}}_{\rm S}^{\rm mrg}$
accounts for the remaining spin-dependence.
We assume the following functional dependence for the
orbital contribution
\begin{align}
\label{eq:amp_zero_old}
\hat{A}_{\rm orb}^{\rm mrg}&=c_0^{\hat{A}^{\rm mrg}_{\rm orb}}+  c^{\hat{A}^{\rm mrg}_{\rm orb}}_1\nu +c^{\hat{A}^{\rm mrg}_{\rm orb}}_2\nu^2.
\end{align}
Note that for this fit we do not impose the test-particle limit value
(that is known \cite{Harms:2014dqa}) but we just check the consistency
of the fit a posteriori\footnote{There is however nothing that prevents us
from doing so. As a matter of fact we will explicitly impose the test-mass
limit behavior in updated fits that will appear elsewhere.}.
The coefficients of the fits are listed in the left column of Table~\ref{tab:merger_fit}.
The spin dependence of $A_{\rm LO}^{\rm SO}$ is inspired by the analytically
known spin-dependence of the $\ell=m=2$ amplitude (see e.g. Eq.~(16) of~\cite{Messina:2018ghh})
and we write
\be
\label{eq:A_LO_SO}
\hat{A}^{\rm SO}_{\rm LO}=1- \left(\hat{a}_0+\dfrac{1}{3}X_{AB}\tilde{a}_{AB}\right)x^{3/2}_{\rm mrg},
\ee
where $x_{\rm mrg} \equiv \left(\omega^{\rm mrg}_{22}/2\right)^{2/3}$ and  $\tilde{a}_{AB}\equiv\tilde{a}_{A}-\tilde{a}_{B}$.
Defining $\hat{a}_{\rm eff}\equiv \hat{a}_0+X_{AB}\tilde{a}_{AB}/3$, the residual spin-dependence
is fitted with a rational function
\begin{align}
\label{eq:A_mrg_spin_old}
\hat{\hat{A}}_{\rm S}^{\rm mrg}=\frac{1 - n^{\hat{\hat{A}}}(\nu)\hat{a}_{\rm eff}}{1-d^{\hat{\hat{A}}}(\nu)\hat{a}_{\rm eff}},
\end{align}
where $(n^{\hat{\hat{A}}},d^{\hat{\hat{A}}})$  are quadratic functions of $X_{AB}$ defined as
\begin{align}
\label{eq:spin_dm_funcs_n}
n^{\hat{\hat{A}}}(\nu) &\equiv  n_{\nu=1/4}^{\hat{A}^{\rm mrg}_{\rm spin}} + n_1^{\hat{A}^{\rm mrg}_{\rm spin}}X_{AB} + n_2^{\hat{A}^{\rm mrg}_{\rm spin}}\left(X_{AB}\right)^2,\\
\label{eq:spin_dm_funcs_d}
d^{\hat{\hat{A}}}(\nu) & \equiv d_{\nu=1/4}^{\hat{A}^{\rm mrg}_{\rm spin}} + d_1^{\hat{A}^{\rm mrg}_{\rm spin}}X_{AB} + d_2^{\hat{A}^{\rm mrg}_{\rm spin}}\left(X_{AB}\right)^2.
\end{align}
The coefficients are listed in the left column of Table~\ref{tab:merger_fit}.

\begin{table}
\caption{\label{tab:merger_fit} The left column shows the coefficients of the 
	waveform amplitude at merger, defined in Eq.~\eqref{eq:amp_temp1}~--~\eqref{eq:spin_dm_funcs_d}.
	The right column shows the coefficients of the waveform frequency at merger, defined in 
	Eq.~\eqref{eq:freq_spin_zero_old}~--~\eqref{eq:freq_spin}, relying on 
	\eqref{eq:spin_dm_funcs_n}~--~\eqref{eq:spin_dm_funcs_d}.}
\begin{ruledtabular}
\begin{tabular}{c l c l || l c l c}
&$c_0^{\hat{A}^{\rm mrg}_{\rm orb}}$ & $=$  	& $\;\;\;1.43842$ & $c_0^{\omega^{\rm mrg}_{\rm orb}}$ & $=$  & $\;\;\;0.273813$&\\
&$c_1^{\hat{A}^{\rm mrg}_{\rm orb}}$ & $=$  	& $\;\;\;0.100709$ & $c_1^{\omega^{\rm mrg}_{\rm orb}}$ & $=$  & $\;\;\;0.223977$&\\
&$c_2^{\hat{A}^{\rm mrg}_{\rm orb}}$ & $=$  	& $\;\;\;1.82657$ & $c_2^{\omega^{\rm mrg}_{\rm orb}}$ & $=$  & $\;\;\;0.481959$&\\\hline
&$n_{\nu=1/4}^{\hat{A}^{\rm mrg}_{\rm spin}}$ & $=$   & $-0.293524$ & $n_{\nu=1/4}^{\omega^{\rm mrg}_{\rm spin}}$ & $=$   & $-0.283200$&\\
&$d_{\nu=1/4}^{\hat{A}^{\rm mrg}_{\rm spin}}$ & $=$   & $-0.472871$ & $d_{\nu=1/4}^{\omega^{\rm mrg}_{\rm spin}}$ & $=$   & $-0.696960$&\\\hline
&$n_1^{\hat{A}^{\rm mrg}_{\rm spin}}$ & $=$ 	& $\;\;\;0.176126$ & $n_1^{\omega^{\rm mrg}_{\rm spin}}$ & $=$ & $\;\;\;0.1714956$&\\ 
&$n_2^{\hat{A}^{\rm mrg}_{\rm spin}}$ & $=$ 	& $-0.0820894$ & $n_2^{\omega^{\rm mrg}_{\rm spin}}$ & $=$ & $-0.24547$&\\
&$d_1^{\hat{A}^{\rm mrg}_{\rm spin}}$ & $=$ 	& $\;\;\;0.20491$ & $d_1^{\omega^{\rm mrg}_{\rm spin}}$ & $=$ & $\;\;\;0.1653028$&\\ 
&$d_2^{\hat{A}^{\rm mrg}_{\rm spin}}$ & $=$ 	& $-0.150239$ & $d_2^{\omega^{\rm mrg}_{\rm spin}}$ & $=$ & $-0.1520046$&\\
\end{tabular}
\end{ruledtabular}
\end{table}

\begin{figure}[t]
\center
\includegraphics[width=0.45\textwidth]{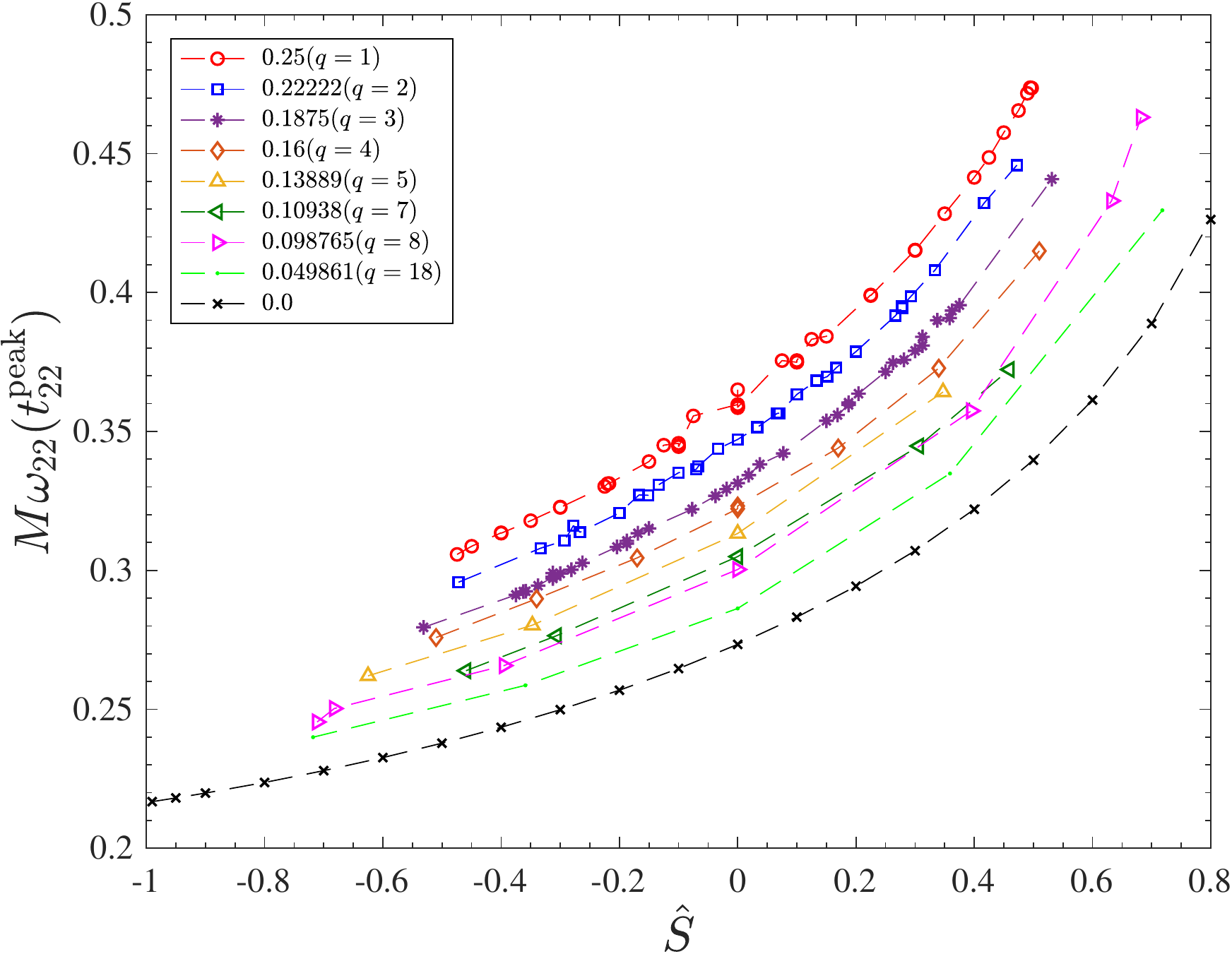}
\caption{\label{fig:omg_peak}Uniform (or quasi-universal) behavior of the NR gravitational wave
frequency at merger time $\omega_{22}^{\rm mrg}\equiv \omega_{22}(t_{22}^{\rm peak})$ for 
various mass ratios when plotted versus $\hat{S}=(S_A+S_B)/M^2$. 
Both SXS and BAM data are shown together. It is remarkable the qualitative consistency
between the test-mass limit curve (black online) and the finite mass-ratio ones.}
\end{figure}

\begin{figure}[t]
\center
\includegraphics[width=0.45\textwidth]{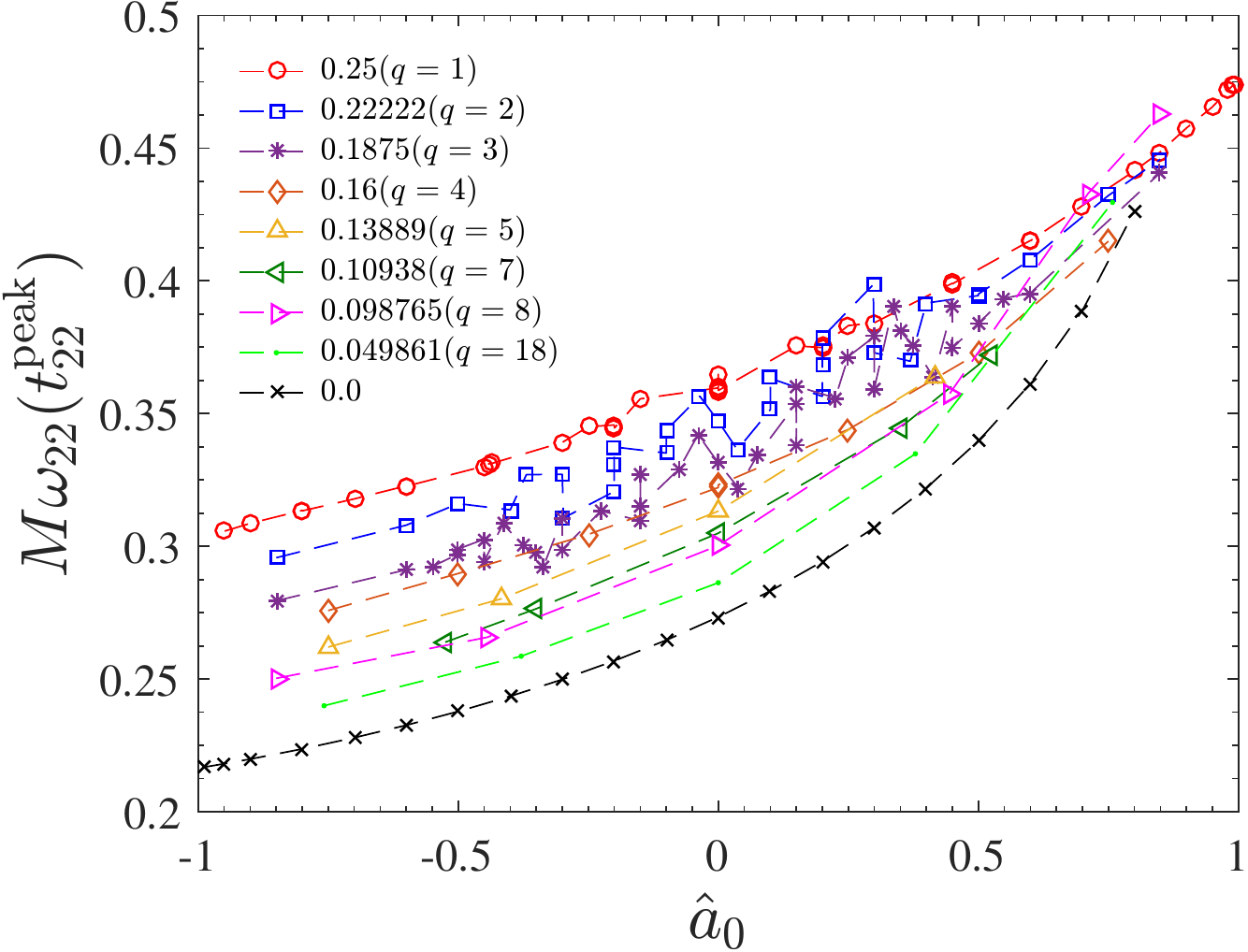}
\caption{\label{fig:omg_peak_a0} Same data of Fig.~\ref{fig:omg_peak}, but plotted versus
  the standard effective Kerr parameter $\hat{a}_0=\hat{S}+\hat{S}_*$.}
\end{figure}

\begin{figure}[t]
  \center
  \includegraphics[width=0.45\textwidth]{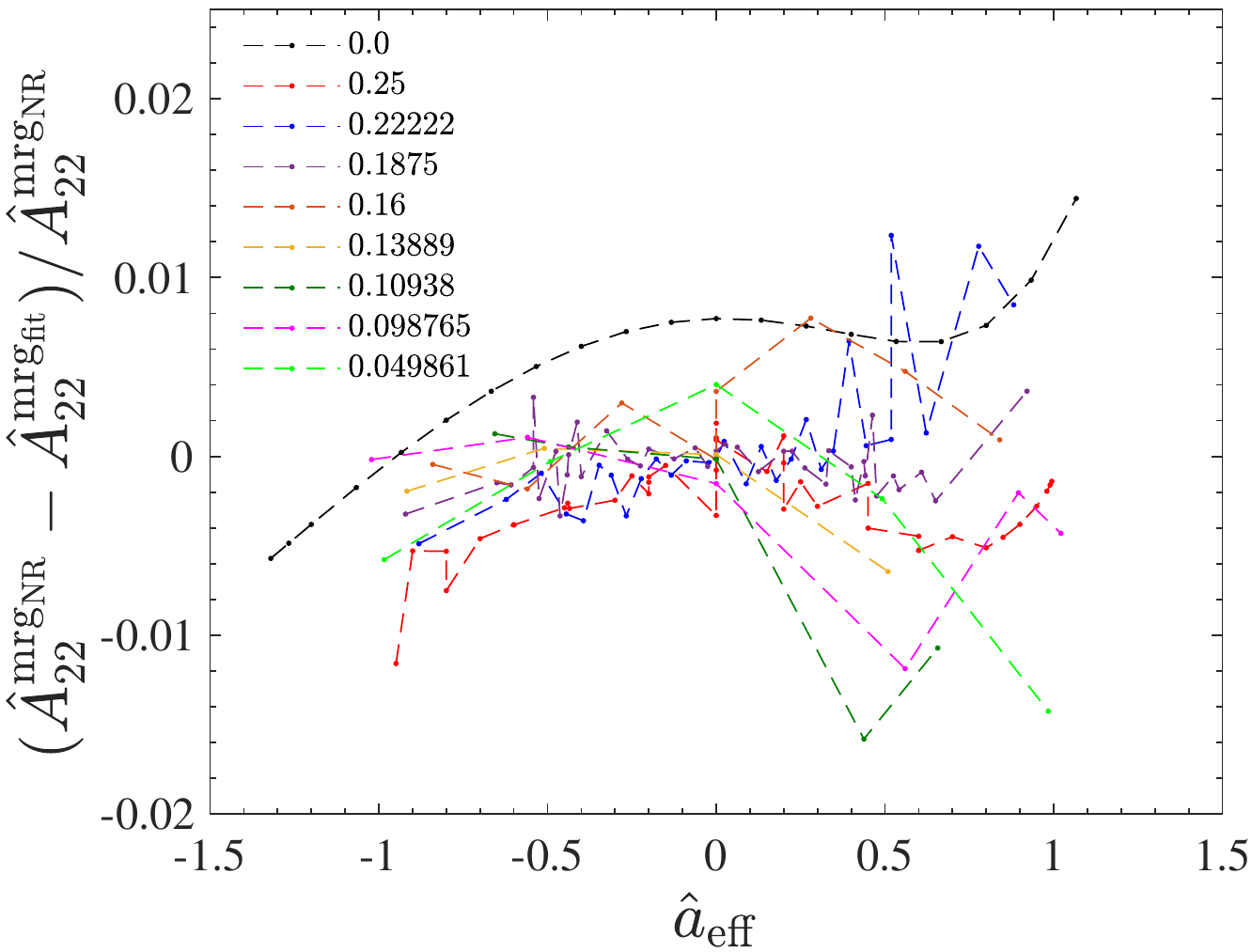}
    \includegraphics[width=0.45\textwidth]{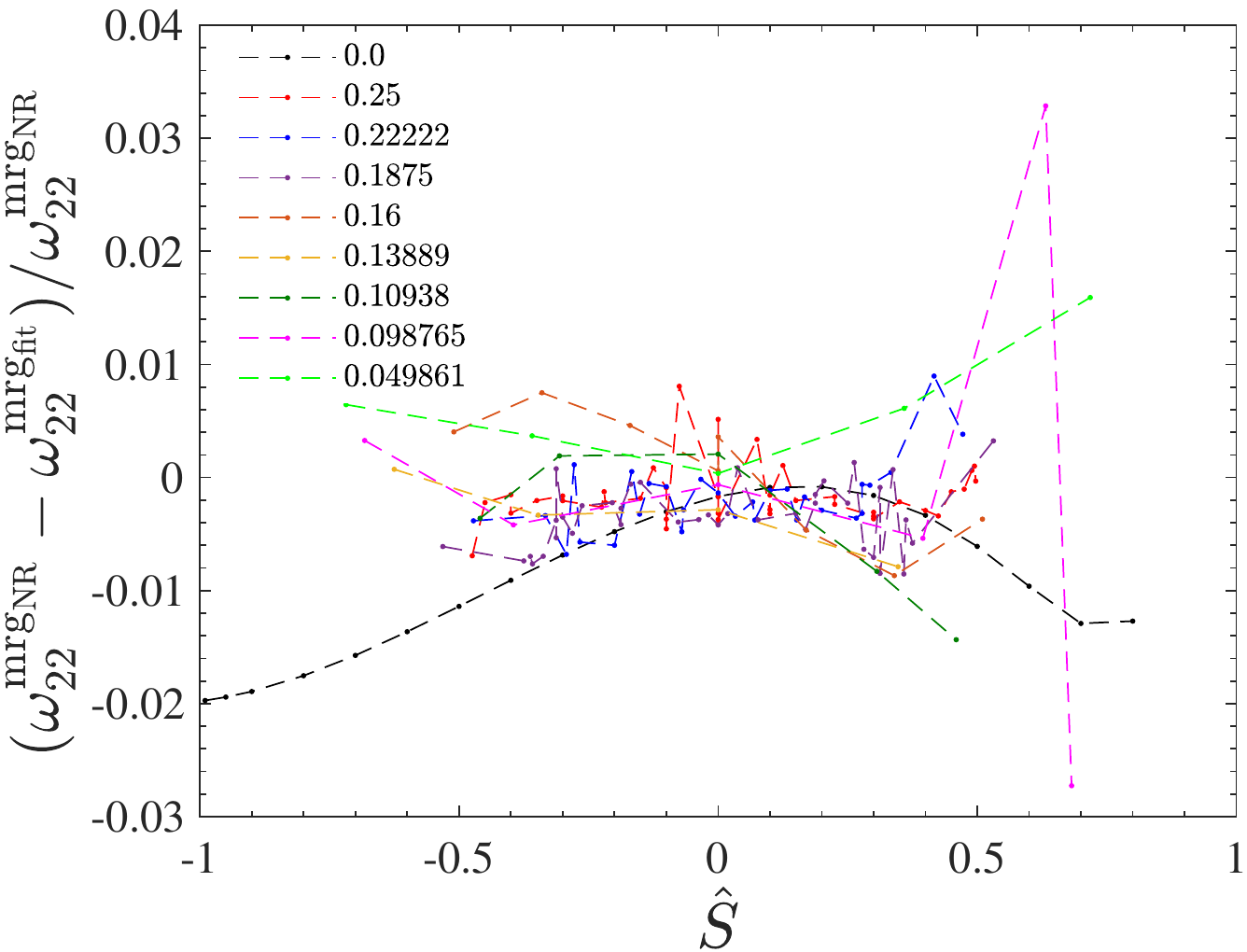}
  \caption{\label{fig:Amp_old_vs_NR}Evaluating the performance of the global fit of
  the merger amplitude given in Table~\ref{tab:merger_fit}.}
\end{figure}

Let us turn now to discussing the merger frequency. In Fig.~\ref{fig:omg_peak} we show
the NR values of $\omega_{22}^{\rm mrg}$ versus the effective spin variable $\hat{S}\equiv (S_A+S_B)/M^2$.
It is interesting to note that, independently of the value of the two spins, there
is one, smooth, curve for each mass ratio. In addition, the behavior in $\hat{S}$,
that is well represented by a fourth-order polynomial in $\hat{S}$, is qualitatively
{\it the same} for each configuration, including the test-mass limit (black curve
in the plot).  Such globally nice and simple behavior is apparent because we are using
$\hat{S}$ as spin variable. In fact, had one used $\hat{a}_0$, that is the other natural
choice of effective spin that one may consider, the curve belonging to each mass ratio
becomes rather complicated, with large oscillations corresponding to those configurations
where the spins are different, see Fig.~\ref{fig:omg_peak_a0}. This is evidently the case
for $q=2$ and $q=3$ data. The complete understanding of such nice property of the merger
state requires a more precise study that goes beyond the purpose of this Appendix.
Let us however put forward a few theoretical arguments to get an intuitive feeling
of what might be going on. By assuming that $\omega_{22}^{\rm merg}\simeq 2 \Omega$
(that is more true for high, positive spins), at some $u=u_{\rm mrg}$, one has
\be
\label{eq:Omg_by_2_mrg}
\omega_{22}^{\rm mrg}\simeq 2\left[\dfrac{A u_c^2 p_\varphi^{\rm mrg} }{H_{\rm EOB}\hat{H}_{\rm eff}^{\rm orb}}+H_{\rm EOB}^{-1}\left(G_S \hat{S}+G_{S_*}\hat{S}_*\right)\right].
\ee
Since $u_c^2$ contains only even-parity powers of $\hat{a}_0=\hat{S}+\hat{S}_*$, one
easily understands where the, seemingly natural, spin-dependence on $\hat{S}$ may come from.
However, if the functional dependence on $\hat{S}$ is obvious for $q=1$,
since $\hat{S}_*=\hat{S}$, it is less obvious when $q\neq 1$ because of the
presence of also $\hat{S}_*$. In fact, Fig.~\ref{fig:omg_peak} seems to suggest
two facts: (i) the $\hat{S}_*$ dependence is subdominant with respect to the $\hat{S}$ one
and (ii) in Eq.~\eqref{eq:Omg_by_2_mrg} one has to worry {\it at most} of terms quartic
in $\hat{S}$, that only originate from the pure orbital part of the frequency because
of $p_\varphi^{\rm mrg}$ and $u_c^2$. The fourth-order spin dependence is, a priori,
not that surprising, because, since $u_c^2=u^2/(1+u^2(\hat{a}_0^2+2u)+\delta\hat{a}^2 u^2)$,
it is the first correction to the leading-order spin-spin term coming from the expansion
of $u_c^2$ and it is there already for a test-particle orbiting a Kerr black hole.
By contrast, there is a priori no reason why $\hat{S}$ should be more important
than $\hat{S}_*$. To get some intuition of why it is so, working in the circular
approximation, we can take $u^{\rm mrg}=1/3$ as a ``fiducial'' merger
of a binary with $q=2$, and expand Eq.~\eqref{eq:Omg_by_2_mrg} in both $u$ and the spins.
One then verifies that the spin dependence of the quantity $u_c^2/u^2 p_\varphi(u)$
at $u=u^{\rm mrg}$ is such that the coefficients of the highest
powers of $\hat{S}_*$ (e.g. $\hat{S}_*^4$, $\hat{S}_*^3$) are {\it numerically smaller}
than the corresponding ones for $\hat{S}$. By contrast, this does not happen
for $(\hat{S}^2,\hat{S}_*^2)$, which indicates that the lack of symmetry between
$(\hat{S},\hat{S}_*)$ seen in Fig.~\ref{fig:omg_peak} may just crucially stem from
NLO (and higher) spin-spin effects. These facts intuitively suggest (though certainly
do not explain it quantitatively) what might be the origin of the simple scaling with $\hat{S}$
illustrated in Fig.~\ref{fig:omg_peak}. This will deserve a more dedicated and extensive
study on its own to be fully understood. For the moment, we content ourselves of
having identified the simple structure of $\omega_{22}^{\rm mrg}$ of Fig.~\ref{fig:omg_peak}
and exploit it at best to obtain its global fit.
To do so, we again assume a template in factorized form as 
\begin{align}
\omega_{22}^{\rm mrg}=\omega_{\rm orb}^{\rm mrg}\left(\nu\right) \omega_{\rm S}^{\rm mrg}\left(\hat{S},X_{AB}\right),
\end{align}
where the orbital factor $\omega_{\rm orb}^{\rm mrg}$ is fitted with a quadratic function in $\nu$,
\begin{align}
\label{eq:freq_spin_zero_old}
\omega_{\rm orb}^{\rm mrg}\left(\nu\right) = c_0^{\omega^{\rm mrg}_{\rm orb}} + c_1^{\omega^{\rm mrg}_{\rm orb}}\nu + c_2^{\omega^{\rm mrg}_{\rm orb}}\nu^2.
\end{align} 
The functional form of $\omega_{\rm S}^{\rm mrg}$ is identical
to $\hat{\hat{A}}_{\rm S}^{\rm mrg}$ (Eq.~\eqref{eq:A_mrg_spin_old}),
though one is using now $\hat{S}$ as spin variable so that
\begin{align}
\label{eq:freq_spin}
\omega_{\rm S}^{\rm mrg}=\frac{1 - n^\omega(\nu) \hat{S} }{1-d^\omega(\nu) \hat{S} }
\end{align}
where the functions $(n^\omega,d^\omega)$ have the same functional form as stated previously 
in equations~\eqref{eq:spin_dm_funcs_n}~--~\eqref{eq:spin_dm_funcs_d}.
Coefficients of both the nonspinning and spinning parts of the fits are shown
in the right column of Table~\ref{tab:merger_fit}. We found the use of the spin
variable $\hat{S}$ simplified the fitting of the 
waveform frequency at merger. The precise analytical reason behind such simple
spin dependence is currently not deeply understood. 
The estimate its performance, the result of the fit was compared with the full set of data available.
The fractional differences are displayed in Fig.~\ref{fig:Amp_old_vs_NR}.
The largest differences are found for $(8,0.8,0)$ and $(8,0.85,0.85)$. Since the corresponding frequency
values seem to be slightly inconsistent with the expected $q=8$ trend of the frequency
(see  the magenta line in Fig.~\ref{fig:omg_peak}) we have conservatively preferred not 
to use them in the global fit.

\subsubsection{Fits of the QNMs parameter of the final black hole}
\label{sec:fit_qnms}
\begin{table}
  \caption{\label{tab:QNM_fits} Coefficients of the fits of the
    fundamental QNM frequency and inverse damping time of the final
    remnant $(\omega_1,\alpha_1)$ as well as the difference $\alpha_{21}=\alpha_2-\alpha_1$
    of the inverse damping times of the first two modes. See Eq.~\eqref{eq:QNM_g}
    for definitions.}
\begin{ruledtabular}
\begin{tabular}{c c | l | l | l c}
 & & $\;\;\;Y=\omega^{22}_1$      & $\;\;\;Y=\alpha^{22}_1$      & $\;\;\;Y=\alpha^{22}_{21}$& \\ \hline
& $Y_0$   & $\;\;\;0.373672$ & $\;\;\;0.0889623$ & $\;\;\;0.184953$ & \\
& $b_1^Y$ & $-1.74085$ & $-1.82261$ & $-1.41681$ & \\
& $b_2^Y$ & $\;\;\;0.808214$ & $\;\;\;0.701584$ & $-0.0593166$ & \\
& $b_3^Y$ & $-0.0598838$ & $\;\;\;0.121126$ & $\;\;\;0.476420$ & \\
& $c_1^Y$ & $-2.07641$ & $-1.80020$ & $-1.35955$ & \\
& $c_2^Y$ & $\;\;\;1.31524$ & $\;\;\;0.720117$ & $-0.0763529$ & \\
& $c_3^Y$ & $-0.235896$ & $\;\;\;0.0811633$ & $\;\;\;0.438558$ & \\
\end{tabular}
\end{ruledtabular}
\end{table}
\begin{figure*}[t]
  \center
  \includegraphics[width=0.30\textwidth]{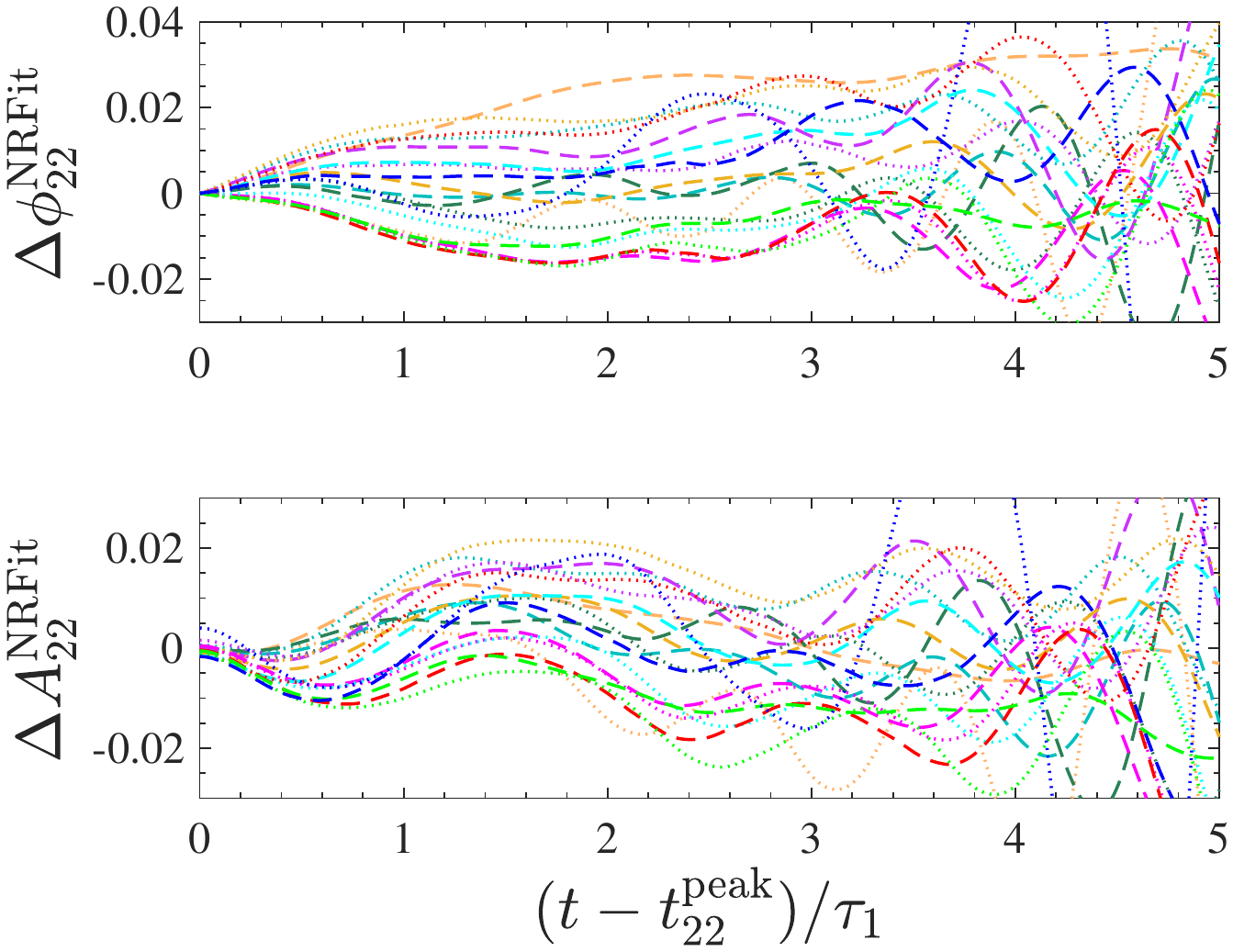}
  \includegraphics[width=0.30\textwidth]{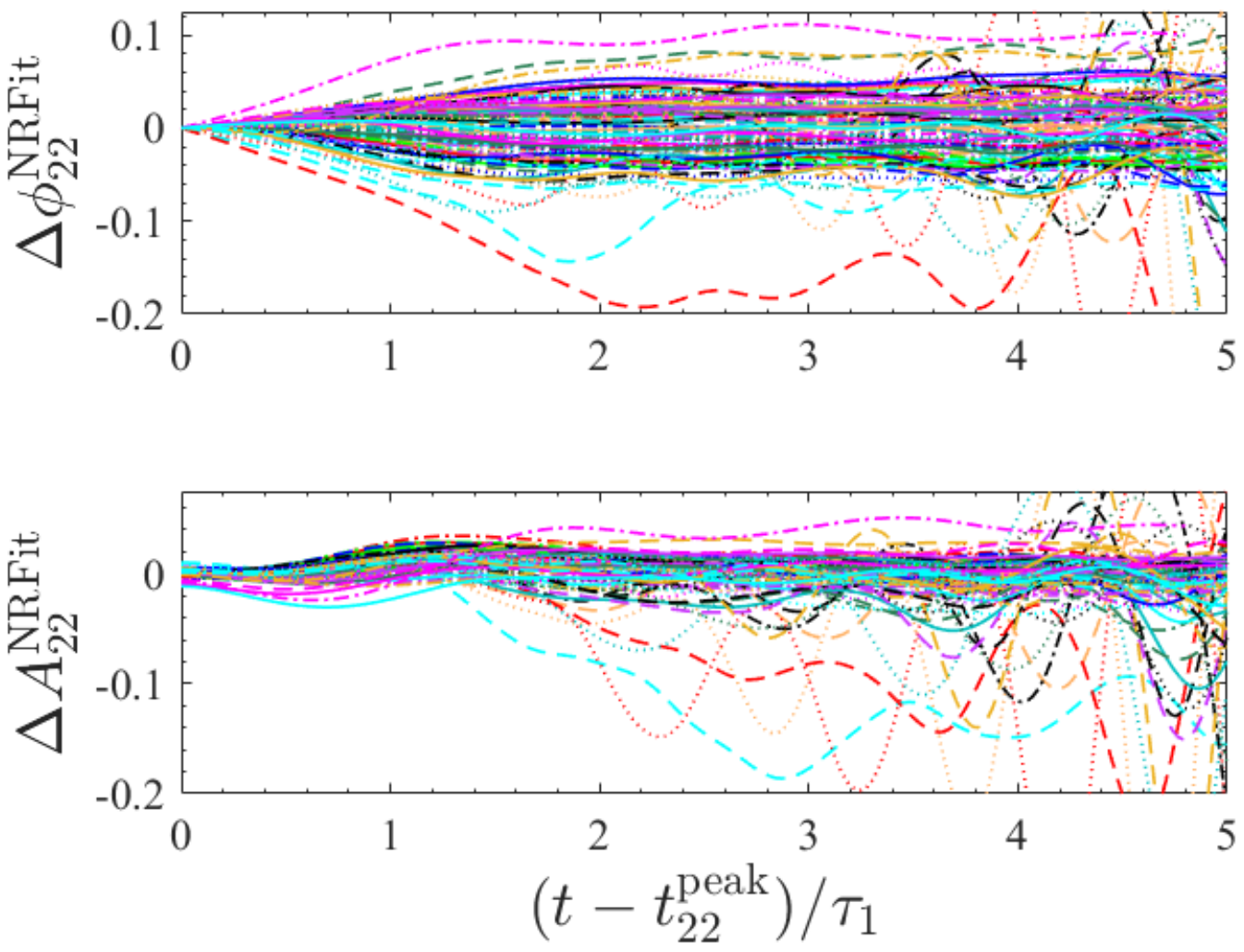}
  \includegraphics[width=0.30\textwidth]{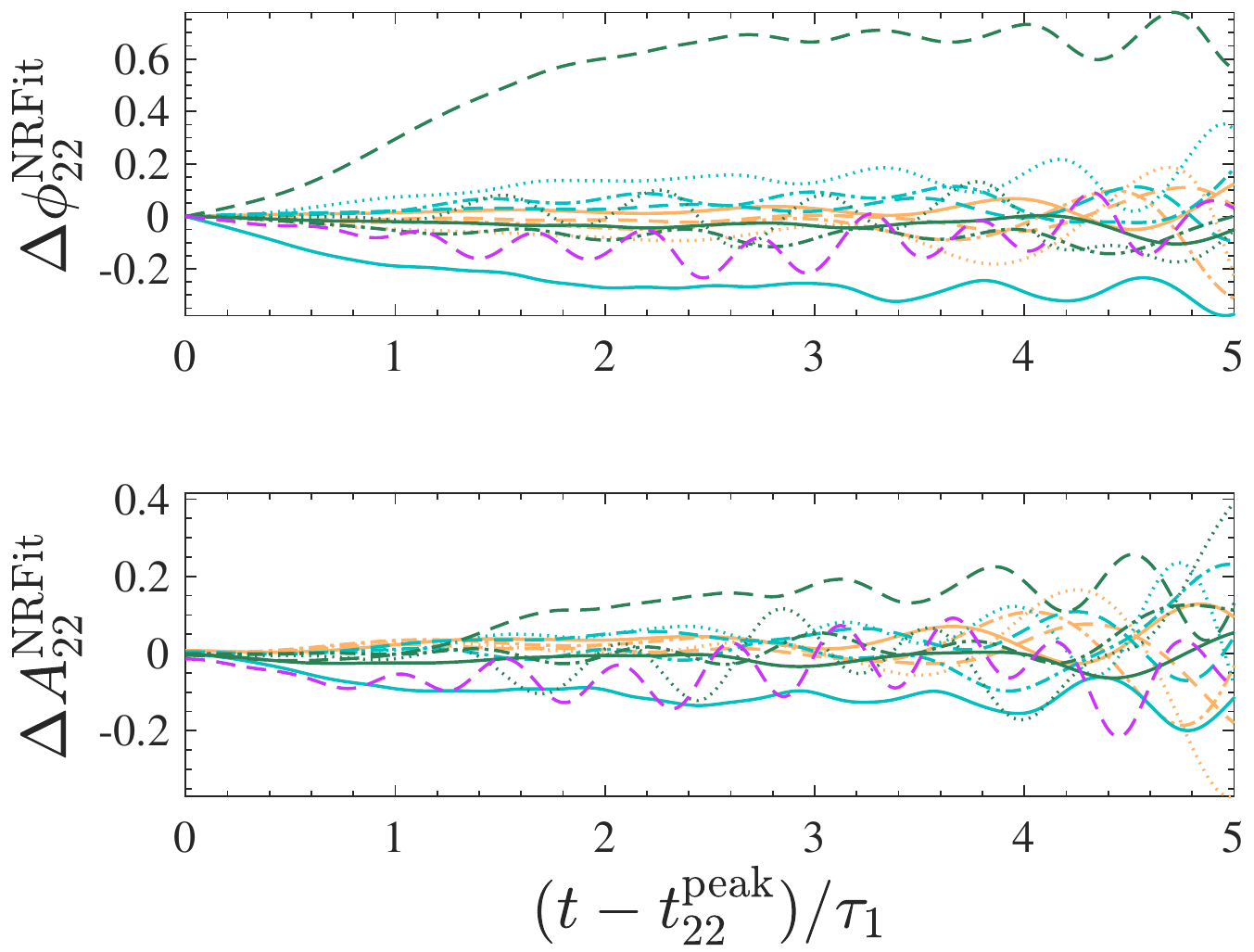}
    \caption{\label{fig:RD_perform} Left panel: The $\ell=m=2$ phasing and amplitude performance for
      nonspinning waveforms. Middle panel: the phasing and amplitude performance for all spinning SXS waveforms.
      Right panel: BAM waveforms. Top row: Phase error $\Delta\phi_{22}^{\rm NRFit}\equiv \phi^{\rm NR}_{22}-\phi^{\rm fit}_{22}$. 
      Bottom row: Fractional amplitude difference $\Delta A^{\rm NRFit}_{22}\equiv (A^{\rm NR}_{22}-A^{\rm fit}_{22})/A^{\rm NR}_{22}$.
      The time is given in units of $\tau_1\equiv M_{\rm BH}/\alpha_1$.}
\end{figure*}
The fits of $Y= \left\lbrace\omega_1^{22},\alpha_1^{22},\alpha_2^{22}\right\rbrace$ presented 
here were informed using data obtained using publicly available tables of Berti et al.~\cite{Berti:2005ys,Berti:2009kk}. 
The fits are done versus the dimensionless spin parameter 
$\chi_{\rm BH}\equiv J_{\rm BH}/M_{\rm BH}^2$ of the final black hole.
$\chi_{\rm BH}$ is computed for a given set of initial conditions with 
the fits of Jimenez-Forteza et al.~\cite{Jimenez-Forteza:2016oae}.
The three parameters are fitted with rational functions after 
the $\chi_{\rm BH}=0$ limit was factorized
\begin{align}
\label{eq:QNM_g}
Y\left(\hat{a}_f\right)  = Y_0\,\frac{1+b_1^Y\chi_{\rm BH}+b_2^Y\chi_{\rm BH}^2+b_3^Y\chi_{\rm BH}^3}{1+c_1^Y\chi_{\rm BH}+c_2^Y\chi_{\rm BH}^2+c_3^Y\chi_{\rm BH}^3},
\end{align}
The fitted coefficients are listed in Table~\ref{tab:QNM_fits}.

\subsubsection{Fits of the additional parameters}
\label{sec:fit_evol}

The three parameters $Y_{\rm orb}=\left\lbrace c_3^A,c_3^\phi,c_4^\phi\right\rbrace$
are first obtained by fitting each QNM-rescaled ringdown NR waveform
on a time interval of length $4\tau_1$ after the amplitude peak,
using the functional form of Eq.~\eqref{eq:amp_temp} for the
QNM-rescaled amplitude, and of Eq.~\eqref{eq:phi_temp}
for the QNM-rescaled phase. The result of this primary fit, done
for each NR dataset considered in the main text, is then globally
fitted using the following functional form
\begin{align}
\label{eq:evol_spin}
Y(\nu;\,\hat{S}) = b^Y_0(\nu)+ & b^Y_1\left(X_{AB}\right) \hat{S}
  + b^Y_2\left(X_{AB}\right) \hat{S}^2 \nonumber\\
& + b^Y_3\left(X_{AB}\right) \hat{S}^3 + b^Y_4\left(X_{AB}\right) \hat{S}^4.
\end{align}
The coefficients of the fit are listed in Table~\ref{tab:rng}.
Note that the fit of $Y=c_3^A$ is done using $\hat{a}_{\rm eff}$ instead of $\hat{S}$.  
To demonstrate the accuracy of the fits, Fig.~\ref{fig:RD_perform} shows
the direct comparison of the globally interpolating fits with the available NR information.
The comparison is done through the difference of the waveform phase and 
the fractional amplitude difference. Note that two BAM waveforms show a strong dephasing.
These are $(8,0.8,0)$ (light blue, solid) and $(8,0.85,0.85)$ (green, dashed), 
and is due to the error of the merger frequency.
As mentioned above, we leave it to future work to resolve these differences.

\begin{table*}
  \caption{\label{tab:rng} The fitted coefficients of $\left\lbrace c_3^A,c_3^\phi,c_4^\phi\right\rbrace$
  as defined in Eq.~\eqref{eq:evol_spin}.}
\begin{ruledtabular}
\begin{tabular}{c l c l l | l c l l | l c l l  c}
& \multicolumn{4}{c|}{$Y=c_3^A$} & \multicolumn{4}{c|}{$Y=c_3^\phi$} 	& \multicolumn{4}{c}{$Y=c_4^\phi$} &\\\hline
&  $b^{c_3^A}_0(\nu)$ 	 & $=$ & $-0.561584$ & $+0.829868\nu$ 			&  $b^{c_3^\phi}_0(\nu)$ 	 & $=$ & $\;\;\;3.88838$ 	& $+0.455847\nu$ &  $b^{c_4^\phi}_0(\nu)$ 	 & $=$	& $1.49969$ & $+2.08223\nu$ &\\ 
&  $b^{c_3^A}_1(X_{AB})$ & $=$ & $-0.199494$ & $+0.0169543X_{AB}$ 		&  $b^{c_3^\phi}_1(X_{AB})$ & $=$& $\;\;\;5.11992$ 	& $-0.924642X_{AB}$ &  $b^{c_4^\phi}_1(X_{AB})$ & $=$& $8.26248$ & $-0.899952X_{AB}$ &\\
&  $b^{c_3^A}_2(X_{AB})$ & $=$ & $\;\;\;0.0227344$ & $-0.0799343X_{AB}$&  $b^{c_3^\phi}_2(X_{AB})$ & $=$& $\;\;\;10.29692$ 	& $-3.618048X_{AB}$ &  $b^{c_4^\phi}_2(X_{AB})$ & $=$& $14.27808$ & $-3.923652X_{AB}$ &\\
&  $b^{c_3^A}_3(X_{AB})$ & $=$ & $\;\;\;0.0907477$ & $-0.115928X_{AB}$ &  $b^{c_3^\phi}_3(X_{AB})$ & $=$& $-4.041224$ 		& $+3.501976X_{AB}$ &  $b^{c_4^\phi}_3(X_{AB})$ & $=$&\multicolumn{2}{c}{$0$} &\\
&  $b^{c_3^A}_4(X_{AB})$ & $=$ & \multicolumn{2}{c|}{$0$} 				&  $b^{c_3^\phi}_4(X_{AB})$ & $=$& $-32.92144$ 		& $+29.24000X_{AB}$ 	&  $b^{c_4^\phi}_4(X_{AB})$ & $=$ &\multicolumn{2}{c}{$0$} &\\
\end{tabular}
\end{ruledtabular}
\end{table*}

\subsection{Fits of the NQC point}
\label{sec:NQC}

In this Section we present fits of the values of the NR waveform taken at
the point $t_{\rm NQC}\equiv t_{\rm mrg}+2M$. On each SXS NR data set one
measures the quantities $\left\lbrace\hat{A}_{22}^{\rm NQC},\dot{\hat{A}}_{22}^{\rm NQC},\omega_{22}^{\rm NQC},\dot{\omega}_{22}^{\rm NQC}\right\rbrace$
that are then properly fitted. These are then used to determine the NQC parameters 
defined in Sec.~\ref{sec:mainfeats}. 
Note that the results of this Section refer to the Regge-Wheeler normalized strain
waveform $\hat{\Psi}_{22}\equiv \hat{h}_{22}/\sqrt{24}$ already used in the main text.
With a slight abuse of notation, we will refer here to the amplitude (and time derivative)
of this quantity at $t_{\rm NQC}$ as $(\hat{A}^{\rm NQC},\dot{\hat{A}}^{\rm NQC})$ where
$\hat{A}^{\rm NQC}\equiv \left|\Psi_{22}(t_{\rm NQC})\right|$.

\begin{table*}[t]
  \caption{\label{tab:NQC} Coefficients of the NQC extraction points defined in Eqs.~\eqref{eq:amp_ans_NQC}-\eqref{eq:df_spin_NQC}.
  From left to right the columns show $\left\lbrace\hat{A}_{22}^{\rm NQC},\dot{\hat{A}}_{22}^{\rm NQC},
  \omega_{22}^{\rm NQC},\dot{\omega}_{22}^{\rm NQC}\right\rbrace$.}
\begin{ruledtabular}
\begin{tabular}{c lll | lll | lll | lll c }
	 &\multicolumn{3}{c|}{$\hat{A}^{\rm NQC}_{22}$} & \multicolumn{3}{c|}{$\dot{A}^{\rm NQC}_{22}/\nu$} & \multicolumn{3}{c|}{$\omega^{\rm NQC}_{22}$} & \multicolumn{3}{c}{$\dot{\omega}^{\rm NQC}_{22}$} &\\\hline
  & $c_0^{\hat{A}^{\rm NQC}_{\rm orb}}$ & $=$ & $\;\;\;0.294888$ & $N_0^{\dot{A}^{\rm NQC}_{\rm orb}}$ & $=$ & $-0.00421428$ 
  & $c_0^{\omega^{\rm NQC}_{\rm orb}}$ & $=$ & $\;\;\;0.286399$ & $N_0^{\dot{\omega}^{\rm NQC}_{\rm orb}}$ & $=$ & $\;\;\;0.00649349$& \\
  & $c_1^{\hat{A}^{\rm NQC}_{\rm orb}}$ & $=$ & $-0.0427442$ & $N_1^{\dot{A}^{\rm NQC}_{\rm orb}}$ & $=$ & $-0.0847947$ 
  & $c_1^{\omega^{\rm NQC}_{\rm orb}}$ & $=$ & $\;\;\;0.251240$ & $N_1^{\dot{\omega}^{\rm NQC}_{\rm orb}}$ & $=$ & $\;\;\;0.00452138$& \\
  & $c_2^{\hat{A}^{\rm NQC}_{\rm orb}}$ & $=$ & $\;\;\;0.816756$ & $D_1^{\dot{A}^{\rm NQC}_{\rm orb}}$ & $=$ & $\;\,16.1559$  
  & $c_2^{\omega^{\rm NQC}_{\rm orb}}$ & $=$ & $\;\;\;0.542717$ & $D_1^{\dot{\omega}^{\rm NQC}_{\rm orb}}$ & $=$ & $-1.44664$& \\
  & $c_3^{\hat{A}^{\rm NQC}_{\rm orb}}$  & $=$ & $-0.986204$ & & & & & & & & & & \\\hline
  & $n_{1/4}^{\hat{A}^{\rm NQC}_{\rm spin}}$ & $=$ & $-0.275052$ & $n_{1/4}^{\dot{A}^{\rm NQC}_{\rm spin}}$ & $=$ & $\;\;\;0.00374616$ 
  & $n_{1/4}^{\omega^{\rm NQC}_{\rm spin}}$ & $=$ & $-0.292192$ &$a_{1/4}^{\dot{\omega}^{\rm NQC}_{\rm spin}}$ & $=$ & $\;\;\;0.1209112$& \\
  & $d_{1/4}^{\hat{A}^{\rm NQC}_{\rm spin}}$ & $=$ & $-0.469378$ & $d_{1/4}^{\dot{A}^{\rm NQC}_{\rm spin}}$ & $=$ & $\;\;\;0.0636083$ 
  & $d_{1/4}^{\omega^{\rm NQC}_{\rm spin}}$ & $=$ & $-0.686036$ &$b_{1/4}^{\dot{\omega}^{\rm NQC}_{\rm spin}}$ & $=$ & $-0.1198332$& \\\hline
  & $n_1^{\hat{A}^{\rm NQC}_{\rm spin}}$ & $=$ & $\;\;\;0.143066$ & $n_1^{\dot{A}^{\rm NQC}_{\rm spin}}$ & $=$ & $\;\;\;0.00129393$ 
  & $n_1^{\omega^{\rm NQC}_{\rm spin}}$ & $=$ & $\;\;\;0.1996112$ &$a_1^{\dot{\omega}^{\rm NQC}_{\rm spin}}$ & $=$ & $\;\;\;0.142343$& \\
  & $n_2^{\hat{A}^{\rm NQC}_{\rm spin}}$ & $=$ & $-0.0425947$ & $n_2^{\dot{A}^{\rm NQC}_{\rm spin}}$ & $=$ & $-0.00239069$ 
  & $n_2^{\omega^{\rm NQC}_{\rm spin}}$ & $=$ & $-0.236196$ &$a_2^{\dot{\omega}^{\rm NQC}_{\rm spin}}$ & $=$ & $-0.1001772$& \\ 
  & $d_1^{\hat{A}^{\rm NQC}_{\rm spin}}$ & $=$ & $\;\;\;0.176955$ & $d_1^{\dot{A}^{\rm NQC}_{\rm spin}}$ & $=$ & $-0.0534209$ 
  & $d_1^{\omega^{\rm NQC}_{\rm spin}}$ & $=$ & $\;\;\;0.1843102$ &$b_1^{\dot{\omega}^{\rm NQC}_{\rm spin}}$ & $=$ & $\;\;\;0.1844956$& \\
  & $d_2^{\hat{A}^{\rm NQC}_{\rm spin}}$ & $=$ & $-0.111902$ & $d_2^{\dot{A}^{\rm NQC}_{\rm spin}}$ & $=$ & $-0.186101$ 
  & $d_2^{\omega^{\rm NQC}_{\rm spin}}$ & $=$ & $-0.148057$ &$b_2^{\dot{\omega}^{\rm NQC}_{\rm spin}}$ & $=$ & $-0.0612272$& \\
\end{tabular}
\end{ruledtabular}
\end{table*}
$\hat{A}_{22}^{\rm NQC}$ is fitted similarly to the amplitude at merger by
factoring it as
\begin{align}
\label{eq:amp_ans_NQC}
\hat{A}_{22}^{\rm NQC}&=\hat{A}_{\rm orb}^{\rm NQC}\hat{A}_{\rm LO}^{\rm SO} \hat{\hat{A}}_{\rm S}^{\rm NQC},
\end{align}
with $\hat{A}_{\rm LO}^{\rm SO}$ similar to Eq.~\eqref{eq:A_LO_SO}, however evaluated using 
$x_{\rm NQC} \equiv \left(\omega^{\rm NQC}_{22}/2\right)^{2/3}$.
The nonspinning (orbital) contribution $\hat{A}_{\rm orb}^{\rm NQC}$ is fitted with
\begin{align}
\label{eq:amp_zero_NQC}
\hat{A}_{\rm orb}^{\rm NQC}&= c_3^{\hat{A}^{\rm NQC}_{\rm orb}}\nu^3 +c_2^{\hat{A}^{\rm NQC}_{\rm orb}}\nu^2 + c_1^{\hat{A}^{\rm NQC}_{\rm orb}}\nu +c_0^{\hat{A}^{\rm NQC}_{\rm orb}}.
\end{align}
The residual spin dependence is represented as 
\begin{align}
\label{eq:A_NQC_spin}
\hat{\hat{A}}_{\rm S}^{\rm NQC}=\frac{1 - n^{\rm NQC}_S \hat{a}_{\rm eff} }{1-d^{\rm NQC}_S\hat{a}_{\rm eff}  }.
\end{align}
where $(n^{\rm NQC}_S,d^{\rm NQC}_S)$ are both second-order polynomial in $X_{AB}$ 
as defined in Eq.~\eqref{eq:spin_dm_funcs_n}~--~\eqref{eq:spin_dm_funcs_d}. 
All coefficients are listed in the first column of Table~\ref{tab:NQC}.

To fit the  time derivative of the amplitude at $t_{\rm NQC}$ we found it
useful to assume the following behavior
\begin{align}
\label{eq:fit_derv_amp}
\dot{\hat{A}}_{22}^{\rm NQC}&=\omega_{22}^{\rm NQC} \left[\dot{A}_{\rm orb}^{\rm NQC}\left(\nu\right)+\dot{A}_{S}^{\rm NQC}\left(\hat{a}_{\rm eff},X_{12}\right)\right].
\end{align}
The nonspinning contribution is fitted using the following rational function
\begin{align}
\label{eq:fit_derv_amp_orb}
\dot{A}_{\rm orb}^{\rm NQC}\left(\nu\right)= -\frac{N_0^{\dot{A}^{\rm NQC}_{\rm orb}} + N_1^{\dot{A}^{\rm NQC}_{\rm orb}}\nu}{1+D^{\dot{A}^{\rm NQC}_{\rm orb}}_1\nu}.
\end{align}
The spin-dependence is similarly fitted with a rational function of the form
\begin{align}
\label{eq:A_NQC_spin}
\dot{A}_{S}^{\rm NQC}=\frac{n^{\dot{A}_{\rm NQC}}\hat{a}_{\rm eff} }{1+d^{\dot{A}_{\rm NQC}}\hat{a}_{\rm eff} },
\end{align}
where the $\nu$ dependence is encoded in the
functions $({n^{\dot{A}_{\rm NQC}},d^{\dot{A}_{\rm NQC}}})$
as second-order polynomials in $X_{AB}$ as defined in 
Eq.~\eqref{eq:spin_dm_funcs_n}~--~\eqref{eq:spin_dm_funcs_d}.
The explicit values of the coefficients are listed in the
second column of Table~\ref{tab:NQC}.

We now turn our attention to the NQC frequency.
We again consider a factorization as
\begin{align}
\label{eq:fits_freq_full2}
M\omega_{22}^{\rm NQC}\left(\nu ;\hat{S}\right)&=M\omega_{\rm orb}^{\rm NQC}\left(\nu\right)\omega_{S}^{\rm NQC}\left(\hat{S},X_{AB}\right),
\end{align}
in which the nonspinning contribution is given by
\begin{align}
\label{eq:freq_orb_NQC}
M\omega_{\rm orb}^{\rm NQC}\left(\nu\right) = c_0^{\omega^{\rm NQC}_{\rm orb}} + c_1^{\omega^{\rm NQC}_{\rm orb}}\nu + c_2^{\omega^{\rm NQC}_{\rm orb}}\nu^2.
\end{align} 
The spin factor is represented with the usual rational function
\begin{align}
\label{eq:freq_spin_NQC}
\omega_{S}^{\rm NQC}=\frac{1 - n^{\omega^{\rm NQC}}(\nu) \hat{S}}{1-d^{\omega^{\rm NQC}}(\nu) \hat{S}},
\end{align}
where the functions $(n^{\omega^{\rm NQC}},d^{\omega^{\rm NQC}})$ are, as for the amplitude, quadratic
functions of $X_{AB}$, as defined in 
Eq.~\eqref{eq:spin_dm_funcs_n}~--~\eqref{eq:spin_dm_funcs_d}.
The coefficients are listed in the third column of Table~\ref{tab:NQC}.

The time derivative of the frequency is fitted from the following factorized ansatz
\begin{align}
  \dot{\omega}_{22}^{\rm NQC}&=\dot{\omega}_{\rm orb}^{\rm NQC}\left(\nu\right)
  \dot{\omega}_{S}^{\rm NQC}\left(\hat{S},X_{AB}\right),
\end{align}
where the nonspinning part is
\begin{align}
\label{eq:df_orb_NQC}
\dot{\omega}_{\rm orb}^{\rm NQC}\left(\nu\right)= \frac{N_0^{\dot{\omega}^{\rm NQC}_{\rm orb}} +N_1^{\dot{\omega}^{\rm NQC}_{\rm orb}}\nu}{1+D_1^{\dot{\omega}^{\rm NQC}_{\rm orb}}\nu}.
\end{align}
Finally, the spin-dependent correction is fitted with
a quadratic polynomial in $\hat{S}$ as
\begin{align}
\label{eq:df_spin_NQC}
\dot{\omega}_{S}^{\rm NQC}\left(\hat{S};\,X_{AB}\right)=1+ a_{\dot{\omega}^{\rm NQC}}\left(\nu\right)\hat{S} + b_{\dot{\omega}^{\rm NQC}}\left(\nu\right)\hat{S}^2,
\end{align}
where the coefficients $(a_{\dot{\omega}^{\rm NQC}}, b_{\dot{\omega}^{\rm NQC}})$ are represented, as above, with
quadratic functions of $X_{AB}$.  The corresponding coefficients are listed in the fourth column of Table~\ref{tab:NQC}. 


 \bibliography{references,local,spec_refs}

\end{document}